\documentclass[twocolumn,trackchanges]{aastex631}
\newcommand{\teff}{$T_{\rm eff}$}
\newcommand{\bprp}{$G_{\rm BP}-G_{\rm RP}$}
\newcommand{\prot}{$P_{\rm rot}$}
\newcommand{\sigmavz}{$\sigma_{vz}$}

\usepackage{amsmath}
\usepackage{comment}
\begin{document}

\title{Anchoring Stellar Age Indicators: A Cross-Calibration of [C/N] and Gyrochronology Ages via the Age-Velocity-Dispersion Relation}

\correspondingauthor{Yuxi(Lucy) Lu}
\email{lucylulu12311@gmail.com}

\newcommand{\osu}{Department of Astronomy, The Ohio State University, Columbus, 140 W 18th Ave, OH 43210, USA}
\newcommand{\ccapp}{Center for Cosmology and Astroparticle Physics (CCAPP), The Ohio State University, 191 W. Woodruff Ave., Columbus, OH 43210, USA}

\author[0000-0003-4769-3273]{Yuxi(Lucy) Lu}
\affiliation{\osu}
\affiliation{\ccapp}

\author[0000-0002-7549-7766]{Marc H. Pinsonneault}
\affiliation{\osu}
\affiliation{\ccapp}

\author[0000-0001-5082-9536]{Yuan-Sen Ting}
\affiliation{\osu}
\affiliation{\ccapp}

\author[0009-0009-4567-9946]{Phil R. Van-Lane}
\affiliation{David A. Dunlap Department of Astronomy \& Astrophysics, University of Toronto, Toronto, ON, Canada}
\affiliation{Dunlap Institute of Astronomy \& Astrophysics, University of Toronto, Toronto, ON, Canada}

\author[0000-0002-2854-5796]{John D Roberts}
\affiliation{\osu}
\affiliation{\ccapp}

\author[0000-0002-4818-7885]{Jamie Tayar}
\affiliation{Department of Astronomy, University of Florida, Gainesville, FL 32611, USA}

\author[0000-0003-4761-9305]{Alexander Stone-Martinez}
\affiliation{Department of Astronomy, New Mexico State University, P.O.Box 30001, MSC 4500, Las Cruces, NM, 88033, USA}



\begin{abstract}
Determining stellar ages is challenging, as it depends on other stellar parameters in a non-linear way and often relies on stellar evolution models to infer the underlying relation between these parameters and age. 
This complexity increases when comparing different age-dating methods, as they rely on distinct indicators and are often applicable to non-overlapping regions of the color-magnitude diagram. 
Moreover, many empirical calibration methods rely on pre-determined ages, often from open clusters or asteroseismology, which only cover a limited parameter space.
Fortunately, the age-velocity-dispersion relation (AVR), in which the velocity dispersion increases with age, is a universal feature among stars of all evolutionary stages. 
In this paper, we 1) explore the parameter space in which [C/N] and gyrochronology are applicable, extending beyond the domains probed by asteroseismology and open clusters, and 2) assess whether the traditionally assumed [C/N] and gyrochronology relations yield ages on a consistent physical scale, after calibrating both using the same AVR.
We find gyrochronology can be applied to all partially convective stars after they have converged onto the slow rotating sequence and before they experience weakened magnetic braking; [C/N] can be used to infer ages for all giants with metallicity $>$ -0.8 dex and [C/N] $<$ -0.05 dex, and can be used as an age-indicator down to [Fe/H] of -1 dex if only selecting the low-$\alpha$ disk.
Lastly, ages obtained from [C/N] and gyrochronology agree within uncertainty after accounting for systematic offsets. 

\end{abstract}



\section{Introduction} \label{sec:intro}
In the recent decade, astronomers have started to obtain large quantities of stellar ages for individual field stars thanks to large spectroscopy and photometric surveys such as Gaia \citep{gaia}, Kepler \citep{kepler}, K2 \citep{K2}, TESS \citep{TESS}, APOGEE \citep{apogee}, GALAH \citep{galah}. 
With current and near-future surveys such as Milky-Way-Mapper \citep{MWM}, LSST \citep{LSST}, and Roman \citep{Spergel2015}, stellar ages will continue to be one of the most important stellar parameters we can infer to better our understanding of stellar physics, planet, galaxy formation, and galaxy evolution.

With the new Gaia Data Release 3, about 30 million stars now have reliable 6--D kinematic information, and with the final Gaia data release, this number will reach about 150 million.
As a result, kinematics is becoming an increasingly viable tool for stellar population studies. 
In particular, the age--velocity--relation \citep[AVR; e.g.,][]{Spitzer1951, Lacey1984,sellwood2014, Bird2013}, the fact that observationally, the velocity dispersion of stars increases with age, can be used to determine ages for populations of stars.
Since the AVR is consistent for stars of all evolutionary stages, it can be used to estimate ages for any population, especially for when other stellar age estimates are not possible or reliable. 
For example, kinematic ages have been used to understand the true nature of young high--$\alpha$ stars \citep[e.g.,][]{Zhang2021, Jofre2023, Lu2024_yhalpha} and intermediate-age RR Lyrae stars \citep[e.g.,][]{Iorio2021}, or to understand the spin-down of M dwarfs \citep[e.g.,][]{Irwin2011, CortesContreras2024, Lu2024_abrupt}.
Stellar kinematics have also been used to estimate ages for exoplanet host stars \citep{Sagear2024} and calibrate gyrochronology \citep{Lu2020_kinage, Lu2023gyro}.

Besides kinematic ages, other popular age-dating methods suitable for single field stars in smaller parameter space include but are not limited to asteroseismology \cite[e.g.][]{Chaplin2014, Silva2017, Pinsonneault2018, Pinsonneault2024}, isochrone fitting \cite[e.g.][]{Berger2020, Xiang2022, Kordopatis2023, Queiroz2023}, spectra fitting \cite[e.g.][]{Martig2016, Ness2016, Mackereth2019, Casali2019, Lu2021_disk, Spoo2022}, and gyrochronology \cite[e.g.][]{Barnes2010, Angus2015, Lu2020_kinage, Bouma2023, Gaidos2023, Lu2023gyro, VanLane2024}.
These age-dating methods are in general, more accurate and precise than kinematic ages, but each of them relies on different physical processes, and is not applicable for stars of all evolutionary stages.
Moreover, age catalogs from various methods can display significant variance and systematic offsets \citep[e.g.,][]{Lu2020_kinage}. 
A key question, then, is whether age estimates from different methods can be brought onto a consistent scale—particularly if all methods are calibrated to the same underlying assumption --- stars of all evolutionary stages should exhibit the same AVR. 
Even though the AVR is intrinsically a loose relation, we propose that it can serve as a common physical anchor to cross-calibrate tighter empirical age relations, such as those derived from gyrochronology and [C/N] abundances in giants.

A further challenge with many age-dating methods is their strong dependence on empirical calibration, especially in the cases of gyrochronology and [C/N] age-dating.
Both calibration of gyrochronology and [C/N]-age relations are mostly done using open clusters \citep[e.g.,][]{Spoo2022, Bouma2023, VanLane2024} and asteroseismic ages \citep[e.g.,][]{Angus2015, Saunders2024, StoneMartinez2024, Roberts2024}.
Kinematic ages have also recently been used to calibrate gyrochronology \citep[e.g.,][]{Angus2020, Lu2023gyro, Lu2020_kinage}, and have already yielded huge success beyond the parameter spaces where clusters and asteroseismic ages live.
However, the limitations of these empirical-calibrated methods can only be explored in the parameter space that the age calibrators live.
In this paper, we examine the extent to which kinematic ages can reveal the limitations and enable the expansion of these empirical age relations into previously inaccessible parts of parameter space.

Much work has already been devoted to testing and cross-calibrating age-dating methods and identifying their systematic uncertainties.
For example, \cite{Byrom2024} show the limitation of stellar models for low-mass, main-sequence stars when they find a large uncertainty and strong bias compared to the cluster age for the low-mass dwarfs if they infer ages as they are individual field stars, despite the underlying stellar structure model.
\cite{Tayar2025} also recently find that asteroseismic ages for individual stars are in tension with the cluster ages.
Other studies have incorporated stellar rotation or asteroseismology into stellar interior models \citep[e.g.,][]{Townsend2013, vansaders2016, Angus2019, Amard2019}, or combined lithium depletion and gyrochronology for young stars \citep[e.g.,][]{Bouma2024}.
There have also been many other important studies on understanding the limitation and offsets of different stellar interior models \citep[e.g.,][]{Tayar2017, SilvaAguirre2020, Byrom2024}, and in particular, their effect on asteroseismic mass or age determinations \citep[e.g.,][]{Silva2017, Pinsonneault2024}.
Collectively, these studies form the foundation for efforts to unify age-dating methods across evolutionary phases.

In this paper, we will continue this line of work. 
We aim to 1) explore the parameter space in which [C/N] and gyrochronology are applicable, extending beyond the domains of asteroseismology and open clusters, and 2) assess whether the traditionally assumed [C/N] and gyrochronology relations yield ages on a consistent physical scale, after calibrating both using the same AVR obtained from APOKASC--3 \citep{Pinsonneault2024}.
We describe the data used in Section~\ref{sec:data}, and our methodology to calibrate both [C/N] and gyrochronology using the AVR in Section~\ref{sec:method}. 
In Section~\ref{sec:stelphys}, we apply our method to observational data to address point 1), and in Section~\ref{sec:comp}, we evaluate point 2) by comparing calibrated ages against asteroseismic benchmarks, wide binaries, and open clusters. 
We discuss limitations and future extensions in Section~\ref{sec:future}.

\section{Data} \label{sec:data}
The asteroseismic data was obtained from the APOKASC--3 catalog \citep{Pinsonneault2024}.
We used the recommended ages provided in Table 4 of their paper. 
The [C/N] and [Fe/H] abundances, log$g$, and temperature measurements are taken from APOGEE DR17 \citep{Abdurrouf2022} to be mostly consistent with the APOKASC--3 catalog. 
Period measurements are taken from various publicly available catalogs in the literature \citep{McQuillan2014, Santos2021, Holcomb2022, Lu2022_prot, Lu2023gyro, Colman2024}, which include measurements from Kepler, the Zwicky Transient Facility \citep[ZTF;][]{Bellm2019, ztfdata, ztftime}, and TESS. 

The 6-D kinematic properties, including the Galactocentric velocities ($v_r$, $v_\phi$, and $v_z$), are calculated from Gaia DR3 measurements (RA, Dec., parallax, proper motions, and radial velocity) \citep{Gaiadr3} by transforming from the solar system barycentric ICRS reference frame to the Galactocentric Cartesian and cylindrical coordinates using \texttt{astropy} \citep{astropy:2013, astropy:2018, astropy2022}, using updated solar motion parameters from \citet{Hunt:2022}.

\subsection{Selection for Gyrochronology Age Calibration}
To calibrate gyrochronology against APOKASC--3, we cross-matched the rotation period sample with 6-D kinematic properties from Gaia and performed the following selection:
\begin{itemize}
    \item Absolute $G$-band magnitude, $M_G >$ 4.2 dex to mainly select dwarf stars where gyrochronology is known to be applicable.
    \item RUWE $<$ 1.4 to exclude binaries.
    \item Excluding stars with \prot\ between 28-30 days to remove those affected by the systematic signal introduced by the Moon in ZTF. 
    \footnote{Excluding or including these periods does not introduce  significant alteration on the trends observed.}
\end{itemize}
This left us with $\sim$86,000 stars with rotation measurements from Kepler and ZTF and kinematic measurements from Gaia DR3. 
One may also consider further excluding the evolved stars and the equal-mass binaries using the CMD by fitting a polynomial \citep[see][for detailed steps in how we could perform this cut]{Lu2023gyro}, but doing so we risk excluding the pre-main-sequence stars, which we were interested in. 
We tested excluding stars with this method, which leaves us with $\sim$78,000 stars, and found no significant deviation from those without the cut.
We did not correct for reddening as most of these stars are close-by ($<$ 0.5 kpc), and reddening values are small for the majority of our samples.

\subsection{Selection for APOKASC--3}
The APOKASC--3 sample is used to calibrate the AVR.
We cross-matched APOKASC--3 with Gaia DR3 kinematics and obtained 15,808 stars with 6-D kinematics, from which we selected 8,037 high-quality disk stars using the following criteria:
\begin{itemize}
    \item [Fe/H] $>$ -1 dex to select disk stars.
    \item Relative mass uncertainty $<$ 20\%, temperature uncertainty $<$ 50 K, [Fe/H] uncertainty $<$ 0.02 dex, and [$\alpha$/Fe] uncertainty $<$ 0.05 dex to ensure we are only using precise ages.
\end{itemize}.

\subsection{Selection for [C/N] Age Calibration}
To calibrate [C/N] ages against APOKASC--3, we cross-matched APOGEE DR17 with stars with 6-D kinematic properties from Gaia and performed the following selection:
\begin{itemize}
    \item log$g <$ 3.2 dex, \teff $<$ 5,500 K. To only select giant stars since [C/N] ages only work for stars that have gone through the first dredge-up.
    
    \item No temperature, log$g$, [Fe/H], [C/Fe], and [N/Fe] flags reported by APOGEE DR17 to only include stars with reliable stellar parameters.
    
    \item {\rm [}Fe/H] $>$ -1 dex to exclude halo stars and outliers. 

    \item {\rm [}C/N] measurements $>$ -0.5 dex to exclude stars below due to insufficient sample sizes for statistically robust conclusions.
    
    \item With scattering in multiple radial velocity (RV) measurements $<$ 1 km/s reported by APOGEE DR17 to exclude binaries. 
    
    \item The Renormalised Unit Weight Error (RUWE) $<$ 1.4. To exclude other binaries as not all stars in our sample have multiple RV measurements from APOGEE.

\end{itemize}
This left us with $\sim$110,000 giant stars from APOGEE DR17 with kinematic measurements from Gaia DR3. 

\section{Method for cross calibrating [C/N]-age and gyrochronology using the AVR} \label{sec:method}
To anchor both relations to the AVR, we started with modeling the vertical velocity distribution conditioned on age ($\tau$) and metallicity ([Fe/H]).
We assumes the vertical velocity for each star is drawn from a Gaussian distribution (denoted by $\mathcal{N}$) with a mean of 0 and a standard deviation following the vertical velocity-dispersion relation (AVR).
This can be written as:
\begin{equation}\label{eq:P_avr}
    P(v_z|\tau, {\rm [Fe/H]}) = \mathcal{N}(0, \sigma_{vz}^2(\tau, {\rm [Fe/H]}))
\end{equation}

Since the classic empirical gyrochronology relations assumes a unique mapping between rotation period (\prot), color (\bprp) and age, we can also model the vertical velocity distribution for the sample of stars with period measurements to be:
\begin{equation}\label{eq:P_gyro}
    P(v_z|P_{\rm rot}, {\rm G_{\rm BP}-G_{\rm RP}}) = \mathcal{N}(0, \sigma_{vz}^2(P_{\rm rot}, {\rm G_{\rm BP}-G_{\rm RP}}))
\end{equation}

Similarly, [C/N]-age relation assumes a unique mapping between [C/N], metallicity, and age, and thus, we can also write down the vertical velocity distribution as:
\begin{equation}\label{eq:P_cn}
    P(v_z|{\rm [C/N]}, {\rm [Fe/H]}) = \mathcal{N}(0, \sigma_{vz}^2({\rm [C/N]}, {\rm [Fe/H]})
\end{equation}

Assuming the vertical velocity distributions are the same across the parameter space of interest, we can equate the standard deviations:

\begin{equation}\notag
    \sigma_{vz}^2(\tau) =\sigma_{vz}^2(P_{\rm rot}, {\rm G_{\rm BP}-G_{\rm RP}})
\end{equation}
and
\begin{equation}\notag
\sigma_{vz}^2(\tau)=\sigma_{vz}^2({\rm [C/N]}, {\rm [Fe/H]})
\end{equation}
This will provide us with calibrated ages that are hopefully on the same physical scale.
This is essentially a coordinate transfer and the biggest underlying assumption is that all our datasets covers similar Galactic region where the averaged AVR produced with these different datasets are similar.
This might not be true while transferring from the AVR using APOKASC--3 sample to that using the full APOGEE DR17 catalog, and future work should include better selection of the data.
Using the AVR as an anchor point, we are also hoping to extrapolate the relations to where the traditional fitting methods are not suitable. 

One important assumption for this method is that it requires a unique mapping between (\prot, \bprp) and age for gyrochronology and ([C/N], [Fe/H]) and age for [C/N] age-dating.
When this holds true, there should be a monotonic increase in \sigmavz\ with increasing \prot\ for a fixed \bprp, and a monotonic increase in \sigmavz\ with increasing [C/N] for a fixed [Fe/H].
We can even leverage this assumption to find parameter spaces where the unique mapping breaks down, and thus, when the age-dating method is not applicable (see examples for this in gyrochronology in Section~\ref{subsubsec:WMB}-\ref{subsubsec:convergence})

Since gyrochronology mainly works for main-sequence stars, and [C/N] ages are only obtainable for giants, there does not exist a set of stars with all physical quantities ($v_z$, [Fe/H], \prot, \bprp, [C/N], $\tau$) available.
As a result, we will fit these distributions separately.
We fit Equation~\ref{eq:P_avr} with the APOKASC--3 sample, as it provides the most accurate ages from asteroseismology (see Section~\ref{subsec:avr}); 
We fit Equation~\ref{eq:P_gyro} with the period sample combining TESS, ZTF, and Kepler, as described in Section~\ref{sec:data} (see Section~\ref{subsec:gyro});
And we fit Equation~\ref{eq:P_cn} with the APOGEE DR17 sample with quality cuts described in Section~\ref{sec:data} (see Section~\ref{subsec:cn}).

The calibration result will be described in Section~\ref{sec:comp}.
We can also use \sigmavz\ as an age proxy to understand the limitations of gyrochronology and [C/N]-ages (see Section~\ref{sec:stelphys}) in the parameter space (e.g., slow-rotating stars, low-metallicity giants) where no other reliable ages are available.

\subsection{Fitting the Age-Velocity-Relation to APOKASC--3}\label{subsec:avr}
Observationally, the increase in velocity dispersion of stars with age has been documented for decades, but the physical origin of the age-velocity relation is most likely a complicated process that involves a combination of effects caused by interactions with giant molecular cloud \citep[GMC; e.g.,][]{Lacey1984} and dwarf satellites \citep[e.g.,][]{Quinn1993, Abadi2003, Brook2004}, upside-down formation \citep[e.g.,][]{Bird2013}, radial migration \citep[e.g.,][]{Loebman2011, Minchev2014, sellwood2014}, and other mechanisms \citep[e.g.,][]{Grand2016, Buck2020}.
The velocity dispersion increases in all three dimensions ($r, \phi, z$) in the Milky Way (MW), but the increase in the vertical direction is the most prominent. 
Predictions from heating by stars scattering off GMC takes the form of a power law, in which $\sigma_{vz} \propto \sqrt{\tau}$, in which $\sigma_{vz}$ and $\tau$ are the vertical velocity dispersion and age, respectively. 
With an increasing amount of observations in recent years, the index in the power law has been revised, and dependence on angular momentum and metallicity have also been found \citep[e.g.,][]{Yu2018, Ting2019, Sharma2021}. 

For this work, we used the vertical velocity dispersion relation that included the metallicity dependence.
Metallicity is important in both stellar spin-down \citep[e.g.,][]{Amard2020, Claytor2020, See2024} and mass determination using [C/N] abundances \citep[e.g.,][]{Martig2016}. 
However, we acknowledge the caveat that observational data has shown evidence that the AVR deviates from a simple power law for older stars $>\sim 7-10$ Gyr \citep[e.g.,][]{Mackereth2019, Sun2024}, and is a strong function of Galactic height and radius besides metallicity.
Using APOKASC--3, we could not obtain a significant number of stars $>$ 10 Gyr to confirm a deviation from a simple power law, and there is a possible mild flattening of the AVR for younger stars (see Figure~\ref{fig:7}).
We also did not include the effect of Galactic location, as the sample size of APOKASC--3 did not allow us to make reliable inferences with these extra dimensions. 
Future work should also include the dependencies on Galactic height and radius once more asteroseismic ages are available. 

As we wanted to calibrate both [C/N] ages and gyrochronology to the APOKASC--3 sample, we first determined the vertical velocity dispersion relation and adopted a similar functional form as \cite{Sharma2021}:
\begin{equation}
    \sigma_{vz}(\tau, \rm{[Fe/H]}) = a\tau^b(1+\gamma \rm{[Fe/H]})
    \label{eq:1}
\end{equation}
In which $\sigma_{vz}(\tau, \rm{[Fe/H]})$ is the vertical velocity dispersion in Equation~\ref{eq:P_avr}, $\tau$ is the stellar age, and [Fe/H] is the metallicity of the star. 

Assuming the vertical velocity of each star is independent, the log-likelihood can then be written as:
\begin{equation}\notag
\log \mathcal{L}(P(v_{z,i})) = -\frac{1}{2} \sum_{i=1}^N \left[ \log(2\pi) + \log(\sigma_{vz,i}^2) + \frac{v_{vz,i}^2}{\sigma_{vz,i}^2} \right]
\end{equation}

We then fitted for $a$, $b$, and $c$ by maximizing the log-likelihood function with Markov chain Monte Carlo using the \texttt{PYTHON} package \texttt{emcee} \citep{emcee}, assuming a uniform prior for all parameters. 
The final result yields $\mathrm{a = 10.54^{+0.42}_{-0.41}}$, $\mathrm{b = 0.44^{+0.02}_{-0.02}}$, and $\mathrm{\gamma = -0.92^{+0.04}_{-0.04}}$.
This fit resulted in a similar dependency on age, but a stronger dependency on metallicity compared to the results in \cite{Sharma2021}.

\subsection{Calibrating Gyrochronology with the Age-Velocity Dispersion Relation}\label{subsec:gyro}
Since our main goal is to determine whether the ages agree between gyrochronology and [C/N]-age relations after calibrating both using the same underlying AVR, we will assume a classical gyrochronology functional form similar to \cite{Angus2015}:
\begin{equation}
    P_{\rm rot} = c\tau^d((G_{\rm BP}-G_{\rm RP})-0.55)^f
    \label{eq:2}
\end{equation}
In which $P_{\rm rot}$ is the rotation period, $(G_{\rm BP}-G_{\rm RP})$ is the Gaia BP-RP color.
Eliminating the unknown variable, age, with Equation~\ref{eq:1}, \sigmavz\ in Equation~\ref{eq:P_gyro} can then be written as: 
\begin{multline}\notag
    {\rm ln}\sigma_{vz}(P_{\rm rot}, G_{\rm BP}-G_{\rm RP}) = \frac{b}{d}({\rm ln}(P_{\rm rot})-{\rm ln}(c)-\\
    f{\rm ln}((G_{\rm BP}-G_{\rm RP})-0.55))+{\rm ln}(a)+{\rm ln}(1+\gamma{\rm [Fe/H]})    
\end{multline}

Since the majority of our stars with period measurements do not have available APOGEE DR17 metallicity, and most of the stars with period measurements are in the solar vicinity, we chose to assume all stars have close to solar metallicity. 
Metallicity measurements from Gaia XP spectra could potentially be used, however, crosshatch with Gaia XP measurements from \cite{Andrae2023} only leave us with about 1/4 of the original sample, and it is unclear how accurate the metallicity can be for dwarf stars, especially for M dwarfs \citep[e.g.,][]{Behmard2025}.
As a result, we decided not including the metallicity in our main analysis, and instead, we used the Gaia XP metallicity measurements to validate our result.
However, in the future, we plan to add metallicity once more reliable measurements are available. 
We can then re-parameterize the equation, eliminate the metallicity term and get:
\begin{multline}\label{eq:3}
    {\rm ln}\sigma_{vz}(P_{\rm rot}, G_{\rm BP}-G_{\rm RP}) = 
    s_0+s_1{\rm ln}(P_{\rm rot})+\\
    s_2{\rm ln}(G_{\rm BP}-G_{\rm RP}-0.55))
\end{multline}
In which $s_0 = \ln a -(b/d)\ln c$, $s_1 = b/d$, and $s_2 = bf/d$.

We first attempted a joint fit with individual \bprp\ and \prot\ measurements, similar to our approach with APOKASC--3 in the previous section, but were unable to achieve a reasonable result that reproduces the AVR seen in APOKASC--3 for [Fe/H] between –0.2 and 0.2 dex.
As a result, we chose to bin the data in \prot\ and \bprp\ and calculate \sigmavz\ for stars with similar rotation and color.
To do this, we divided the rotation period data into 8 bins in \bprp\ between 0.65 and 3.
For the first 6 bins, most stars are partially convective, and we used bin edges equally spaced in logarithmic scale between 0.65 and 2.6, which is similar to equally spaced linearly in temperature. 
For the last bin, we picked the bin edges in \bprp\ to be 3 to ensure enough stars are included in each bin.
There is no significant number of stars cooler than \bprp\ $>$ 3 to extend the relation further.
We excluded stars with \bprp $<$ 0.65 as these stars correspond to a temperature of $\sim$6500K, which do not spin down much throughout their main-sequence lifetime \citep{Kraft1967,Beyer2024}. 
We then divided stars in each color bin into 49 period bins equally spaced in (${P_{\rm rot}}^2$) between the 5$^{\rm th}$-percentile and 95$^{\rm th}$-percentile of the periods in that color bin to exclude any outliers.
We chose to bin in ${P_{\rm rot}}^2$ as it is close to binning equally in age with the ``Skumanich spin-down'' \citep{Skumanich1972}.
For each period bin, we calculated \sigmavz\ by taking 1.5*MAD and only keeping bins with more than 20 stars, same as what was done in Section~\ref{subsec:avr}. 
To account for measurement uncertainty, we perturbed \bprp\ measurements by their uncertainty reported by Gaia and \prot\ measurements by 10\% and recalculated \sigmavz. 
We did this 100 times and recorded the centers of the \bprp\ and \prot\ bins, as well as the corresponding \sigmavz\ measurements.

For each color bin, we perform a linear fit with deming regression between \prot\ and \sigmavz\ in logarithmic space to minimize the bias in the linear fits caused by measurement uncertainties.
Deming regression is a technique for fitting a straight line to 2-D data where both variables are measured with error, for derivation for an analytical form and more details, see \cite{Ting2025}.
The uncertainty on the fits are determined with bootstrapping. 
Since gyrochronology only works for stars that have already converged onto the slow rotating sequence, and have not experienced weakened magnetic braking \citep{vansaders2016}, we excluded data points with period $<$ 10 days\footnote{Although some stars, especially G dwarfs, do converge $<$ 10 days.} and those with Rossby number $> 1.866$,  \citep[][]{Saunders2024}. 
Synchronized binaries also make up a significant fraction of stars with rotation periods below 10 days \citep{Lurie2017}, and these do not follow the gyrochronology relation we aim to recover. 
That said, we expect to be able to model them as a separate population, which we plan to pursue in future work.

Note that the method described here is a similar approach as described in \cite{Angus2020} and \cite{Lu2020_kinage}, which was then used to calibrate a gyrochronology relation in \cite{Lu2023gyro}.
However, in this case, do not assume a specific functional form while fitting for the gyrochronology relations for individual \bprp\ bins. 
This enable us to compare a classical gyrochronology model with a model with color-dependent power law indexes between \prot, \bprp, and age (i.e., $d$, and $f$ in Equation~\ref{eq:2} can be a function of color while fitting the relations). 

\begin{figure*}
    \centering
    \includegraphics[width=\textwidth]{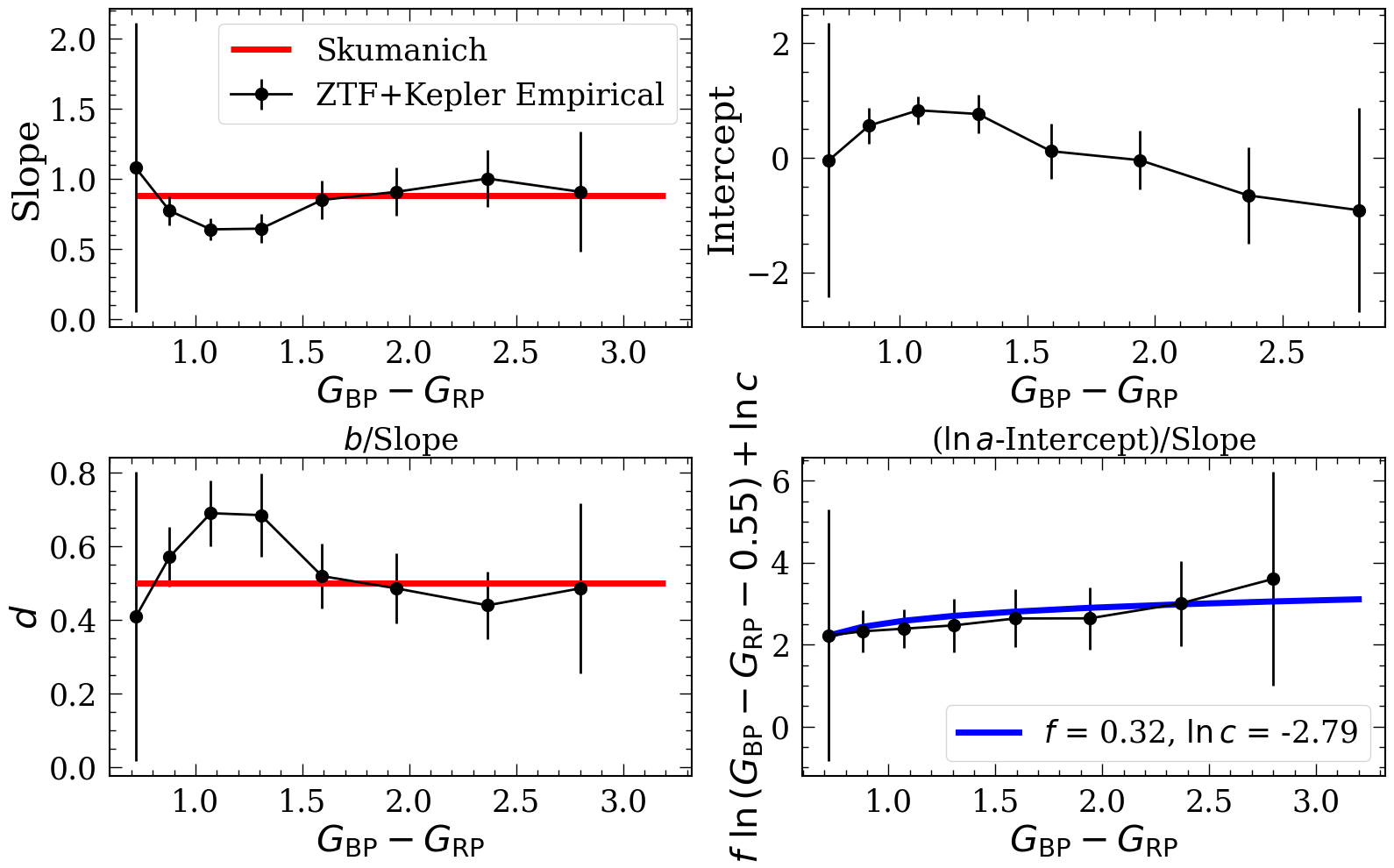}
    \caption{Top row shows the slopes and intercepts for the fit to the \sigmavz-\prot\ relations in logarithmic space for each \bprp\ bin. 
    Bottom row shows the physical parameters associated with the spin-down law in Equation~\ref{eq:2}, derived from the slopes and intercepts.
    The red solid line shows the classic ``Skumanich'' spin-down, where $d$ = 0.5, and the blue solid line shows the deming regression result for partially convective stars with \bprp\ $<$ 2.6.
    Our fitting result suggests G dwarf possibly exhibit a faster spin-down compared to KM dwarfs. 
    We also find the traditional power-law form for the color-dependency of stellar spin-down agrees with observational data within uncertainty.}
    \label{fig:1}
\end{figure*}

From Equation~\ref{eq:3}, the slope for each color bin is $s_1 = b/d$, which contains information on the power index for age \citep[e.g., $d$ = 0.5 is the classic ``Skumanich'' spin-down][]{Skumanich1972}.
The intercept is then $s_0+s_2\ln{(G_{\rm BP}-G_{\rm RP}-0.55)}$, which contains the normalization constant and information on the power index for temperature.
Figure~\ref{fig:1} top row shows the slopes and intercepts for the fits, and the values are shown in Table~\ref{tab:1}.
The bottom row in Figure~\ref{fig:1} shows the parameters derived using the slopes and intercepts.
If the functional form (Equation~\ref{eq:2}) we assumed is correct, then $d$ should be constant across all color bins, and a linear relation should exist between $\ln$(\bprp) and ($\ln(a)$-Intercept)/Slope, which is shown on the $y$-axis of the bottom right plot.
The age power-law index for gyrochronology, $d$, the color power-law index for gyrochronology, $f$, and the normalization constant $\ln{c}$, can be calculated using the equations shown in the titles of the plots in the bottom row, in which $b$ and $\ln{a}$ are parameters determined for the AVR in Section~\ref{subsec:avr}.

\begin{table}[h!]
    \centering
    \caption{Table for the slopes and intercepts obtained from the \prot-\sigmavz\ relation in individual \bprp\ bins. 
    The values in this table can be used to infer stellar ages with equation~\ref{eq:8_gyro}.
    The ages in this work are inferred using the optimized values and performing linear interpolation between the color bins.}
    \begin{tabular}{c|c|c}
    \hline
    \hline
        \bprp\ center & Slope & Intercept \\
        \hline
        0.72 & $1.08^{+1.03}_{-1.03}$ & $-0.04^{+2.39}_{-2.39}$ \\
        0.88 & $0.77^{+0.10}_{-0.10}$ & $0.56^{+0.31}_{-0.31}$ \\
        1.07 & $0.64^{+0.08}_{-0.08}$ & $0.83^{+0.24}_{-0.24}$ \\
        1.31 & $0.64^{+0.10}_{-0.10}$ & $0.77^{+0.33}_{-0.33}$ \\
        1.59 & $0.85^{+0.14}_{-0.14}$ & $0.12^{+0.48}_{-0.48}$ \\
        1.94 & $0.91^{+0.17}_{-0.17}$ & $-0.04^{+0.52}_{-0.52}$ \\
        2.37 & $1.00^{+0.21}_{-0.21}$ & $-0.66^{+ 0.84}_{- 0.84}$ \\
        2.80 & $0.91^{+0.43}_{-0.43}$ & $-0.91^{+1.79}_{-1.79}$ \\
    \hline
    \end{tabular}
    \label{tab:1}
\end{table}

The bottom left plot in Figure~\ref{fig:1} suggests the age-dependent power-law index is not constant across all temperatures.
K and M dwarfs follow a ``Skumanich'' spin-down while G dwarfs spin down faster than K and M dwarfs.
This could suggest a transition between how the angular momentum is transported between the core and the envelope from G to K dwarf.
The bottom right plot in the same figure suggests the traditional power-law form for the color-dependency of stellar spin-down agrees with observational data within uncertainty.

With the slope and intercepts for each color bin, the age can then be inferred using:
\begin{multline}\label{eq:8_gyro}
    \ln{(\tau\ {\rm [Gyr]})} = \frac{({\rm Inter.}+{\rm Slope}\times\ln{(P_{\rm rot})}-\ln{(a)}}{b}
\end{multline}
In which $\mathrm{a = 10.54^{+0.42}_{-0.41}}$ and $\mathrm{b = 0.44^{+0.02}_{-0.02}}$ are the optimized parameters in equation~\ref{eq:1} from fitting the AVR using APOKASC--3, as described in section~\ref{subsec:avr}.

The ages are obtained with the optimized values, with a linear interpolation between the slope and intercepts shown in Table~\ref{tab:1}.
The uncertainties are the standard deviation from recalculating the ages 100 times with the parameters perturbed within their respective uncertainties. 
We provide a step-by-step age inference for one star using this method in the section below.

\subsubsection{Step-by-step Inference for an Example Star with Gyrochronology}\label{sec:A1}
Here we gave an example of how we infer the age and associated uncertainty for a star with rotation period of 21$\pm$2.1 (10\% uncertainty) and \bprp\ of 1.306$\pm$0.05 (middle of a \bprp\ bin with a typical color uncertainty from Gaia).

\begin{figure*}
    \centering
    \includegraphics[width=\textwidth]{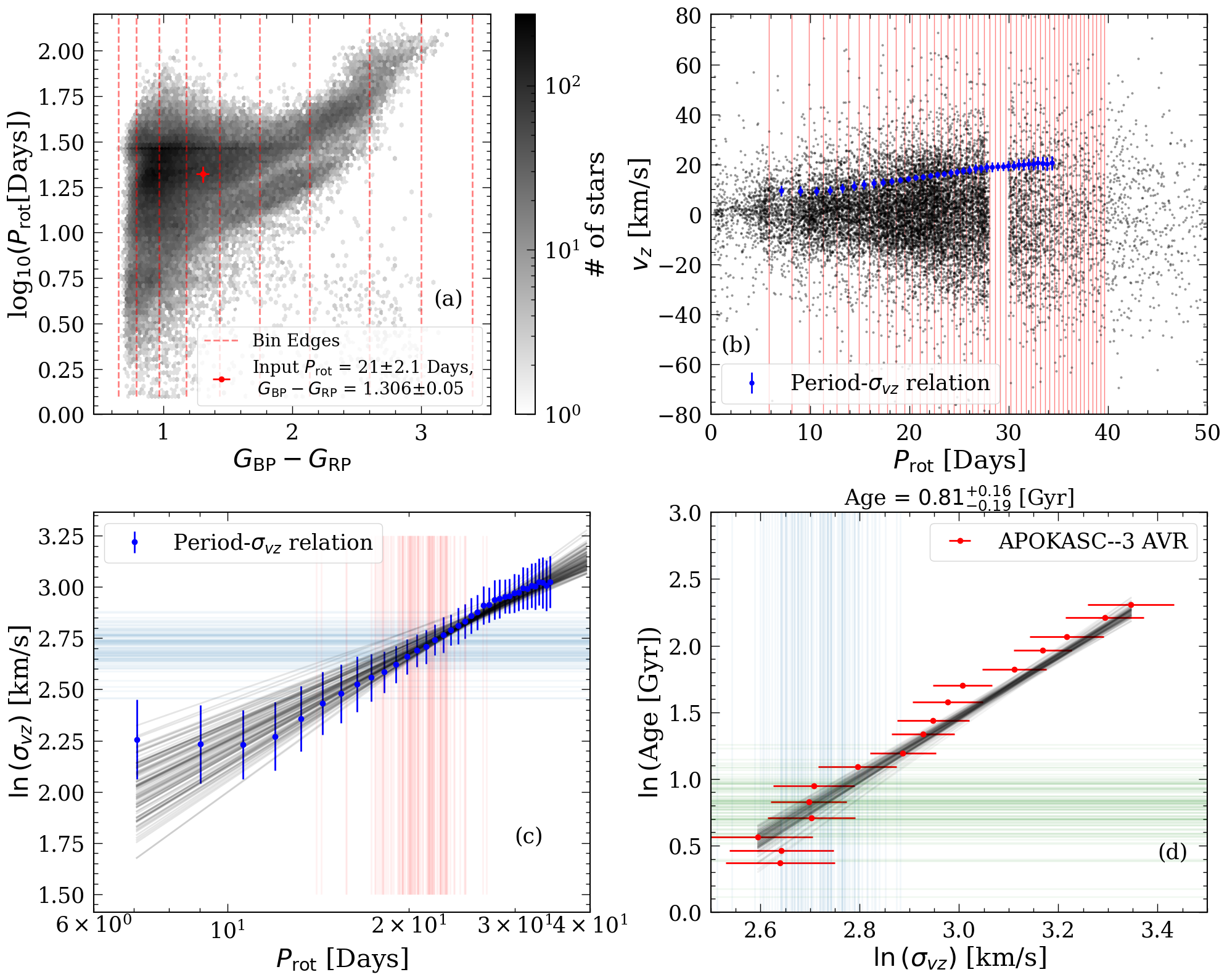}
    \caption{(a): \bprp-\prot\ histogram of the full sample of rotation periods obtained from Kepler, ZTF, and TESS.
    The red dashed line shows the bin edges used to obtain the \prot-\sigmavz\ relations.
    The red point shows the example star to infer age.
    (b): the \prot-$v_z$ scatter plot for the color bin the example star is in. 
    The red lines show the bin edges used to obtain the \prot-\sigmavz\ relation for this color bin.
    The blue points with errorbars in both plot (b) and (c) show the resulting \prot-\sigmavz\ relation for this color bin.
    \sigmavz\ is calculated for each narrow \prot\ bin (as indicated in the red vertical lines) by taking the 1.5*(Median Absolute Deviation) of the vertical velocities (as shown as the underlying black points) in each bin.
    (c): the black lines show the deming regression results. 
    The 100 realizations of the period perturbed within its uncertainty for the example star are shown in red vertical lines, and the corresponding \sigmavz\ taking into account the uncertainty of the fit is shown as the blue horizontal lines.
    This is the expected \sigmavz\ for this star.
    (d): the red points show the AVR from APOKASC--3, and the black lines show the 100 realization of the relation, assuming solar metallicity.
    The blue vertical lines show the 100 realization obtained from plot (c), and the green horizontal lines show the final inferred age posterior using the method described in this work.}
    \label{fig:A1}
\end{figure*}

To infer the age, we first obtain the \prot-\sigmavz\ relation for each color bin according to the description in Section~\ref{subsec:gyro}.
Plot (b)-(c) in Figure~\ref{fig:A1} shows one of the examples to obtain such a relation for stars with \bprp\ between 1.18 and 1.44.
The color bin edges are shown in red dashed line in plot (a), and the rotation edges to obtain the \prot-\sigmavz\ relation are shown in red lines in plot (b).
The blue points in (b) and (c) show the resulting relation, and the black lines in plot (c) show the deming regression results for this color bin to fit the \prot-\sigmavz\ relation in logarithmic space.
Plot (d) red points and black lines show the AVR and fits obtained with APOKASC--3, as described in Section~\ref{subsec:avr}.
With these relations, we identify the \bprp\ bin the star is in.
In this case, the star is at the center of the bin so no linear interpolation between the \prot-\sigmavz\ relation is needed.
The 100 realizations of the period are shown in red vertical lines in plot (c), and the corresponding \sigmavz\ taking into account the uncertainty of the fit is shown as the blue horizontal lines.
This is the expected \sigmavz\ for this star.
We then move to plot (d) where the 100 realizations of the expected \sigmavz\ posteriors are shown as the blue vertical lines, and the corresponding age posteriors are shown in the green horizontal lines, taking into account the uncertainty in the AVR best-fit parameters.

\subsection{Calibrating [C/N] ages with the Age-Velocity Dispersion Relation}\label{subsec:cn}
\begin{figure*}
    \centering
    \includegraphics[width=\textwidth]{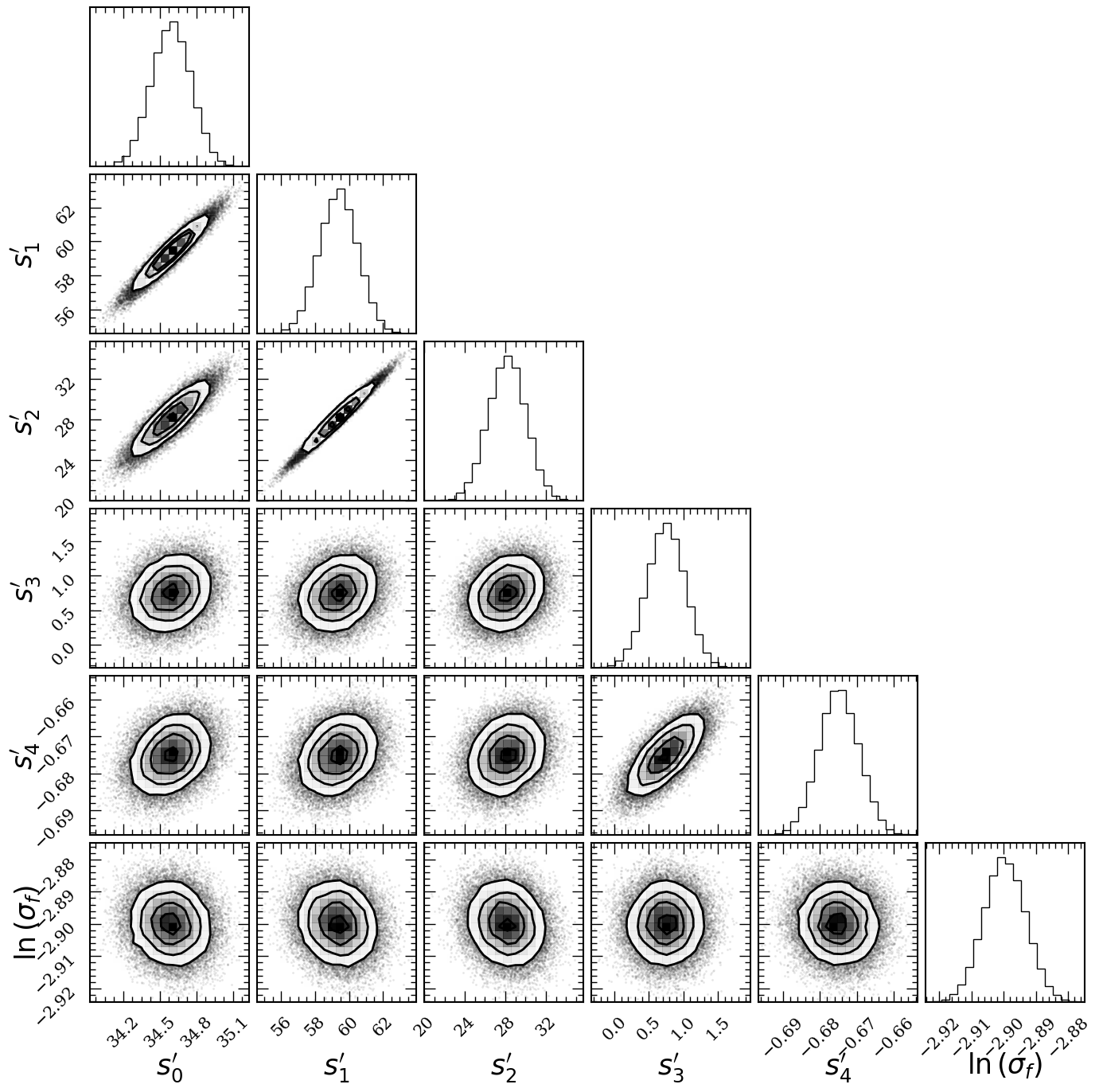}
    \caption{The posterior distribution for the optimized parameters in Equation~\ref{eq:8} used to obtain the [C/N]-age relations.
    $\sigma_f$ is a parameter that accounts for the underestimated variance fraction in the likelihood.
    The data we performed the fitting to can also be visualized in Figure~\ref{fig:4} plot (a).}
    \label{fig:2}
\end{figure*}

Similar to calibrating gyrochronology with the AVR, we also assume a polynomial functional form for the [C/N] age relation:
\begin{multline}\label{eq:5}
    \tau = s_0+s_1{\rm[C/N]}+s_2{\rm[C/N]}^2+s_3{\rm[C/N]}{\rm[Fe/H]}
\end{multline}
We adopted this functional form from \cite{Roberts2024} after eliminating terms that are small for simplicity.
We also fitted age in linear space in attempt to increase the dynamically range as [C/N] are known to mostly applicable for intermediate-age stars \citep[between about 2-10 Gyr,][]{Roberts2024}. 
Combining with Equation~\ref{eq:1}, we can obtain:
\begin{multline}\notag
    {\rm ln}\sigma_{vz} = \\
    b\ln{(s_0+s_1{\rm[C/N]}+s_2{\rm[C/N]}^2+
    s_3{\rm[C/N]}{\rm[Fe/H]})}+
    \\(1+\gamma{\rm [Fe/H]})+{\rm ln}(a)
\end{multline}
We can re-parameterizing the equation to ensure no strong dependencies exist for the parameters in use, and we get:
\begin{multline}\label{eq:8}
    \ln\sigma_{vz} =\\
\ln{(s_0'+s_1'{\rm[C/N]}+s_2'{\rm[C/N]}^2+s_3
'{\rm[C/N]}{\rm[Fe/H]})}\\
    +\ln{(1+s_4'{\rm[Fe/H]})}
\end{multline}
We performed the same procedure as Section~\ref{subsec:gyro}, but instead of separating stars in \bprp\ bins, we separated them in 16 metallicity bins, equally spaced between -1 dex and 0.45 dex.
We then measured the velocity dispersion with the width of the bin in [C/N] to be 0.1 dex, spanning between -0.5 dex to 0.1 dex, matching the uncertainty cut we performed in Section~\ref{sec:data}.
We performed the fit for all giants, as this enabled us to have more stars in each metallicity bin.
We also did this because, as will show in Section~\ref{subsec:cn-vz-rels}, we see no significant difference between the [C/N]-\sigmavz\ relations for red clump stars (RC) and lower red giant branch (RGB) stars except in the lowest metallicity bin.
Figure~\ref{fig:2} shows the resulting parameters from fitting Equation~\ref{eq:8}, using the MCMC code, \texttt{emcee} \citep{emcee}.
$\sigma_f$ is a parameter that accounts for the underestimated variance fraction in the likelihood, assuming its distribution is Gaussian. 

We can then infer ages for the giant stars using the optimized parameters:
\begin{multline}\label{eq:10}
    \tau\ {\rm [Gyr]} = \\
(\tfrac{(s_0'+s_1'{\rm[C/N]}+s_2'{\rm[C/N]}^2+s_3
'{\rm[C/N]}{\rm[Fe/H]})(1+s_4'{\rm[Fe/H]})}
    {a(1+\gamma*{\rm [Fe/H]})})^{1/b}
\end{multline}
In which $s_0'=34.59^{+0.16}_{-0.17}$, $s_1'=59.31^{+1.16}_{-1.15}$, $s_2'=28.07^{+1.86}_{-1.85}$, $s_3'=0.75^{+0.27}_{-0.28}$, $s_4'=-0.675^{+0.005}_{-0.005}$.
$b$, $a$, and $\gamma$ are parameters determined for the AVR in Section~\ref{subsec:avr}.
The ages and the associated uncertainties are obtained from recalculating the ages 100 times with the parameters and measurements perturbed within their respective uncertainties and taking the median as the age, and the 84th- and 16th-percentile as the uncertainties. 

\section{Understanding the limitation of age-dating methods with kinematics}
\label{sec:stelphys}
For this section, we will analyze the relation between \prot\ (or [C/N]) as a function of \sigmavz\ in different \bprp\ (or [Fe/H]) bins.
We will show how these relations, purely obtained from observables, can uncover interesting stellar physics and be used to understand the physical limitations of these age-dating methods, assuming the AVR does not vary strongly across \bprp\ bins for gyrochronology, and the metallicity dependency is accounted for using the APOKASC--3.
We will also show the calibrated age-\sigmavz\ relations to understand any deviation from the assumed model.

\subsection{\prot--\sigmavz\ Relations in Individual \bprp\ Bins}
\begin{figure*}
    \centering
    \includegraphics[width=\textwidth]{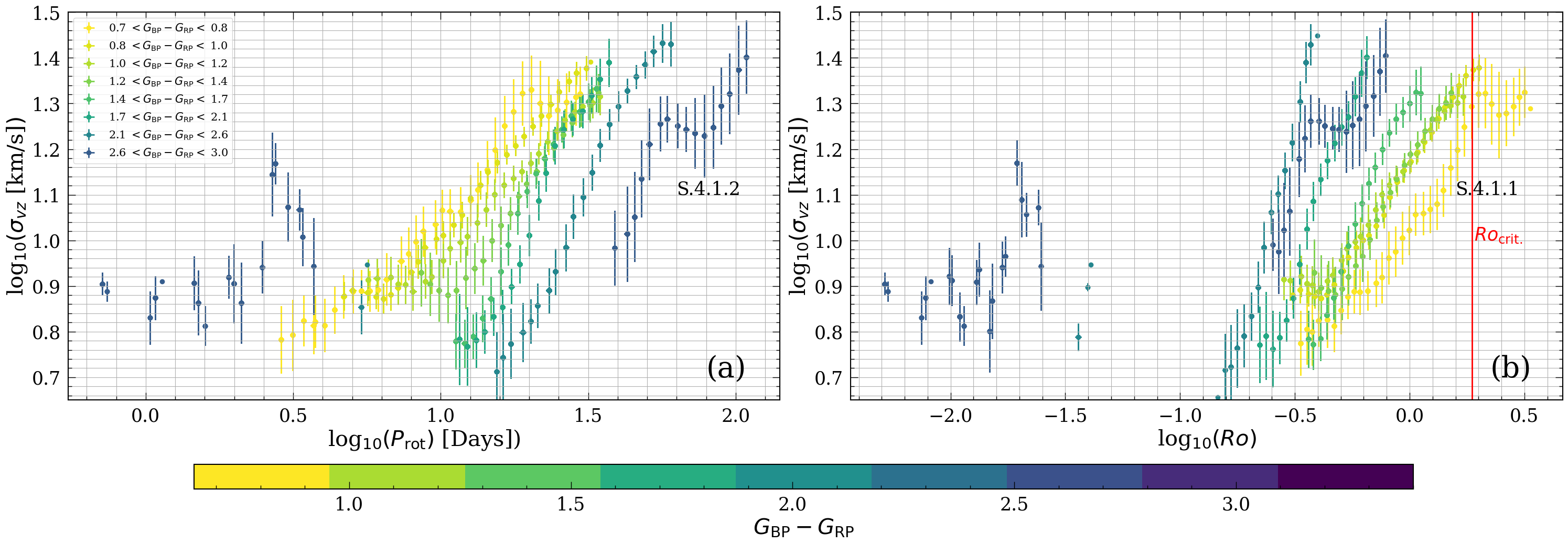}
    \caption{(a): \prot--\sigmavz\ relations for individual \bprp\ bins using the method described in Section~\ref{subsec:gyro}.
    (b): $Ro$--\sigmavz\ relations, where $Ro$:=\prot$\tau_c$.
    $Ro$ is converted from \prot\ in (a) using the convective turnover time, $\tau_c$, calculated with a relation described in \cite{See2024}.
    The red solid line shows the critical Rossby number where weakened magnetic braking is in effect \citep[$Ro_{\rm crit}$=1.866][]{vansaders2016, Saunders2024}.
    The sections corresponding to the the features are shown on the figures.}
    \label{fig:3}
\end{figure*}

We follow the same procedure as described in Section~\ref{subsec:gyro} to calculate the  \prot--\sigmavz\ relations for individual \bprp\ bins. 
Figure~\ref{fig:3} shows the result, where (a) shows the \prot--\sigmavz\ relations colored by the \bprp\ of the center of the color bins, (b) shows the $Ro$--\sigmavz\ relations.
$Ro$ is the Rossby number, which is defined as the rotation period, normalized by the convective overturn timescale ($\tau_c$), which is a proxy for magnetic activity and convection zone depth. 
We calculated the $Ro$ using the equation stated in \cite{Lu2024_abrupt}, taken from See et al. in prep.
$Ro$ is typically used as a direct indicator of stellar spin-down \citep[e.g.,][]{Kawaler1988}.

We excluded stars with age $<$ 0.01 Gyr, as those stars correspond to the fast rotating M dwarfs with Rossby number $<$ 0.1 in plot (b) that has not converged.
The points in these plots are calculated with a running median after taking into account measurement uncertainties in \prot\ \citep[assuming 10\% uncertainty,][]{Epstein2014} and color, and the error bars show the standard deviation of the measurements in each \prot\ bin while calculating the running median. 
We converted \prot\ into $Ro$ after measuring the \prot--\sigmavz\ relation in individual \bprp\ bins instead of recalculating the relations in $Ro$--\sigmavz\ space so that we can keep the results mostly model independent.
We want to analyze the results in $Ro$--space as $Ro$ is more directly associated with the magnetic dynamo and the angular momentum transport of the stars \citep[e.g.,][]{Kawaler1988, vansaders2016}.

\begin{figure*}
    \includegraphics[width=\textwidth]{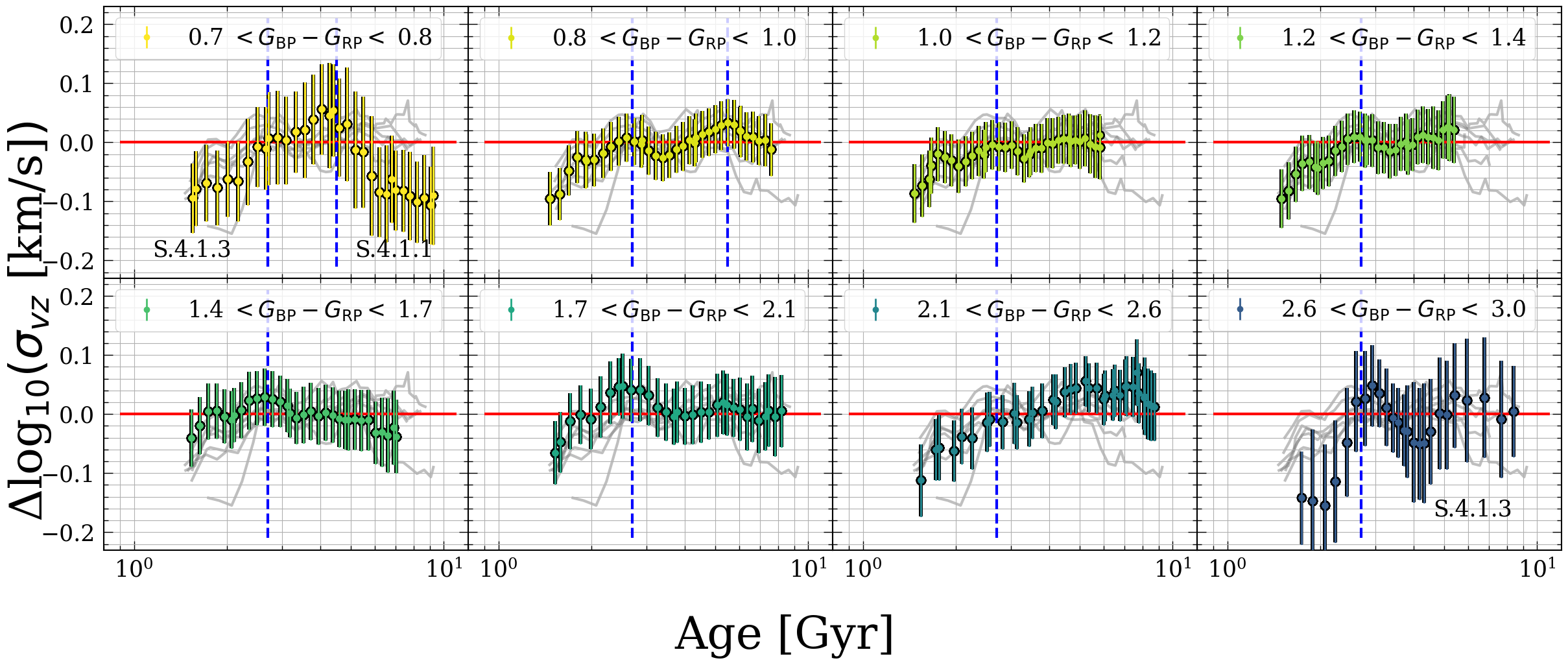}
    \caption{The difference in \sigmavz\ in various color bins between the \sigmavz-age relation determined from our gyrochronology ages in this work and that from APOKASC--3.
    This is similar to first getting ages with the steps described in Figure~\ref{fig:A1}, then calculating the age-velocity-dispersion relation with the ages inferred, then subtracting the AVR from APOKASC--3 shown in red in plot (d) of Figure~\ref{fig:A1}, which is {\it not} a linear relation as assumed.
    This shows how well our method can be applied to infer gyrochronology ages. 
    The grey lines in the background shows the difference for all color bins for better comparison. 
    The vertical blue dashed lines at 2.7 Gyr shows the location where a excess in \sigmavz\ can be seen using our gyrochronology ages compared to using the ages from APOKASC--3. 
    This excess becomes more significant as moving towards lower-mass stars.
    The location of this bump is close to the intermediate period gap.
    The second vertical blue dashed lines at older ages for \bprp\ $<$ 1 shows the location near $Ro = Ro_{\rm crit}$ in plot (b) of Figure~\ref{fig:3}. 
    The errorbars combine the uncertainties in the \sigmavz-gyrochronology age relation and the \sigmavz-APOKASC--3 age relation.
    The combined uncertainty is calculated by addition in quadrature.
    The sections corresponding to the the features are shown on the figures.}
    \label{fig:3_2}
\end{figure*}

Figure~\ref{fig:3_2} shows the difference in \sigmavz\ using the age-\sigmavz\ relations determined with our gyrochronology ages calibrated in the last section and that from APOKASC--3 for various color bins.

In the next few sessions, we will examine these relations to understand the limitation of gyrochronology and their associated physical implications.

\subsubsection{Weakened Magnetic Braking}\label{subsubsec:WMB}
The most obvious feature is the dramatic upturn right before and a flattening after the hottest stars at the critical Rossby number \citep[$Ro_{\rm crit}$=1.866;][]{Saunders2024} in the $Ro$--\sigmavz\ relation shown in plot (b) in Figure~\ref{fig:3}.
$Ro_{\rm crit}$ is where stars experience weakened magnetic braking and stop spinning down \citep[e.g.,][]{vansaders2016, Saunders2024}.
This process seems to be associated with a fast decrease of the wind torque around $Ro_{\rm crit}$ \citep[e.g.,][]{Metcalfe2024, Metcalfe2025}.
If this is the case, stars with $Ro>Ro_{\rm crit}$ will be a mix of ages that are older than the age that requires a star to reach $Ro_{\rm crit}$.
Since stars with the same Rossby number will have a range of ages, and the stalling of spin-down causes stars with older ages to pile up at similar periods, this will manifest as a jump in the $Ro$--\sigmavz\ relation, as the older stars (associated with higher \sigmavz) will still have the same $Ro$ number.
This is exactly what we see at $Ro_{\rm crit}$ in plot (b) in Figure~\ref{fig:3}.
According to the AVR calibrated gyrochronology relation on individual color bins, the age of the peak at $Ro_{\rm crit}$ for the highest temperature color bin (yellow lines centered at $\sim$6,100 K in Figure~\ref{fig:3}) is about 3 Gyr, this is about 6 Gyr for a star of 5,740 K (similar to the Sun).
However, even though we have cut all stars with $M_G < 4.2$ mag to exclude giants and sub-giants, the cut could be incomplete, and the bump could be due to contamination from the rotation measurements of sub-giant stars. 

\subsubsection{The Fully Convective Boundary}\label{subsubsec:FCB}
Another obvious feature lies at the fully convective boundary, where a plateau or even decrease exists for stars rotating between 50-80 days and with 2.6 $<$ \bprp\ $<$ 3.0, which can be seen in plot (a) of Figure~\ref{fig:3}.
This is likely associated with an abrupt change in spin-down across the fully convective boundary \citep{Lu2024_abrupt} or the dramatic increase in the convective turnover time across this boundary \citep{Chiti2024, Gossage2024}.
This can be explained when the fully convective stars with long rotation periods are younger than partially convective stars with the same period, either because they exhibit a different spin-down law or it can be caused by the jump in convective turnover time across the boundary.

\subsubsection{The Intermediate Period Gap}\label{subsubsec:IPG}
The intermediate period gap is a void in the \prot--\bprp\ space (shown as the red region in Figure~\ref{fig:conc} top row), and is thought to be associated with core-envelope decoupling \citep[e.g.,][]{Curtis2020, Lu2022_prot, Cao2023} where stars temporally stopped spinning down right below the gap due to the core spinning faster than the convective envelope, and the angular momentum loss on the surface is replenished by the transfer of momentum from the core.
If this is true, similar to the weakened magnetic braking case, there will be a bump in the \prot--\sigmavz\ space around the location of the gap in each color bin. 
Potential deviation from the AVR for stars around or above the gap can be seen where the dashed blue lines in Figure~\ref{fig:3_2} at 2.7 Gyr, where stars around the gap deviates from the AVR, and the differences increase with increasing \bprp, suggesting the timescale of stalling is larger for lower-mass stars, as pointed out in \cite{Curtis2020}.
Comparing the kinematic relations obtained from this work with predictions from spin-down models with stalling can potentially put constraints on the mass-dependent stalling timescale. 

\begin{figure*}
    \centering
    \includegraphics[width=\textwidth]{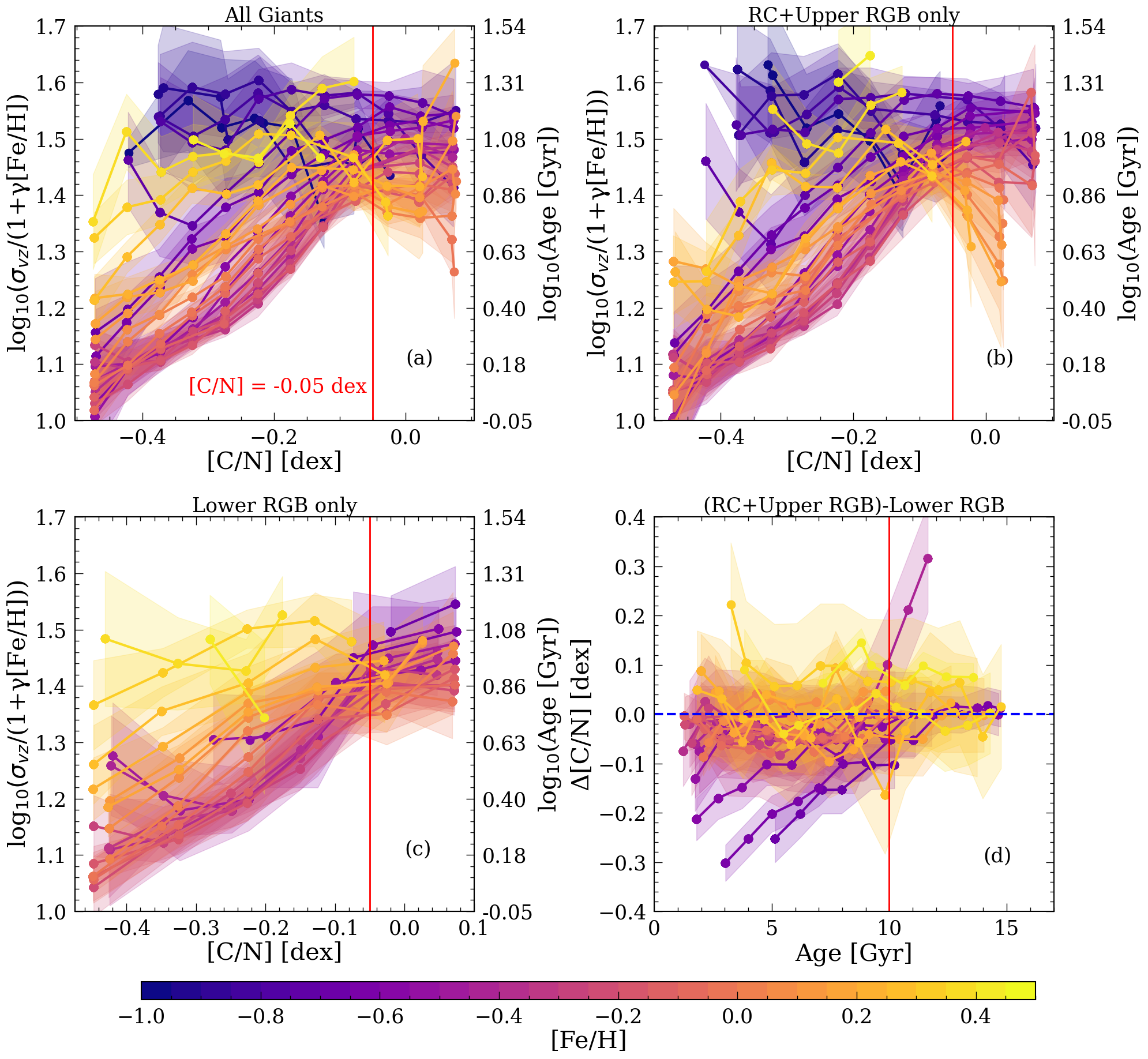}
    \caption{[C/N]-\sigmavz\ relations in metallicity bins for all giants in plot (a), RC in plot (b), and lower RGB in plot (c).
    The $y$-axises are normalized by the metallicity dependence of the AVR to separate the metallicity effect of the [C/N] relations.
    Plot (d) shows the difference in [C/N] between the RC and lower RGB stars as a function of the velocity dispersion of the lower RGB stars. 
    The red vertical lines show [C/N]=-0.05 dex, where [C/N] cannot be used to derive stellar age since the velocity dispersion of these stars do not change with [C/N].
    There do not exist a relation where the velocity dispersion increases with increasing [C/N] for stars with [Fe/H] $<$ -0.8 dex, mostly due to the fact that they exhibit halo-like kinematic, and thus, are similar in age.
    If we separate stars and look at stars in the low-$\alpha$ disk where an AVR should exist and a larger age range is available, there exist a positive correlation between \sigmavz\ and [C/N] down to [Fe/H] = -1 dex, suggesting it is possible to infer ages using [C/N] even for low-metallicity stars.}
    \label{fig:4}
\end{figure*}

\subsubsection{The Convergence from Fast to Slow Rotating Sequence}\label{subsubsec:convergence}
The convergence from the fast to the slow rotating sequence would show as a flattening of the AVR for the fastest rotating stars, as stars that have not converged can have a range of rotation periods, meaning rotation is not an age indicator.
According to plot (a) in Figure~\ref{fig:3}, there exists a well-defined age-\sigmavz\ relation for nearly all stars in our sample, even the fast-rotating stars, except for those that are fully convective (with \bprp $>\sim$ 2.6).
This also suggests that the AVR could hold true even for very young stars ($<$ 1 Gyr), which has never been shown in the literature as the AVR are mainly being calculated with ages inferred from methods that are mostly applicable to older stars (e.g., isochrone, asteroseismology, [C/N] ages).

\subsection{[C/N]-\sigmavz\ Relations in Individual [Fe/H] Bins}\label{subsec:cn-vz-rels}
Similar to looking at the \prot-\sigmavz\ relations for individual color bins, we can perform the same procedure for understanding the [C/N]-\sigmavz\ relation for individual [Fe/H] bins.
Figure~\ref{fig:4} shows the results, similar to what was shown in (a) of Figure~\ref{fig:3}, but plotting the velocity dispersion normalized by the metallicity dependence obtained for the AVR on the $y$-axis to better separate the metallicity dependence of the [C/N]-\sigmavz\ relations. 
A different version of Figure~\ref{fig:4}, plotting [C/N]-ages calibrated in Section~\ref{subsec:cn} on the $x$-axis is shown in Figure~\ref{fig:A2}.
The right-hand-side $y$-axis shows the age of the population, converted using the AVR calibrated in Section~\ref{subsec:avr}.
We separated stars into RC+Upper RGB in plot (b) and lower RGB in plot (c) as stars going through first dredge-up are known to change their [C/N] ratios 
\citep[e.g.,][]{Carbon1982, Kraft1994, Gratton2000, Masseron2017, Shetrone2019}.
However, it is known that this effect is not significant for higher metallicity stars \citep[e.g.,][]{Spoo2022}.

RC stars are identified with the criterion provided in \cite{APOK2_1} which is based on the methods of \cite{Warfield_2021}. 
A star is compared with an expected temperature for the RGB according to its log$g$, [Fe/H] and [C/N], and stars that are noticeably hotter than expected of a first-ascent giant are flagged as RC stars. 
To identify lower RGB stars from the remaining stars, a log$g$ threshold of 2.5 was set as the approximate location of the RGB bump. 
Stars with higher surface gravities than 2.5 and not identified as RC stars are considered lower RGB stars.

\subsection{Flattening of [C/N] at old ages}\label{subsec:4.3}
Figure~\ref{fig:4} plot (a)-(c) shows the [C/N]-\sigmavz\ relations flatten significantly or even turnover for stars with [C/N] $>\sim$ -0.05 dex (red vertical lines).
This means [C/N] cannot be used as an age indicator beyond -0.05 dex, which corresponds to $\sim$ 8-10 Gyr for a star with a metallicity of -0.2 dex.
This is also seen in previous studies in trying to obtain [C/N] ages for old stars \citep[e.g.,][]{StoneMartinez2024, StoneMartinez2025}.
The flattening also exists for both lower RGB and RC stars, suggesting mixing is not the cause of such flattening. 

Since almost no giants naturally have [C/N] $\gtrsim$0 dex at super-solar metallicity prior to the first dredge-up \citep[e.g.,][]{Roberts2024}, the downturn for stars with [C/N] $>$ -0.05 dex and [Fe/H] $>$ 0 dex suggests the inclusion of merger products, where younger stars gained carbon from merger, which is also seen in merger-induced high-$\alpha$ stars \citep[e.g.,][]{Hekker2019, Jofre2023}.
This is more evident for RC+Upper RGB stars in plot (b) compared to Lower RGB stars in (c) as the merger rate increases as stars ascend up the giant branch \citep[e.g.,][]{Miglio2021}, in which the star grow bigger and mergers have longer time to merge.

\subsection{[C/N] as an age indicator at low metallicity}
It has been unclear whether we can infer [C/N] ages for low metallicity stars, as the calibration sample has mostly been against open clusters \citep[e.g.,][]{Spoo2022} or asteroseismic ages \citep[e.g.,][]{StoneMartinez2024}, where the sample size is extremely small below [Fe/H] -0.5 dex.
Moreover, low metallicity stars not only experience more mixing on the giant branch, they could also represent a population with similar age purely from a chemical evolution stand point. 
As a result, it is unclear whether we can or at what metallicity we can infer ages for metal-poor giants.
With kinematic ages, we have a large sample size for low-metallicity stars, and we can understand whether [C/N] can be used as an age indicator for parameter spaces where other calibration samples are limited.

Looking at plot (b), \sigmavz\ increases as [C/N] increases for [Fe/H] $>$ -0.8 dex, which means we should be able to determine [C/N] ages down to a metallicity of -0.8 dex. 
Stars with [Fe/H] $<$ -0.8 dex exhibit halo-like kinematics, and \sigmavz\ does not change strongly with [C/N].
If we separate stars and look at stars in the low-$\alpha$ disk where an AVR should exist and a larger age range is available, there exist a positive correlation between \sigmavz\ and [C/N] down to [Fe/H] = -1 dex, suggesting it is possible to infer ages using [C/N] even for low-metallicity stars.
As a result, we conclude that [C/N] can be used as an age indicator down to [Fe/H] of -1 dex, and the reason low-metallicity stars exhibit a flat \sigmavz-[C/N] relation is because the small age range for low-metallicity stars.

\subsubsection{Effect of mixing}
To understand the effect of mixing, we subtracted the average [C/N] of lower RGB from that of RC for age bins, with the ages determined following Section~\ref{subsec:cn}.
The result is shown in plot (d) in Figure~\ref{fig:4}.
We excluded stars with [C/N] $<$ -0.05 dex and [Fe/H] $<$ -0.8 dex so we are staying in the parameter space where [C/N] can be used to determine age.
Our result shows for most metallicity bins, the differences are small and agree within the measurement uncertainties.
The difference in [C/N] increases as metallicity decreases, suggesting increasing mixing with decreasing metallicity, which is also seen in \cite{Shetrone2019}.
Similar to \cite{Shetrone2019}, we also see a continuous mixing after the first dredge-up for low metallicity stars.
This continuous mixing is a function of age, where lower metallicity stars experience extra mixing and have a longer timescale.

\section{Age Validation} \label{sec:comp}
From Section~\ref{sec:stelphys}, we only include stars in our gyrochronology age sample where \prot\ $<$ $Ro_{\rm crit}\tau_c$ as slower rotating stars have experienced weakened magnetic braking and thus, gyrochronology cannot be applied to.
We also only included stars in our [C/N] ages sample where [C/N] $<$ -0.05 dex and [Fe/H] $>$ -0.8 dex as [C/N] can only be used as an age indicator in these physical regime. 

To validate our ages, we first check if we can recover the AVR obtained from APOKASC--3 using the ages we calibrated in Section~\ref{subsec:AVR_cali}.
We then check the ages against asteroseismic ages in Section~\ref{subsec:astero}.
Finally, we check our ages against wide binaries in Section~\ref{subsec:WB} and open clusters in Section~\ref{subsec:OC}.

\subsection{Recovering the Age-Velocity-Dispersion Relations}\label{subsec:AVR_cali}
\begin{figure*}[ht!]
\centering
\includegraphics[width=\textwidth]{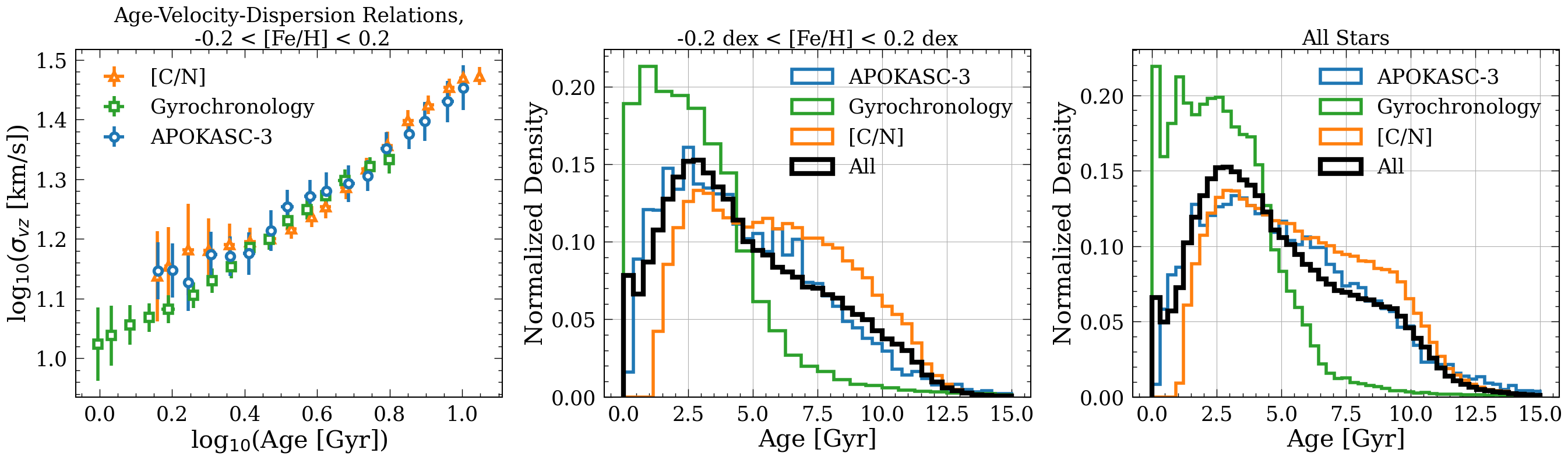}
\caption{Left: AVRs for stars with solar metallicity (-0.2 dex $<$ [Fe/H] $<$ 0.2 dex) using individual ages obtained using our method.
Disagreements could exist where the assumption that there exists an unique age with unique [C/N] and [Fe/H] measurements or unique \bprp\ and \prot\ measurements breaks down.
For example, this assumption breaks down for young stars using [C/N] age-dating method, as [C/N] plateaus towards younger ages \citep[e.g.,][]{Spoo2022}.
This break-down creates a deviation in the AVR at $<$ 2.5 Gyr using [C/N] ages.
Middle: Combined age histograms from APOKASC--3 and AVR calibrated gyrochronology and [C/N] ages for metallicity between -0.2 dex and 0.2 dex for better comparison with the AVR. 
Right: Same as the middle plot but for the full sample.
We can recover the AVR and the peak of the age distribution from APOKASC--3 at solar metallicity with our calibrated ages, suggesting self-consistency between the age catalogs.}
\label{fig:7}
\end{figure*}

Since we calibrated the age methods using the AVR, we wanted to ensure we could reproduce the AVR from APOKASC--3 with our calibrated ages.
Since the AVR changes with metallicity, we selected stars that have metallicity measurements from APOGEE DR17 between -0.2 and 0.2 dex for the APOKASC--3 and the [C/N] age sample. 
For the gyrochronology sample, we cross-matched our data with metallicity from Gaia BP-RP spectra determined from \cite{Andrae2023} and selected metallicity in the same range. 
Figure~\ref{fig:7} left plot shows the AVR obtained from all three sets of ages, and their results agree with each other within uncertainty.
Deviation exist for stars $<$ 2.5 Gyr where both the APOKASC--3 and [C/N]-age AVR plateaus.
This could be due to small sample size in APOKASC--3, and the plateau in [C/N]-\sigmavz\ relation for stars $<$ 2.5 Gyr \citep[See Figure~\ref{fig:A2} and][]{Spoo2022}.

The middle and right plot shows the normalized histogram for each age sample and three samples combined, in which the middle plot shows the stars used to determined the AVR on the left plot only for better comparison.
It is clear that for the solar metallicity sample, all three age catalogs peak around 2.5 Gyr, showing self-consistency. 
The number of [C/N] ages falls off significantly $<$ 1 Gyr as young giants are rare.
The number of gyrochronology ages decreases around 5 Gyr as old main-sequence stars have decreasing activity and often do not show spot modulations needed to obtain period measurements.
However, gyrochronology can provide ages for younger stars as they are more active and thus easier to measure rotation.
This further stresses the importance of cross-calibrating multiple age-dating methods, as doing so will allow us to paint a complete picture of the star formation history and the evolution of the MW, as different age-dating methods are applicable at inferring ages for at evolutionary stages.

\subsection{Comparing with Asteroseismic Ages}\label{subsec:astero}
\begin{figure*}[ht!]
\centering
\includegraphics[width=\textwidth]{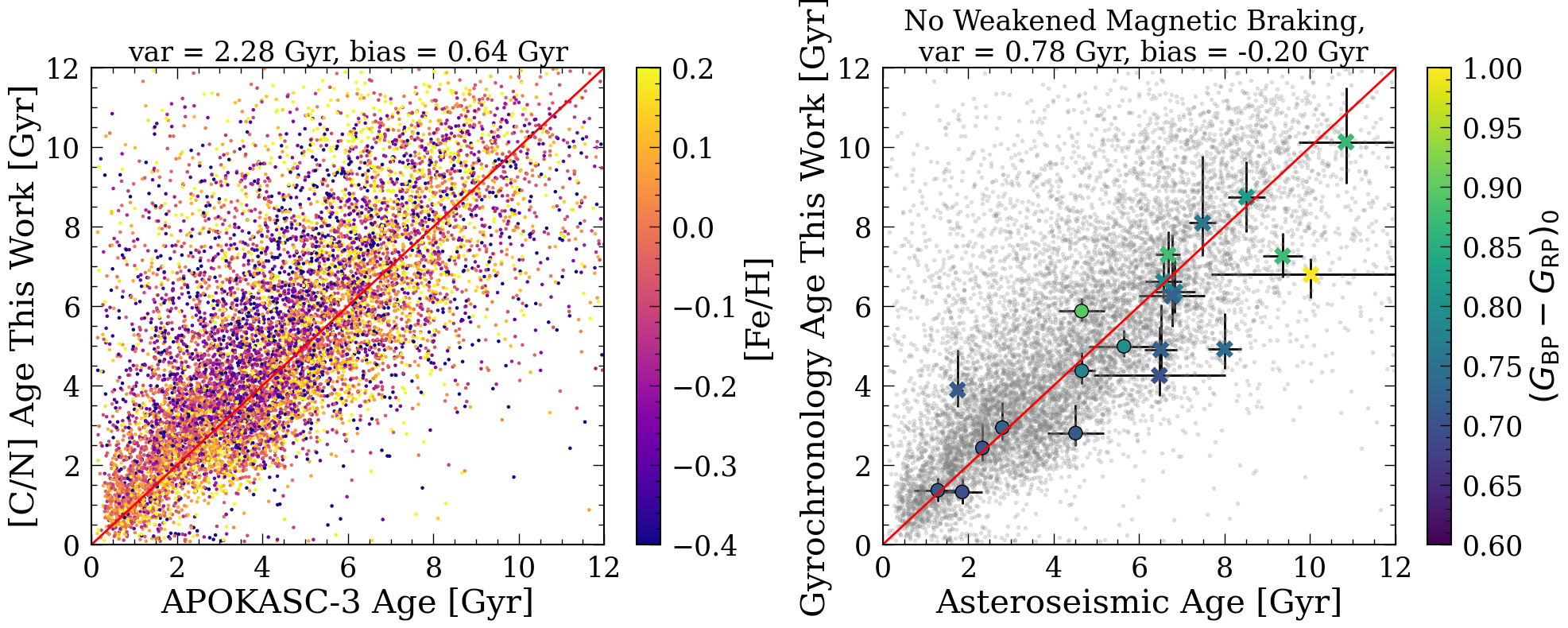}
\caption{[C/N] (left) and gyrochronology (right) ages calibrated in this work compared to asteroseismic ages \citep{Silva2017, Pinsonneault2024, Li2025}.
The grey points in the right plot show the same points as those in the left for better comparison. 
The crosses in the right plot show stars that are going through weakened magnetic braking, in which \prot\ $> Ro_{\rm crit}\tau_c$.
We can recover the asteroseismic ages using both [C/N] and gyrochronology, with a slight positive bias in [C/N] and a slight negative bias in gyrochronology.}
\label{fig:8}
\end{figure*}

We then compared our [C/N] and gyrochronology ages with available asteroseismic ages in the literature.
The results are shown in Figure~\ref{fig:8}.
For the [C/N] ages (left plot), we compared them with the APOKASC--3 sample. 
The points are colored by the metallicity reported in APOGEE DR17. 
Overall, the APOKASC--3 ages and the [C/N] ages calibrated with AVR agree well, and there exists no strong trend with metallicity. 
For the gyrochronology ages (right plot), we compared our ages with the LEGACY sample \citep[colored crosses represent stars experiencing weakened magnetic braking, and the circles represents the rest;][]{Silva2017} and the K dwarf with asteroseismic measurement \cite[colored cross;][]{Li2025}, where the K dwarf is just at the period where it should experience weakened magnetic braking.
The background grey points show the data as in the left plot for better comparison. 
The color shows the \bprp\ color after accounting for extinction with \texttt{dustmap} \citep{Green2018, Green20182}.
Overall, the ages inferred from our gyrochronology model agree well with the asteroseismic sample and the distribution agrees with those from [C/N] ages, with the [C/N] ages have a slight positive bias, and the gyrochronology ages have a slight negative bias.
We also under-predict the K dwarf with asteroseismic age measurement, but the uncertainty is large and it is experiencing weakened magnetic braking, so it is unclear the cause of this discrepancy. 

\subsection{Comparing with Wide-binaries}\label{subsec:WB}
\begin{figure*}[ht!]
\centering
\includegraphics[width=\textwidth]{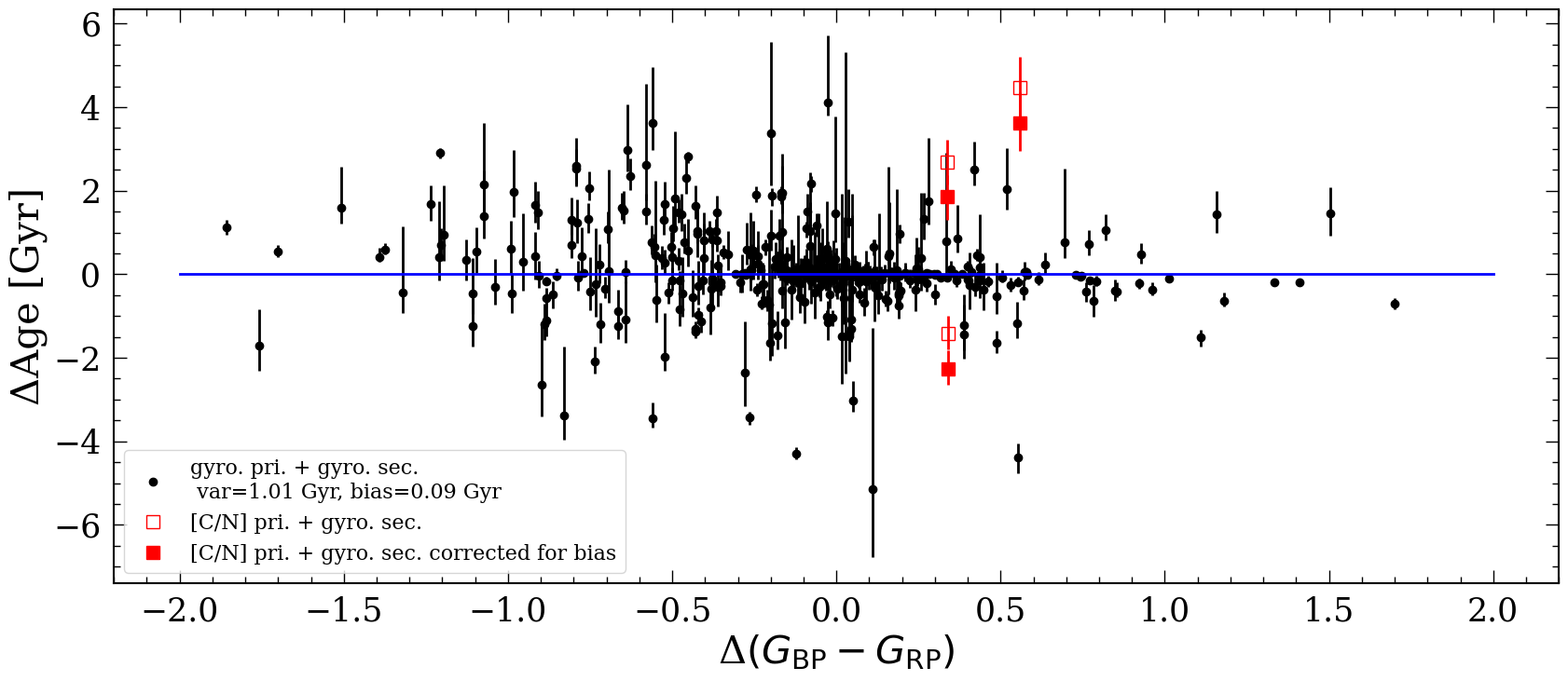}
\caption{Age validation with wide-binaries from \cite{ElBadry2018} and \cite{Gruner2023}. 
The gyrochronology ages inferred from our method for the wide-binary pairs agree well with minimal bias and a variance of $\sim$ 1 Gyr, independent of the differences in \bprp\ between the binary pairs.
More giant--main-sequence wide-binary pairs are needed to draw further conclusion on the cross-calibration between [C/N] and gyrochronology.
However, correcting for the bias where we subtracted 0.64 Gyr for [C/N] ages and -0.20 Gyr for gyrochronology ages, as determined in Figure~\ref{fig:8}, seems to achieve a better agreement between [C/N] and gyrochronology ages.}
\label{fig:9}
\end{figure*}

We cross-matched our age sample with the Gaia DR3 wide-binary sample from \cite{ElBadry2018} and \cite{Gruner2023}. 
We are able to find 417 pairs with gyrochronology ages available for both the primary and the secondary.
We also found 3 pairs with [C/N] ages available for the primary and gyrochronology ages available for the secondary.   
No binary pairs with [C/N] ages for both stars are found.
This is not surprising as [C/N] ages are only obtainable for giant stars, where merger rate is higher and main-sequence binary companions are hard to find around these bright stars. 
The comparison result is shown in Figure~\ref{fig:9}, plotting the difference in the inferred age versus the difference in their \bprp\ color. 

Overall, the gyrochronology ages for the wide-binary pairs agree well with minimal bias and a variance of $\sim$ 1 Gyr, independent of the differences in \bprp\ between the binary pairs.

For the pairs where the primary star is a giant star with [C/N] age and the secondary is a slightly evolved star with gyrochronology age, the result is inconclusive as there are only 3 pairs. 
However, correcting for the bias where we subtracted 0.64 Gyr for [C/N] ages and -0.20 Gyr for gyrochronology ages, as determined in Figure~\ref{fig:8}, is able to achieve better agreement between [C/N] and gyrochronology ages.
More data is needed to draw any firm conclusion. 

\subsection{Comparing with Open Clusters}\label{subsec:OC}
Open clusters (OCs) are often used to validate ages, not only because they are single-age populations, but also because stars that are at different evolutionary stages can be present in a cluster together. 
In this section, we will use the ages derived in this work to compare the median age for individual cluster members with the cluster ages derived from literature \citep{CantatGaudin2020, Hunt2023, Cavallo2024, VanLane2024}.
We selected stars with membership probability $>$ 0.99 using the membership catalog in \cite{CantatGaudin2020}, we then cross-matched the members with our age catalogs.
\cite{VanLane2024} compiled multiple clusters with rotation period measurements available from the literature.
The periods and ages are taken from various sources, described in their Table 1 and 2, respectively.
We selected the rotation period sample following the flags described in \cite{VanLane2024} to exclude binaries, evolved stars, and stars with no available extinction values.
These catalogs provided us with 38 clusters (453 members) with more than one star from each cluster with [C/N] measurements from APOGEE DR17, and 33 clusters (325 members) with more than one star from each cluster with period measurements.
The ``true ages'' for the clusters are obtained using the mean of the ages from different age catalogs, and the uncertainties are taken to be the range of the ages from the literature. 

\begin{figure*}[ht!]
\centering
\includegraphics[width=0.32\textwidth]{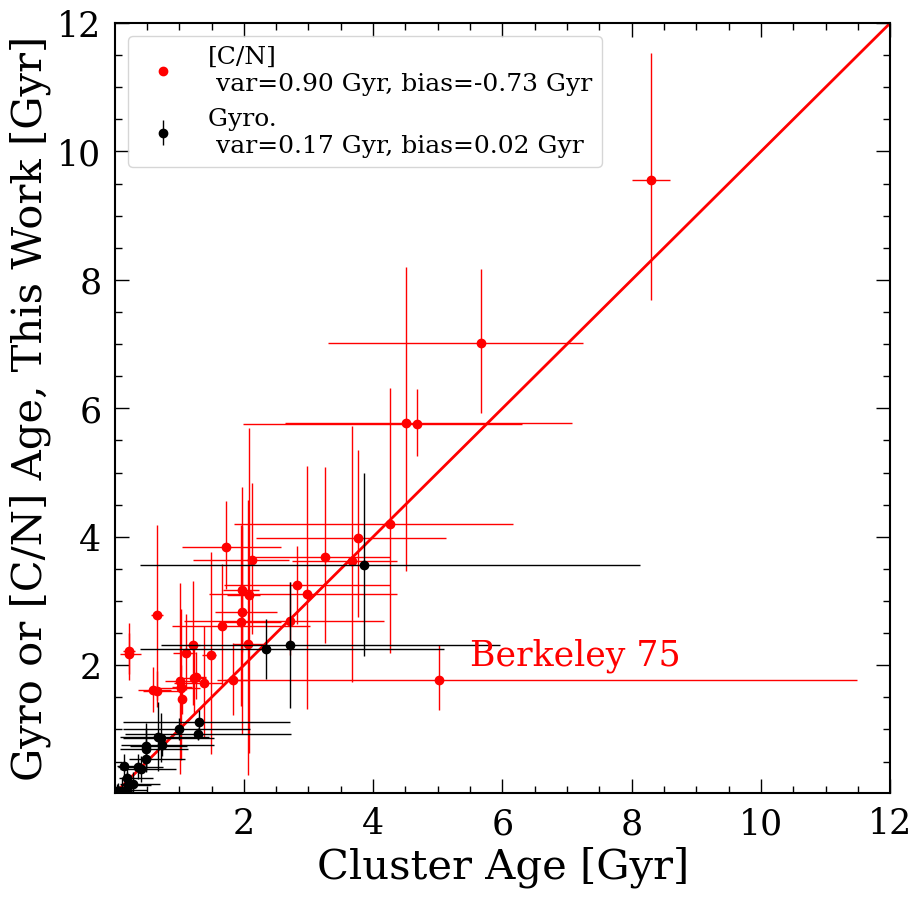}
\includegraphics[width=0.33\textwidth]{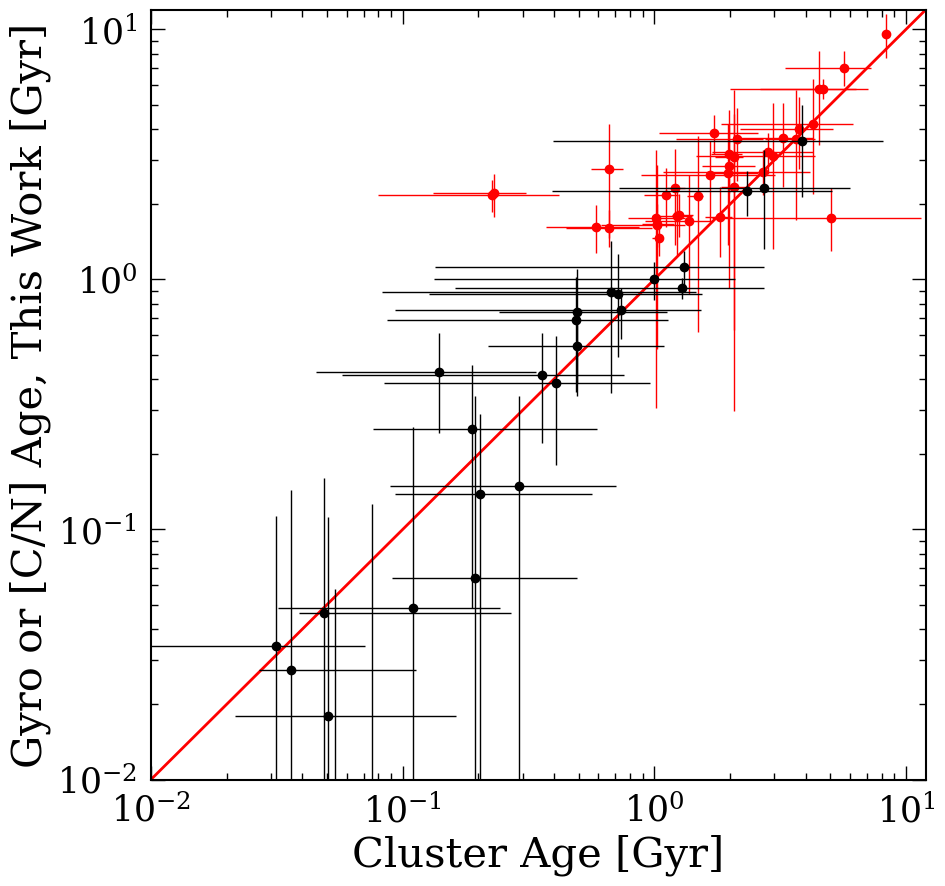}
\includegraphics[width=0.305\textwidth]{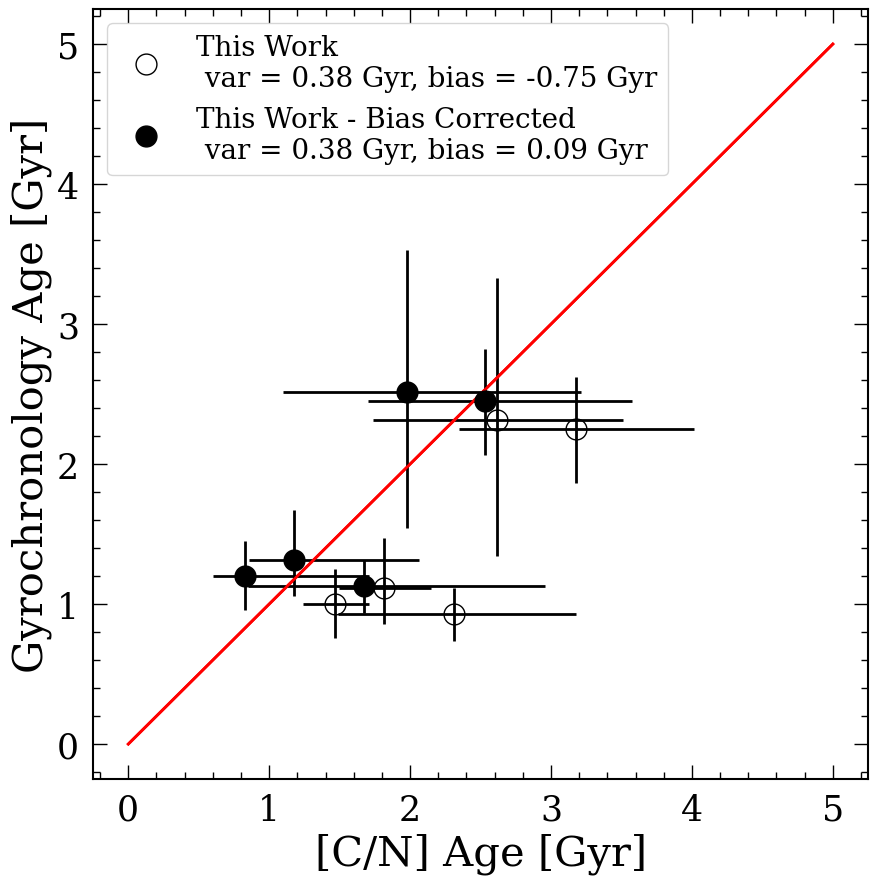}
\caption{Left: Recovery of the cluster samples obtained from the literature \citep{CantatGaudin2020, Hunt2023, Cavallo2024, VanLane2024}. 
Middle: Same as left but in logarithmic scale.
Right: Ages obtained from [C/N] compared to those from rotation periods using the relations calibrated in this work for the same clusters.
After correcting for bias for [C/N] and gyrochronology as shown in Figure~\ref{fig:8}, we are able to achieve agreement between the cluster ages determined from both methods.
However, more data is needed to perform better comparisons.}
\label{fig:10}
\end{figure*}

We inferred the cluster age by measuring the ages for individual stars, then taking the median age of the individual members for each cluster.
The uncertainty is the standard deviation of the individual ages. 
Figure~\ref{fig:10} left and middle plots show the comparison result.
We are able to recover the age of the cluster within uncertainty without including any cluster sample in our training set.
The bias for [C/N] ages is similar to what was found comparing with the APOKASC--3 sample in Figure~\ref{fig:8}.

As mentioned previously, one of the major advantages of using the cluster sample is it provides us the bridge to validate the cross-calibration between [C/N] and gyrochronology ages beyond the three wide-binary pairs shown in Figure~\ref{fig:9}.
However, the youngest OCs do not contain giant stars, and the main-sequence stars of old OCs are too faint to detect spot modulations. 
As a result, there are only 5 clusters (Ruprecht 147, NGC 1817, NGC 752, NGC 6819, and NGC 6811) that we are able to obtain both [C/N] and gyrochronology ages.
The comparison using the raw ages derived in this work is shown in Figure~\ref{fig:10} right plot as hallow black points.
The solid black points are cluster ages after correcting for bias between [C/N] and gyrochronology as shown in Figure~\ref{fig:8} by subtracting 0.64 Gyr in [C/N] cluster ages and adding 0.2 Gyr in gyrochronology cluster ages.
By doing so, we are able to achieve agreement between the cluster ages determined from both methods with minimal bias.
However, more data, perhaps co-moving pairs, is needed to investigate the true bias and dependent factors further as a single correction might not be sufficient. 

\section{Limitations \& future work} \label{sec:future}
We explored the possibility of cross-calibrating various age-dating methods by leveraging the fact that the age-velocity-dispersion relation should be universal across all evolutionary stages. 
However, there is still much to be done.

The biggest caveat for this work is we did not take into account the intrinsic scatters around the relations.
For example, we assume there is a unique relation between \prot, \bprp, and $\ln$(age) for gyrochronology; [C/N], [Fe/H], and age for the [C/N]-age relation. 
This could introduce bias if the scatter around the \prot-$\ln$(age) or [C/N]-age at fixed \bprp\ or [Fe/H] is not constant.
For [C/N], looking at small bins of [Fe/H], the  [C/N]-age relation in APOKASC--3 seems to have a near-constant scatter. 
For gyrochronology, even though the clusters show a very tight relation between \prot-$\ln$(age) at fixed colors after the sequence has converged, it is unclear how the differential rotation can affect this result, especially at old ages. 
This could be why we under-predict the age of the old K dwarf with asteroseismic age. 
Future work should include investigations of the parameters that give rise to the scatter around the relations, and to better take into account such scatter in the fitting procedure.

Another source of uncertainty came from the AVR.
We assumed a simple age velocity relation that is a power-law that only depends on metallicity.
However, there is evidence that the AVR deviates from a simple power law for old ages and the power index is location-dependent \citep[e.g.,][]{Ting2019, Sharma2021}.
Since APOKASC--3 is a local sample that is relatively small, the effect of Galactic radii is not being taken into account. 
This is not a huge problem for the gyrochronology sample as stars with period measurements are all local.
Since [C/N] ages cover the entire disk, this could potentially bias the ages for these stars beyond the solar vicinity.
The future Nancy Romen Space Telescope \citep{Spergel2015, Akeson2019} could potentially resolve this issue as it is predicted to provide asteroseismic ages for more than 100,000 stars in the Galactic bulge \citep{Weiss2025}, providing us with a wider spatial coverage. 
One could also potentially switch to using the AVR relation obtained using a more completed age sample, however, many of these samples are calibrated on spectroscopic ages, thus making the problem more complicated.

Despite all the limitations described in this section, we were still able to minimize the bias between the ages inferred from [C/N] and rotation just by anchoring both on the AVR calibrated with APOKASC--3.
This highlights the huge potential of this method and the significant impact that future efforts to refine its details could have.

\begin{figure*}
\centering
\includegraphics[width=\textwidth]{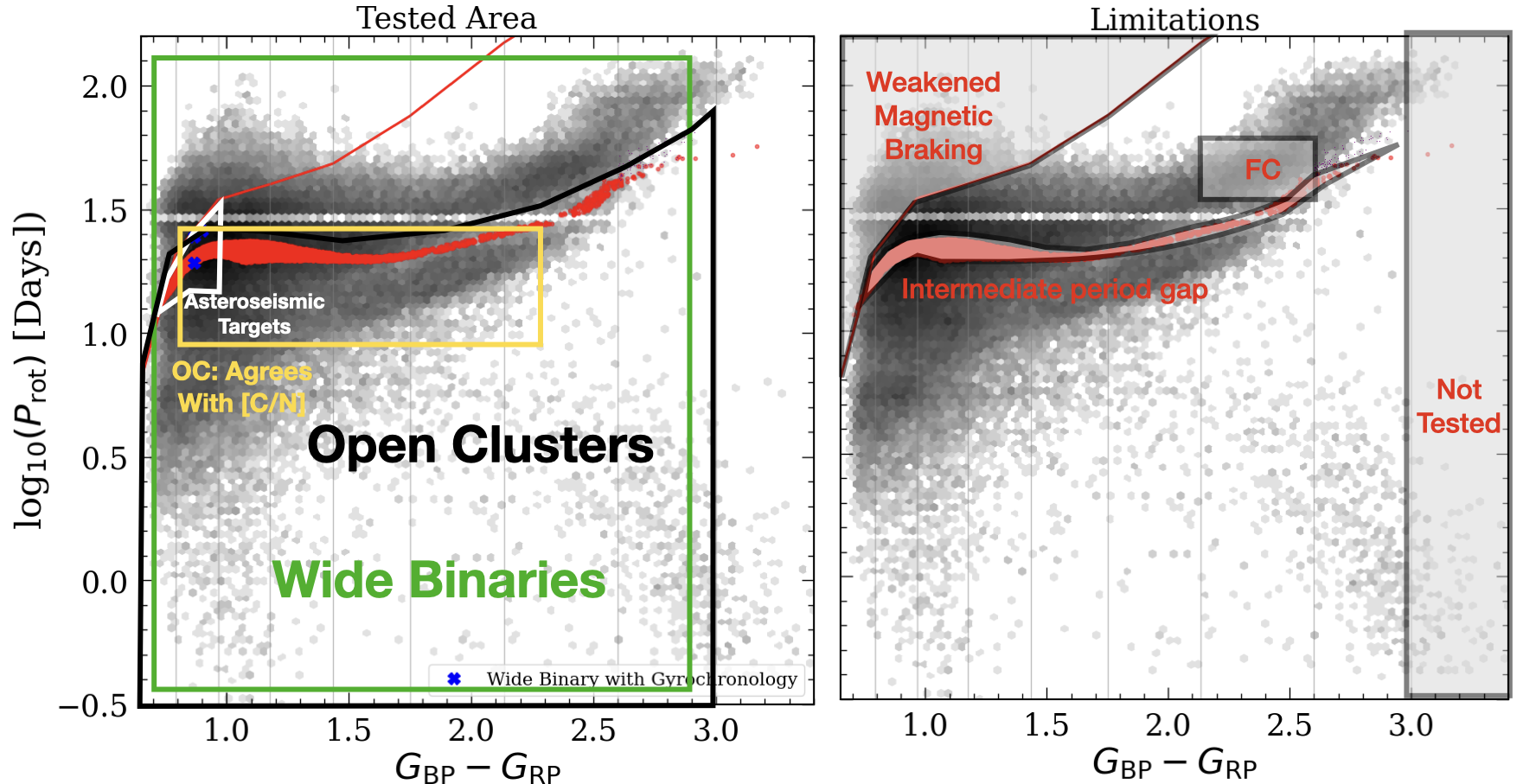}
\includegraphics[width=\textwidth]{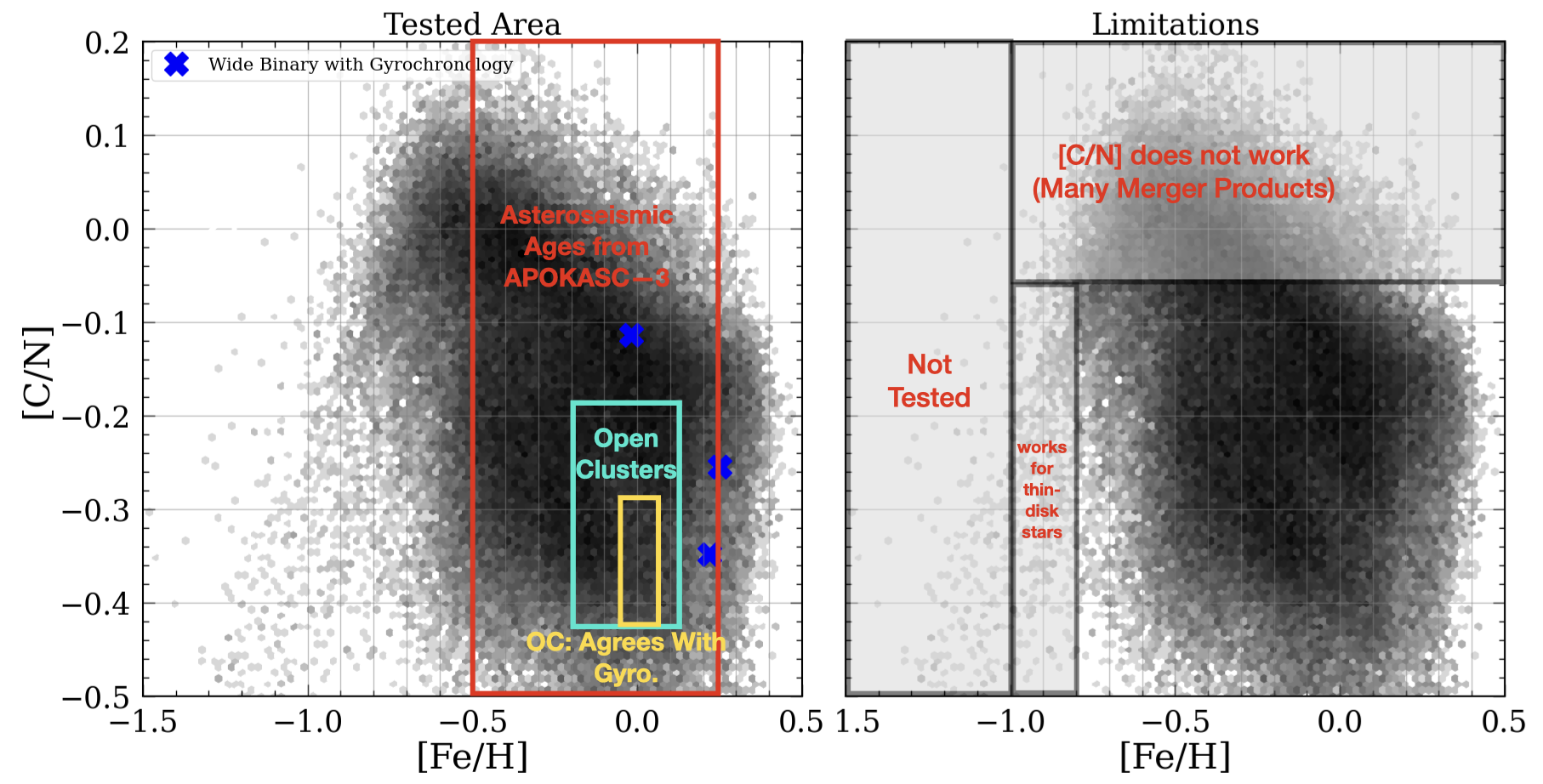}
\caption{Summary plots for the tested areas (left column) and limitations (right column) for gyrochronology (top row) and [C/N] ages (bottom row) obtained from this work.
The background histograms are the full sample of Kepler+TESS+ZTF periods and APOGEE DR17 selection of giants as described in Section~\ref{sec:data}.
The age-velocity-dispersion relations also agree for stars of solar metallicity between the two methods and that of APOKASC--3, which is not plotted. 
We recommend excluding the grey shaded areas in parameter space shown in the right column when applying each method. 
When applying [C/N] relations to low-metallicity stars ([Fe/H] $<$ -0.8 dex), we recommend only selecting the low-$\alpha$ stars. 
Low-metallicity stars in the high-$\alpha$ disk exhibit mostly halo-like kinematic, and does not show a strong [C/N]-\sigmavz\ relation, suggesting their similarity in masses, and that [C/N] cannot be used to distinguish ages of these old stars.}
\label{fig:conc}
\end{figure*}

We provide summary plots for the parameter space we recommend applying gyrochronology and [C/N] ages, as well as the parameter space where we tested these methods in Figure~\ref{fig:conc}.

\section{Conclusion} \label{sec:conclu}
In this work, we investigated the possibility of cross-calibrating [C/N] and gyrochronology ages using the age-velocity-dispersion relation (AVR).
We obtained the parameters for the metallicity-dependent AVR using APOKASC--3, the state-of-the-art asteroseismic sample. 
Using this AVR, we can understand the physical limitations of [C/N] and gyrochronology.
We aimed at: 1) explore the parameter space in which [C/N] and gyrochronology are applicable, extending beyond the domains of asteroseismology and open clusters, and 2) assess whether the traditionally assumed [C/N] and gyrochronology relations yield ages on a consistent physical scale, after calibrating both using the same AVR obtained from APOKASC--3 \citep{Pinsonneault2024}.

For 1), we learn that: 
\begin{itemize}
    \item For gyrochronology ages:
    \begin{itemize}
        \item K and M dwarfs with \bprp\ $>\sim$ 1.5 spin down, on average, slower than G dwarfs. 
        A fast transition happens around \bprp\ = 1.5, which is where the intermediate period gap starts to become apparent (Figure~\ref{fig:1}). 
        \item Weakened magnetic breaking creates a strong feature for stars with Rossby number $>$ 1.866 (Figure~\ref{fig:3} (b)).
        This is the first time this effect is seen in field star populations with an independent age-indicator, placing strong observational constraint on spin-down models.
        \item Deviation exist and becomes stronger towards less massive stars between the APOKSAC--3 AVR and the gyrochronology predicted AVR at the location of K-dwarf stalling (Figure~\ref{fig:3_2}).
        \item Gyrochronology can be used to determine ages for fully convective stars after they have converged onto the slow-rotating sequence (Figure~\ref{fig:3_2}).
    \end{itemize}
    
    \item For [C/N] ages:
    \begin{itemize}
        \item ${\rm[}$C/N] is likely not an age-indicator for giant stars with [C/N] $>$ -0.05 dex (Figure~\ref{fig:4}, (a)-(c)). 
        \item Extra mixing in RC stars is not obvious for [Fe/H] $>$ 0 dex and increases for [Fe/H] $<$ 0 dex where lower metallicity stars experience longer and more mixing that can last more than a few gigayears (Figure~\ref{fig:3} (d)).
        \item ${\rm[}$C/N] is not an age-indicator for stars with [Fe/H] $<$ -0.8 dex (Figure~\ref{fig:3} (a)-(b)), most likely due to the similarity in age for lower-metallicity stars. ${\rm[}$C/N] can be used as an age-indicator for -0.8 dex $<$ [Fe/H] $<$ -1 dex if only low-$\alpha$ disk stars are selected.
        \item Binarity dominates the population with [Fe/H] $>$ 0 and [C/N] $>$ -0.05, and the kinematic relations place observational constraints for future models trying to understand merger rates as a function of metallicity and age (Figure~\ref{fig:3} (b)).
    \end{itemize}
\end{itemize}
These relations obtained with vertical velocity dispersion are purely observational and can be used to test angular momentum transport or [C/N] productions in stellar evolution models.

For 2), we assumed a traditional [C/N]-age and gyrochronology relations and anchored both on the AVR calibrated with APOKASC--3.
By doing so, the calibrated relations can reproduce the same AVR and the mode of the age distribution for solar metallicity stars (Figure~\ref{fig:7}). 
Our calibrated ages also agree well with asteroseismic (Figure~\ref{fig:8}) and cluster (Figure~\ref{fig:10}) ages in the literature.
The wide-binary sample suggests a constant variance between the gyrochronology ages of the primaries and secondaries across the difference in temperature within the pair (Figure~\ref{fig:9}).
Finally, tests with wide binaries and clusters (Figure~\ref{fig:9} and Figure~\ref{fig:10}) suggests our cross-calibrated age relations can potentially minimize the bias between [C/N] and gyrochronology ages after correcting for bias.

Future work should take into account the bias induced by the non-constant scatter around the \prot-$\ln$(age) or [C/N]-age relations at fixed \bprp\ or [Fe/H], and calibrate with a location-dependent AVR. 

\begin{acknowledgments}
\section{Acknowledgments}
Y.S.T is supported
by the National Science Foundation under Grant No.
AST-2406729.
This work has made use of data from the European Space Agency (ESA)
mission Gaia,\footnote{\url{https://www.cosmos.esa.int/gaia}} processed by
the Gaia Data Processing and Analysis Consortium (DPAC).\footnote{\url{https://www.cosmos.esa.int/web/gaia/dpac/consortium}} Funding
for the DPAC has been provided by national institutions, in particular
the institutions participating in the Gaia Multilateral Agreement.
This research also made use of public auxiliary data provided by ESA/Gaia/DPAC/CU5 and prepared by Carine Babusiaux. 
This research has also made use of NASA's Astrophysics Data System, 
and the VizieR \citep{vizier} and SIMBAD \citep{simbad} databases, 
operated at CDS, Strasbourg, France.
\end{acknowledgments}

%

\vspace{5mm}
\facilities{Gaia \citep{gaia}, ZTF \citep{ztfdata, ztftime}, Kepler \citep{kepler}, TESS \citep{TESS}, APOGEE \citep{apogee}}


\software{astropy \citep{astropy:2013, astropy:2018, astropy2022},  
          emcee \citep{emcee}, 
          corner \citep{corner}
          \texttt{Matplotlib} \citep{matplotlib}, 
            \texttt{NumPy} \citep{Numpy}, 
            \texttt{Pandas} \citep{pandas}, 
            \texttt{gala} \citep{gala2017},
            \texttt{galpy} \citep{galpy}}


\appendix
\section{The [C/N]-age relations}\label{sec:A2}
\begin{figure*}
    \centering
    \includegraphics[width=0.98\linewidth]{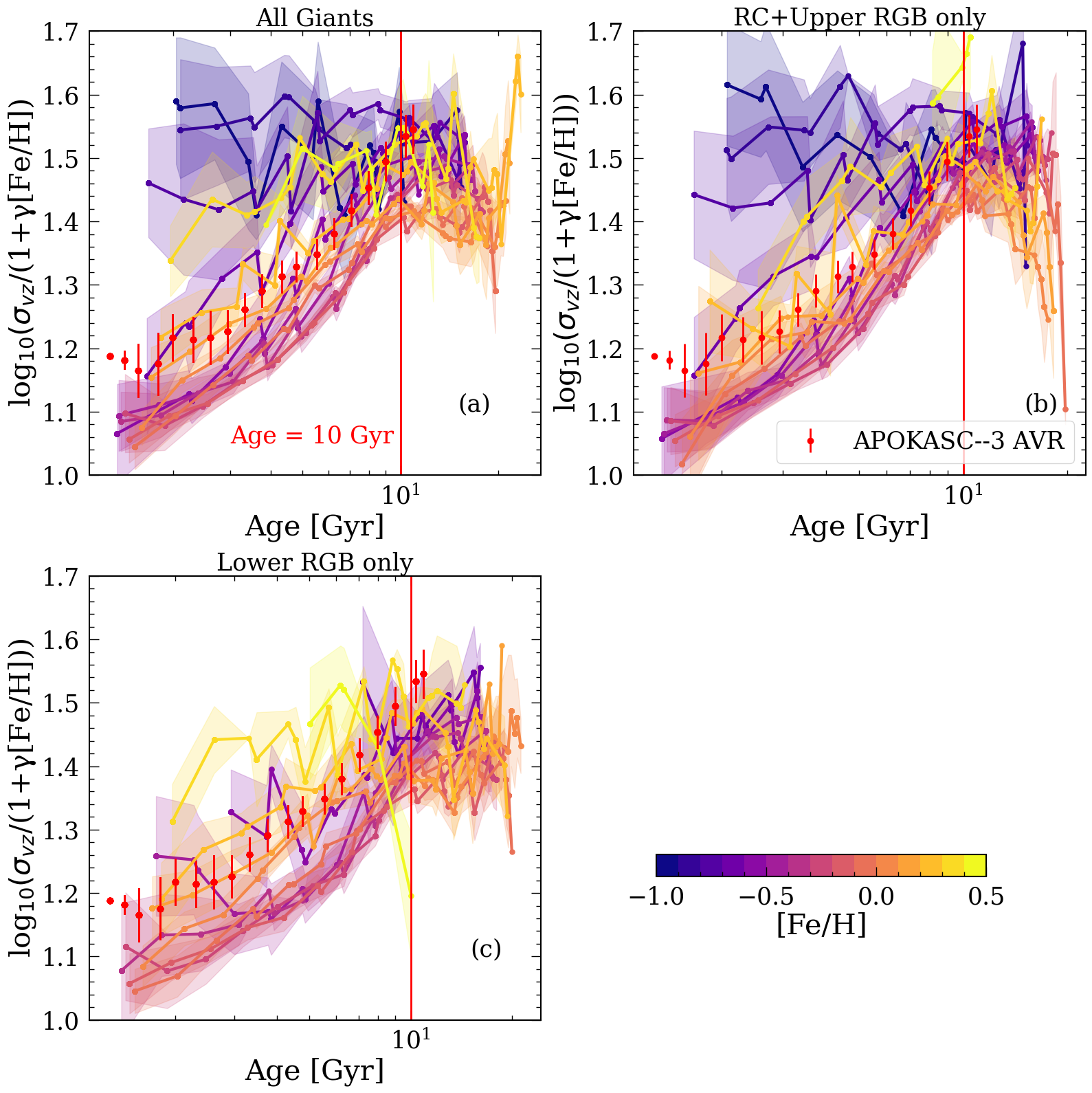}
    \caption{Same as Figure~\ref{fig:4} but plotting ages calibrated in this work on the $x$-axis.
    The red data points with errorbars show the AVR from APOKASC--3.}
    \label{fig:A2}
\end{figure*}

Figure~\ref{fig:A2} shows the same as Figure~\ref{fig:4} but plotting ages calibrated in this work on the $x$-axis.
The red data points with errorbars show the AVR from APOKASC--3.
We agree with the AVR from APOKASC--3 up until 10 Gyr. 
However, we disagree with the AVR for stars with [Fe/H] $<$ -0.8 dex.
As mentioned in the main text, this is because these stars exhibit halo kinematics. 
However, if we select only the low-$\alpha$ stars, the AVR agrees with APOKASC--3 down to [Fe/H] = -1 dex.

There also exist disagreement for the highest metallicity stars in our sample most likely due to binary interactions (see Section~\ref{subsec:4.3}). 

\bibliography{sample631}{}

\begin{thebibliography}{}
\expandafter\ifx\csname natexlab\endcsname\relax\def\natexlab#1{#1}\fi
\providecommand{\url}[1]{\href{#1}{#1}}
\providecommand{\dodoi}[1]{doi:~\href{http://doi.org/#1}{\nolinkurl{#1}}}
\providecommand{\doeprint}[1]{\href{http://ascl.net/#1}{\nolinkurl{http://ascl.net/#1}}}
\providecommand{\doarXiv}[1]{\href{https://arxiv.org/abs/#1}{\nolinkurl{https://arxiv.org/abs/#1}}}

\bibitem[{{Abadi} {et~al.}(2003){Abadi}, {Navarro}, {Steinmetz}, \& {Eke}}]{Abadi2003}
{Abadi}, M.~G., {Navarro}, J.~F., {Steinmetz}, M., \& {Eke}, V.~R. 2003, \apj, 591, 499, \dodoi{10.1086/375512}

\bibitem[{{Abdurro'uf} {et~al.}(2022){Abdurro'uf}, {Accetta}, {Aerts}, {Silva Aguirre}, {Ahumada}, {Ajgaonkar}, {Filiz Ak}, {Alam}, {Allende Prieto}, {Almeida}, {Anders}, {Anderson}, {Andrews}, {Anguiano}, {Aquino-Ort{\'\i}z}, {Arag{\'o}n-Salamanca}, {Argudo-Fern{\'a}ndez}, {Ata}, {Aubert}, {Avila-Reese}, {Badenes}, {Barb{\'a}}, {Barger}, {Barrera-Ballesteros}, {Beaton}, {Beers}, {Belfiore}, {Bender}, {Bernardi}, {Bershady}, {Beutler}, {Bidin}, {Bird}, {Bizyaev}, {Blanc}, {Blanton}, {Boardman}, {Bolton}, {Boquien}, {Borissova}, {Bovy}, {Brandt}, {Brown}, {Brownstein}, {Brusa}, {Buchner}, {Bundy}, {Burchett}, {Bureau}, {Burgasser}, {Cabang}, {Campbell}, {Cappellari}, {Carlberg}, {Wanderley}, {Carrera}, {Cash}, {Chen}, {Chen}, {Cherinka}, {Chiappini}, {Choi}, {Chojnowski}, {Chung}, {Clerc}, {Cohen}, {Comerford}, {Comparat}, {da Costa}, {Covey}, {Crane}, {Cruz-Gonzalez}, {Culhane}, {Cunha}, {Dai}, {Damke}, {Darling}, {Davidson}, {Davies}, {Dawson}, {De Lee}, {Diamond-Stanic}, {Cano-D{\'\i}az}, {S{\'a}nchez},
  {Donor}, {Duckworth}, {Dwelly}, {Eisenstein}, {Elsworth}, {Emsellem}, {Eracleous}, {Escoffier}, {Fan}, {Farr}, {Feng}, {Fern{\'a}ndez-Trincado}, {Feuillet}, {Filipp}, {Fillingham}, {Frinchaboy}, {Fromenteau}, {Galbany}, {Garc{\'\i}a}, {Garc{\'\i}a-Hern{\'a}ndez}, {Ge}, {Geisler}, {Gelfand}, {G{\'e}ron}, {Gibson}, {Goddy}, {Godoy-Rivera}, {Grabowski}, {Green}, {Greener}, {Grier}, {Griffith}, {Guo}, {Guy}, {Hadjara}, {Harding}, {Hasselquist}, {Hayes}, {Hearty}, {Hern{\'a}ndez}, {Hill}, {Hogg}, {Holtzman}, {Horta}, {Hsieh}, {Hsu}, {Hsu}, {Huber}, {Huertas-Company}, {Hutchinson}, {Hwang}, {Ibarra-Medel}, {Chitham}, {Ilha}, {Imig}, {Jaekle}, {Jayasinghe}, {Ji}, {Johnson}, {Jones}, {J{\"o}nsson}, {Katkov}, {Khalatyan}, {Kinemuchi}, {Kisku}, {Knapen}, {Kneib}, {Kollmeier}, {Kong}, {Kounkel}, {Kreckel}, {Krishnarao}, {Lacerna}, {Lane}, {Langgin}, {Lavender}, {Law}, {Lazarz}, {Leung}, {Leung}, {Lewis}, {Li}, {Li}, {Lian}, {Liang}, {Lin}, {Lin}, {Lin}, {Lintott}, {Long}, {Longa-Pe{\~n}a}, {L{\'o}pez-Cob{\'a}}, {Lu},
  {Lundgren}, {Luo}, {Mackereth}, {de la Macorra}, {Mahadevan}, {Majewski}, {Manchado}, {Mandeville}, {Maraston}, {Margalef-Bentabol}, {Masseron}, {Masters}, {Mathur}, {McDermid}, {Mckay}, {Merloni}, {Merrifield}, {Meszaros}, {Miglio}, {Di Mille}, {Minniti}, {Minsley}, \& {Monachesi}}]{Abdurrouf2022}
{Abdurro'uf}, {Accetta}, K., {Aerts}, C., {et~al.} 2022, \apjs, 259, 35, \dodoi{10.3847/1538-4365/ac4414}

\bibitem[{{Akeson} {et~al.}(2019){Akeson}, {Armus}, {Bachelet}, {Bailey}, {Bartusek}, {Bellini}, {Benford}, {Bennett}, {Bhattacharya}, {Bohlin}, {Boyer}, {Bozza}, {Bryden}, {Calchi Novati}, {Carpenter}, {Casertano}, {Choi}, {Content}, {Dayal}, {Dressler}, {Dor{\'e}}, {Fall}, {Fan}, {Fang}, {Filippenko}, {Finkelstein}, {Foley}, {Furlanetto}, {Kalirai}, {Gaudi}, {Gilbert}, {Girard}, {Grady}, {Greene}, {Guhathakurta}, {Heinrich}, {Hemmati}, {Hendel}, {Henderson}, {Henning}, {Hirata}, {Ho}, {Huff}, {Hutter}, {Jansen}, {Jha}, {Johnson}, {Jones}, {Kasdin}, {Kelly}, {Kirshner}, {Koekemoer}, {Kruk}, {Lewis}, {Macintosh}, {Madau}, {Malhotra}, {Mandel}, {Massara}, {Masters}, {McEnery}, {McQuinn}, {Melchior}, {Melton}, {Mennesson}, {Peeples}, {Penny}, {Perlmutter}, {Pisani}, {Plazas}, {Poleski}, {Postman}, {Ranc}, {Rauscher}, {Rest}, {Roberge}, {Robertson}, {Rodney}, {Rhoads}, {Rhodes}, {Ryan}, {Sahu}, {Sand}, {Scolnic}, {Seth}, {Shvartzvald}, {Siellez}, {Smith}, {Spergel}, {Stassun}, {Street}, {Strolger}, {Szalay},
  {Trauger}, {Troxel}, {Turnbull}, {van der Marel}, {von der Linden}, {Wang}, {Weinberg}, {Williams}, {Windhorst}, {Wollack}, {Wu}, {Yee}, \& {Zimmerman}}]{Akeson2019}
{Akeson}, R., {Armus}, L., {Bachelet}, E., {et~al.} 2019, arXiv e-prints, arXiv:1902.05569, \dodoi{10.48550/arXiv.1902.05569}

\bibitem[{{Amard} {et~al.}(2019){Amard}, {Palacios}, {Charbonnel}, {Gallet}, {Georgy}, {Lagarde}, \& {Siess}}]{Amard2019}
{Amard}, L., {Palacios}, A., {Charbonnel}, C., {et~al.} 2019, \aap, 631, A77, \dodoi{10.1051/0004-6361/201935160}

\bibitem[{{Amard} {et~al.}(2020){Amard}, {Roquette}, \& {Matt}}]{Amard2020}
{Amard}, L., {Roquette}, J., \& {Matt}, S.~P. 2020, \mnras, 499, 3481, \dodoi{10.1093/mnras/staa3038}

\bibitem[{{Andrae} {et~al.}(2023){Andrae}, {Rix}, \& {Chandra}}]{Andrae2023}
{Andrae}, R., {Rix}, H.-W., \& {Chandra}, V. 2023, \apjs, 267, 8, \dodoi{10.3847/1538-4365/acd53e}

\bibitem[{{Angus} {et~al.}(2015){Angus}, {Aigrain}, {Foreman-Mackey}, \& {McQuillan}}]{Angus2015}
{Angus}, R., {Aigrain}, S., {Foreman-Mackey}, D., \& {McQuillan}, A. 2015, \mnras, 450, 1787, \dodoi{10.1093/mnras/stv423}

\bibitem[{{Angus} {et~al.}(2019){Angus}, {Morton}, {Foreman-Mackey}, {van Saders}, {Curtis}, {Kane}, {Bedell}, {Kiman}, {Hogg}, \& {Brewer}}]{Angus2019}
{Angus}, R., {Morton}, T.~D., {Foreman-Mackey}, D., {et~al.} 2019, \aj, 158, 173, \dodoi{10.3847/1538-3881/ab3c53}

\bibitem[{{Angus} {et~al.}(2020){Angus}, {Beane}, {Price-Whelan}, {Newton}, {Curtis}, {Berger}, {van Saders}, {Kiman}, {Foreman-Mackey}, {Lu}, {Anderson}, \& {Faherty}}]{Angus2020}
{Angus}, R., {Beane}, A., {Price-Whelan}, A.~M., {et~al.} 2020, \aj, 160, 90, \dodoi{10.3847/1538-3881/ab91b2}

\bibitem[{{Astropy Collaboration} {et~al.}(2013){Astropy Collaboration}, {Robitaille}, {Tollerud}, {Greenfield}, {Droettboom}, {Bray}, {Aldcroft}, {Davis}, {Ginsburg}, {Price-Whelan}, {Kerzendorf}, {Conley}, {Crighton}, {Barbary}, {Muna}, {Ferguson}, {Grollier}, {Parikh}, {Nair}, {Unther}, {Deil}, {Woillez}, {Conseil}, {Kramer}, {Turner}, {Singer}, {Fox}, {Weaver}, {Zabalza}, {Edwards}, {Azalee Bostroem}, {Burke}, {Casey}, {Crawford}, {Dencheva}, {Ely}, {Jenness}, {Labrie}, {Lim}, {Pierfederici}, {Pontzen}, {Ptak}, {Refsdal}, {Servillat}, \& {Streicher}}]{astropy:2013}
{Astropy Collaboration}, {Robitaille}, T.~P., {Tollerud}, E.~J., {et~al.} 2013, \aap, 558, A33, \dodoi{10.1051/0004-6361/201322068}

\bibitem[{{Astropy Collaboration} {et~al.}(2022){Astropy Collaboration}, {Price-Whelan}, {Lim}, {Earl}, {Starkman}, {Bradley}, {Shupe}, {Patil}, {Corrales}, {Brasseur}, {N{\"o}the}, {Donath}, {Tollerud}, {Morris}, {Ginsburg}, {Vaher}, {Weaver}, {Tocknell}, {Jamieson}, {van Kerkwijk}, {Robitaille}, {Merry}, {Bachetti}, {G{\"u}nther}, {Aldcroft}, {Alvarado-Montes}, {Archibald}, {B{\'o}di}, {Bapat}, {Barentsen}, {Baz{\'a}n}, {Biswas}, {Boquien}, {Burke}, {Cara}, {Cara}, {Conroy}, {Conseil}, {Craig}, {Cross}, {Cruz}, {D'Eugenio}, {Dencheva}, {Devillepoix}, {Dietrich}, {Eigenbrot}, {Erben}, {Ferreira}, {Foreman-Mackey}, {Fox}, {Freij}, {Garg}, {Geda}, {Glattly}, {Gondhalekar}, {Gordon}, {Grant}, {Greenfield}, {Groener}, {Guest}, {Gurovich}, {Handberg}, {Hart}, {Hatfield-Dodds}, {Homeier}, {Hosseinzadeh}, {Jenness}, {Jones}, {Joseph}, {Kalmbach}, {Karamehmetoglu}, {Ka{\l}uszy{\'n}ski}, {Kelley}, {Kern}, {Kerzendorf}, {Koch}, {Kulumani}, {Lee}, {Ly}, {Ma}, {MacBride}, {Maljaars}, {Muna}, {Murphy}, {Norman},
  {O'Steen}, {Oman}, {Pacifici}, {Pascual}, {Pascual-Granado}, {Patil}, {Perren}, {Pickering}, {Rastogi}, {Roulston}, {Ryan}, {Rykoff}, {Sabater}, {Sakurikar}, {Salgado}, {Sanghi}, {Saunders}, {Savchenko}, {Schwardt}, {Seifert-Eckert}, {Shih}, {Jain}, {Shukla}, {Sick}, {Simpson}, {Singanamalla}, {Singer}, {Singhal}, {Sinha}, {Sip{\H{o}}cz}, {Spitler}, {Stansby}, {Streicher}, {{\v{S}}umak}, {Swinbank}, {Taranu}, {Tewary}, {Tremblay}, {de Val-Borro}, {Van Kooten}, {Vasovi{\'c}}, {Verma}, {de Miranda Cardoso}, {Williams}, {Wilson}, {Winkel}, {Wood-Vasey}, {Xue}, {Yoachim}, {Zhang}, {Zonca}, \& {Astropy Project Contributors}}]{astropy2022}
{Astropy Collaboration}, {Price-Whelan}, A.~M., {Lim}, P.~L., {et~al.} 2022, \apj, 935, 167, \dodoi{10.3847/1538-4357/ac7c74}

\bibitem[{{Barnes}(2010)}]{Barnes2010}
{Barnes}, S.~A. 2010, \apj, 722, 222, \dodoi{10.1088/0004-637X/722/1/222}

\bibitem[{{Behmard} {et~al.}(2025){Behmard}, {Ness}, {Casey}, {Angus}, {Cunha}, {Souto}, {Lu}, \& {Johnson}}]{Behmard2025}
{Behmard}, A., {Ness}, M.~K., {Casey}, A.~R., {et~al.} 2025, \apj, 982, 13, \dodoi{10.3847/1538-4357/adaf1f}

\bibitem[{{Bellm} {et~al.}(2019){Bellm}, {Kulkarni}, {Graham}, {Dekany}, {Smith}, {Riddle}, {Masci}, {Helou}, {Prince}, {Adams}, {Barbarino}, {Barlow}, {Bauer}, {Beck}, {Belicki}, {Biswas}, {Blagorodnova}, {Bodewits}, {Bolin}, {Brinnel}, {Brooke}, {Bue}, {Bulla}, {Burruss}, {Cenko}, {Chang}, {Connolly}, {Coughlin}, {Cromer}, {Cunningham}, {De}, {Delacroix}, {Desai}, {Duev}, {Eadie}, {Farnham}, {Feeney}, {Feindt}, {Flynn}, {Franckowiak}, {Frederick}, {Fremling}, {Gal-Yam}, {Gezari}, {Giomi}, {Goldstein}, {Golkhou}, {Goobar}, {Groom}, {Hacopians}, {Hale}, {Henning}, {Ho}, {Hover}, {Howell}, {Hung}, {Huppenkothen}, {Imel}, {Ip}, {Ivezi{\'c}}, {Jackson}, {Jones}, {Juric}, {Kasliwal}, {Kaspi}, {Kaye}, {Kelley}, {Kowalski}, {Kramer}, {Kupfer}, {Landry}, {Laher}, {Lee}, {Lin}, {Lin}, {Lunnan}, {Giomi}, {Mahabal}, {Mao}, {Miller}, {Monkewitz}, {Murphy}, {Ngeow}, {Nordin}, {Nugent}, {Ofek}, {Patterson}, {Penprase}, {Porter}, {Rauch}, {Rebbapragada}, {Reiley}, {Rigault}, {Rodriguez}, {van Roestel}, {Rusholme}, {van
  Santen}, {Schulze}, {Shupe}, {Singer}, {Soumagnac}, {Stein}, {Surace}, {Sollerman}, {Szkody}, {Taddia}, {Terek}, {Van Sistine}, {van Velzen}, {Vestrand}, {Walters}, {Ward}, {Ye}, {Yu}, {Yan}, \& {Zolkower}}]{Bellm2019}
{Bellm}, E.~C., {Kulkarni}, S.~R., {Graham}, M.~J., {et~al.} 2019, \pasp, 131, 018002, \dodoi{10.1088/1538-3873/aaecbe}

\bibitem[{{Berger} {et~al.}(2020){Berger}, {Huber}, {Gaidos}, {van Saders}, \& {Weiss}}]{Berger2020}
{Berger}, T.~A., {Huber}, D., {Gaidos}, E., {van Saders}, J.~L., \& {Weiss}, L.~M. 2020, arXiv e-prints, arXiv:2005.14671.
\newblock \doarXiv{2005.14671}

\bibitem[{{Beyer} \& {White}(2024)}]{Beyer2024}
{Beyer}, A.~C., \& {White}, R.~J. 2024, \apj, 973, 28, \dodoi{10.3847/1538-4357/ad6b0d}

\bibitem[{{Bird} {et~al.}(2013){Bird}, {Kazantzidis}, {Weinberg}, {Guedes}, {Callegari}, {Mayer}, \& {Madau}}]{Bird2013}
{Bird}, J.~C., {Kazantzidis}, S., {Weinberg}, D.~H., {et~al.} 2013, \apj, 773, 43, \dodoi{10.1088/0004-637X/773/1/43}

\bibitem[{{Borucki} {et~al.}(2010){Borucki}, {Koch}, {Basri}, {Batalha}, {Brown}, {Caldwell}, {Caldwell}, {Christensen-Dalsgaard}, {Cochran}, {DeVore}, {Dunham}, {Dupree}, {Gautier}, {Geary}, {Gilliland}, {Gould}, {Howell}, {Jenkins}, {Kondo}, {Latham}, {Marcy}, {Meibom}, {Kjeldsen}, {Lissauer}, {Monet}, {Morrison}, {Sasselov}, {Tarter}, {Boss}, {Brownlee}, {Owen}, {Buzasi}, {Charbonneau}, {Doyle}, {Fortney}, {Ford}, {Holman}, {Seager}, {Steffen}, {Welsh}, {Rowe}, {Anderson}, {Buchhave}, {Ciardi}, {Walkowicz}, {Sherry}, {Horch}, {Isaacson}, {Everett}, {Fischer}, {Torres}, {Johnson}, {Endl}, {MacQueen}, {Bryson}, {Dotson}, {Haas}, {Kolodziejczak}, {Van Cleve}, {Chandrasekaran}, {Twicken}, {Quintana}, {Clarke}, {Allen}, {Li}, {Wu}, {Tenenbaum}, {Verner}, {Bruhweiler}, {Barnes}, \& {Prsa}}]{kepler}
{Borucki}, W.~J., {Koch}, D., {Basri}, G., {et~al.} 2010, Science, 327, 977, \dodoi{10.1126/science.1185402}

\bibitem[{{Bouma} {et~al.}(2024){Bouma}, {Hillenbrand}, {Howard}, {Isaacson}, {Masuda}, \& {Palumbo}}]{Bouma2024}
{Bouma}, L.~G., {Hillenbrand}, L.~A., {Howard}, A.~W., {et~al.} 2024, \apj, 976, 234, \dodoi{10.3847/1538-4357/ad855f}

\bibitem[{{Bouma} {et~al.}(2023){Bouma}, {Palumbo}, \& {Hillenbrand}}]{Bouma2023}
{Bouma}, L.~G., {Palumbo}, E.~K., \& {Hillenbrand}, L.~A. 2023, \apjl, 947, L3, \dodoi{10.3847/2041-8213/acc589}

\bibitem[{{Bovy}(2015)}]{galpy}
{Bovy}, J. 2015, \apjs, 216, 29, \dodoi{10.1088/0067-0049/216/2/29}

\bibitem[{{Brook} {et~al.}(2004){Brook}, {Kawata}, {Gibson}, \& {Freeman}}]{Brook2004}
{Brook}, C.~B., {Kawata}, D., {Gibson}, B.~K., \& {Freeman}, K.~C. 2004, \apj, 612, 894, \dodoi{10.1086/422709}

\bibitem[{{Buck} {et~al.}(2020){Buck}, {Obreja}, {Macci{\`o}}, {Minchev}, {Dutton}, \& {Ostriker}}]{Buck2020}
{Buck}, T., {Obreja}, A., {Macci{\`o}}, A.~V., {et~al.} 2020, \mnras, 491, 3461, \dodoi{10.1093/mnras/stz3241}

\bibitem[{{Byrom} \& {Tayar}(2024)}]{Byrom2024}
{Byrom}, S., \& {Tayar}, J. 2024, Research Notes of the American Astronomical Society, 8, 201, \dodoi{10.3847/2515-5172/ad7093}

\bibitem[{{Cantat-Gaudin} {et~al.}(2020){Cantat-Gaudin}, {Anders}, {Castro-Ginard}, {Jordi}, {Romero-G{\'o}mez}, {Soubiran}, {Casamiquela}, {Tarricq}, {Moitinho}, {Vallenari}, {Bragaglia}, {Krone-Martins}, \& {Kounkel}}]{CantatGaudin2020}
{Cantat-Gaudin}, T., {Anders}, F., {Castro-Ginard}, A., {et~al.} 2020, \aap, 640, A1, \dodoi{10.1051/0004-6361/202038192}

\bibitem[{{Cao} {et~al.}(2023){Cao}, {Pinsonneault}, \& {van Saders}}]{Cao2023}
{Cao}, L., {Pinsonneault}, M.~H., \& {van Saders}, J.~L. 2023, \apjl, 951, L49, \dodoi{10.3847/2041-8213/acd780}

\bibitem[{{Carbon} {et~al.}(1982){Carbon}, {Langer}, {Butler}, {Kraft}, {Suntzeff}, {Kemper}, {Trefzger}, \& {Romanishin}}]{Carbon1982}
{Carbon}, D.~F., {Langer}, G.~E., {Butler}, D., {et~al.} 1982, \apjs, 49, 207, \dodoi{10.1086/190796}

\bibitem[{{Casali} {et~al.}(2019){Casali}, {Magrini}, {Tognelli}, {Jackson}, {Jeffries}, {Lagarde}, {Tautvai{\v{s}}ien{\.{e}}}, {Masseron}, {Degl'Innocenti}, {Prada Moroni}, {Kordopatis}, {Pancino}, {Randich}, {Feltzing}, {Sahlholdt}, {Spina}, {Friel}, {Roccatagliata}, {Sanna}, {Bragaglia}, {Drazdauskas}, {Mikolaitis}, {Minkevi{\v{c}}i{\={u}}t{\.{e}}}, {Stonkut{\.{e}}}, {Chorniy}, {Bagdonas}, {Jimenez-Esteban}, {Martell}, {Van der Swaelmen}, {Gilmore}, {Vallenari}, {Bensby}, {Koposov}, {Korn}, {Worley}, {Smiljanic}, {Bergemann}, {Carraro}, {Damiani}, {Prisinzano}, {Bonito}, {Franciosini}, {Gonneau}, {Hourihane}, {Jofre}, {Lewis}, {Morbidelli}, {Sacco}, {Sousa}, {Zaggia}, {Lanzafame}, {Heiter}, {Frasca}, \& {Bayo}}]{Casali2019}
{Casali}, G., {Magrini}, L., {Tognelli}, E., {et~al.} 2019, \aap, 629, A62, \dodoi{10.1051/0004-6361/201935282}

\bibitem[{{Cavallo} {et~al.}(2024){Cavallo}, {Spina}, {Carraro}, {Magrini}, {Poggio}, {Cantat-Gaudin}, {Pasquato}, {Lucatello}, {Ortolani}, \& {Schiappacasse-Ulloa}}]{Cavallo2024}
{Cavallo}, L., {Spina}, L., {Carraro}, G., {et~al.} 2024, \aj, 167, 12, \dodoi{10.3847/1538-3881/ad07e5}

\bibitem[{{Chaplin} {et~al.}(2014){Chaplin}, {Basu}, {Huber}, {Serenelli}, {Casagrande}, {Silva Aguirre}, {Ball}, {Creevey}, {Gizon}, {Handberg}, {Karoff}, {Lutz}, {Marques}, {Miglio}, {Stello}, {Suran}, {Pricopi}, {Metcalfe}, {Monteiro}, {Molenda-{\.Z}akowicz}, {Appourchaux}, {Christensen-Dalsgaard}, {Elsworth}, {Garc{\'\i}a}, {Houdek}, {Kjeldsen}, {Bonanno}, {Campante}, {Corsaro}, {Gaulme}, {Hekker}, {Mathur}, {Mosser}, {R{\'e}gulo}, \& {Salabert}}]{Chaplin2014}
{Chaplin}, W.~J., {Basu}, S., {Huber}, D., {et~al.} 2014, \apjs, 210, 1, \dodoi{10.1088/0067-0049/210/1/1}

\bibitem[{{Chiti} {et~al.}(2024){Chiti}, {van Saders}, {Heintz}, {Hermes}, {Ong}, {Hey}, {Ramirez-Weinhouse}, \& {Dugas}}]{Chiti2024}
{Chiti}, F., {van Saders}, J.~L., {Heintz}, T.~M., {et~al.} 2024, \apj, 977, 15, \dodoi{10.3847/1538-4357/ad856c}

\bibitem[{{Claytor} {et~al.}(2020){Claytor}, {van Saders}, {Santos}, {Garc{\'\i}a}, {Mathur}, {Tayar}, {Pinsonneault}, \& {Shetrone}}]{Claytor2020}
{Claytor}, Z.~R., {van Saders}, J.~L., {Santos}, {\^A}. R.~G., {et~al.} 2020, Astrophysical Journal, 888, 43, \dodoi{10.3847/1538-4357/ab5c24}

\bibitem[{{Colman} {et~al.}(2024){Colman}, {Angus}, {David}, {Curtis}, {Hattori}, \& {Lu}}]{Colman2024}
{Colman}, I.~L., {Angus}, R., {David}, T., {et~al.} 2024, \aj, 167, 189, \dodoi{10.3847/1538-3881/ad2c86}

\bibitem[{{Cort{\'e}s-Contreras} {et~al.}(2024){Cort{\'e}s-Contreras}, {Caballero}, {Montes}, {Cardona-Guill{\'e}n}, {B{\'e}jar}, {Cifuentes}, {Tabernero}, {Zapatero Osorio}, {Amado}, {Jeffers}, {Lafarga}, {Lodieu}, {Quirrenbach}, {Reiners}, {Ribas}, {Sch{\"o}fer}, {Schweitzer}, \& {Seifert}}]{CortesContreras2024}
{Cort{\'e}s-Contreras}, M., {Caballero}, J.~A., {Montes}, D., {et~al.} 2024, \aap, 692, A206, \dodoi{10.1051/0004-6361/202451585}

\bibitem[{{Curtis} {et~al.}(2020){Curtis}, {Ag{\"u}eros}, {Matt}, {Covey}, {Douglas}, {Angus}, {Saar}, {Cody}, {Vanderburg}, {Law}, {Kraus}, {Latham}, {Baranec}, {Riddle}, {Ziegler}, {Lund}, {Torres}, {Meibom}, {Aguirre}, \& {Wright}}]{Curtis2020}
{Curtis}, J.~L., {Ag{\"u}eros}, M.~A., {Matt}, S.~P., {et~al.} 2020, \apj, 904, 140, \dodoi{10.3847/1538-4357/abbf58}

\bibitem[{{De Silva} {et~al.}(2015){De Silva}, {Freeman}, {Bland-Hawthorn}, {Martell}, {de Boer}, {Asplund}, {Keller}, {Sharma}, {Zucker}, {Zwitter}, {Anguiano}, {Bacigalupo}, {Bayliss}, {Beavis}, {Bergemann}, {Campbell}, {Cannon}, {Carollo}, {Casagrande}, {Casey}, {Da Costa}, {D'Orazi}, {Dotter}, {Duong}, {Heger}, {Ireland}, {Kafle}, {Kos}, {Lattanzio}, {Lewis}, {Lin}, {Lind}, {Munari}, {Nataf}, {O'Toole}, {Parker}, {Reid}, {Schlesinger}, {Sheinis}, {Simpson}, {Stello}, {Ting}, {Traven}, {Watson}, {Wittenmyer}, {Yong}, \& {{\v{Z}}erjal}}]{galah}
{De Silva}, G.~M., {Freeman}, K.~C., {Bland-Hawthorn}, J., {et~al.} 2015, \mnras, 449, 2604, \dodoi{10.1093/mnras/stv327}

\bibitem[{{El-Badry} \& {Rix}(2018)}]{ElBadry2018}
{El-Badry}, K., \& {Rix}, H.-W. 2018, \mnras, 480, 4884, \dodoi{10.1093/mnras/sty2186}

\bibitem[{{Epstein} \& {Pinsonneault}(2014)}]{Epstein2014}
{Epstein}, C.~R., \& {Pinsonneault}, M.~H. 2014, \apj, 780, 159, \dodoi{10.1088/0004-637X/780/2/159}

\bibitem[{{Foreman-Mackey}(2016)}]{corner}
{Foreman-Mackey}, D. 2016, The Journal of Open Source Software, 1, 24, \dodoi{10.21105/joss.00024}

\bibitem[{{Foreman-Mackey} {et~al.}(2013){Foreman-Mackey}, {Hogg}, {Lang}, \& {Goodman}}]{emcee}
{Foreman-Mackey}, D., {Hogg}, D.~W., {Lang}, D., \& {Goodman}, J. 2013, \pasp, 125, 306, \dodoi{10.1086/670067}

\bibitem[{{Gaia Collaboration} {et~al.}(2016){Gaia Collaboration}, {Prusti}, {de Bruijne}, {Brown}, {Vallenari}, {Babusiaux}, {Bailer-Jones}, {Bastian}, {Biermann}, {Evans}, {Eyer}, {Jansen}, {Jordi}, {Klioner}, {Lammers}, {Lindegren}, {Luri}, {Mignard}, {Milligan}, {Panem}, {Poinsignon}, {Pourbaix}, {Randich}, {Sarri}, {Sartoretti}, {Siddiqui}, {Soubiran}, {Valette}, {van Leeuwen}, {Walton}, {Aerts}, {Arenou}, {Cropper}, {Drimmel}, {H{\o}g}, {Katz}, {Lattanzi}, {O'Mullane}, {Grebel}, {Holland}, {Huc}, {Passot}, {Bramante}, {Cacciari}, {Casta{\~n}eda}, {Chaoul}, {Cheek}, {De Angeli}, {Fabricius}, {Guerra}, {Hern{\'a}ndez}, {Jean-Antoine-Piccolo}, {Masana}, {Messineo}, {Mowlavi}, {Nienartowicz}, {Ord{\'o}{\~n}ez-Blanco}, {Panuzzo}, {Portell}, {Richards}, {Riello}, {Seabroke}, {Tanga}, {Th{\'e}venin}, {Torra}, {Els}, {Gracia-Abril}, {Comoretto}, {Garcia-Reinaldos}, {Lock}, {Mercier}, {Altmann}, {Andrae}, {Astraatmadja}, {Bellas-Velidis}, {Benson}, {Berthier}, {Blomme}, {Busso}, {Carry}, {Cellino}, {Clementini},
  {Cowell}, {Creevey}, {Cuypers}, {Davidson}, {De Ridder}, {de Torres}, {Delchambre}, {Dell'Oro}, {Ducourant}, {Fr{\'e}mat}, {Garc{\'\i}a-Torres}, {Gosset}, {Halbwachs}, {Hambly}, {Harrison}, {Hauser}, {Hestroffer}, {Hodgkin}, {Huckle}, {Hutton}, {Jasniewicz}, {Jordan}, {Kontizas}, {Korn}, {Lanzafame}, {Manteiga}, {Moitinho}, {Muinonen}, {Osinde}, {Pancino}, {Pauwels}, {Petit}, {Recio-Blanco}, {Robin}, {Sarro}, {Siopis}, {Smith}, {Smith}, {Sozzetti}, {Thuillot}, {van Reeven}, {Viala}, {Abbas}, {Abreu Aramburu}, {Accart}, {Aguado}, {Allan}, {Allasia}, {Altavilla}, {{\'A}lvarez}, {Alves}, {Anderson}, {Andrei}, {Anglada Varela}, {Antiche}, {Antoja}, {Ant{\'o}n}, {Arcay}, {Atzei}, {Ayache}, {Bach}, {Baker}, {Balaguer-N{\'u}{\~n}ez}, {Barache}, {Barata}, {Barbier}, {Barblan}, {Baroni}, {Barrado y Navascu{\'e}s}, {Barros}, {Barstow}, {Becciani}, {Bellazzini}, {Bellei}, {Bello Garc{\'\i}a}, {Belokurov}, {Bendjoya}, {Berihuete}, {Bianchi}, {Bienaym{\'e}}, {Billebaud}, {Blagorodnova}, {Blanco-Cuaresma}, {Boch},
  {Bombrun}, {Borrachero}, {Bouquillon}, {Bourda}, {Bouy}, {Bragaglia}, {Breddels}, {Brouillet}, {Br{\"u}semeister}, {Bucciarelli}, {Budnik}, {Burgess}, {Burgon}, {Burlacu}, {Busonero}, {Buzzi}, {Caffau}, {Cambras}, {Campbell}, {Cancelliere}, {Cantat-Gaudin}, {Carlucci}, {Carrasco}, {Castellani}, {Charlot}, {Charnas}, {Charvet}, {Chassat}, {Chiavassa}, {Clotet}, {Cocozza}, {Collins}, {Collins}, \& {Costigan}}]{gaia}
{Gaia Collaboration}, {Prusti}, T., {de Bruijne}, J.~H.~J., {et~al.} 2016, \aap, 595, A1, \dodoi{10.1051/0004-6361/201629272}

\bibitem[{{Gaia Collaboration} {et~al.}(2023){Gaia Collaboration}, {Vallenari}, {Brown}, {Prusti}, {de Bruijne}, {Arenou}, {Babusiaux}, {Biermann}, {Creevey}, {Ducourant}, {Evans}, {Eyer}, {Guerra}, {Hutton}, {Jordi}, {Klioner}, {Lammers}, {Lindegren}, {Luri}, {Mignard}, {Panem}, {Pourbaix}, {Randich}, {Sartoretti}, {Soubiran}, {Tanga}, {Walton}, {Bailer-Jones}, {Bastian}, {Drimmel}, {Jansen}, {Katz}, {Lattanzi}, {van Leeuwen}, {Bakker}, {Cacciari}, {Casta{\~n}eda}, {De Angeli}, {Fabricius}, {Fouesneau}, {Fr{\'e}mat}, {Galluccio}, {Guerrier}, {Heiter}, {Masana}, {Messineo}, {Mowlavi}, {Nicolas}, {Nienartowicz}, {Pailler}, {Panuzzo}, {Riclet}, {Roux}, {Seabroke}, {Sordo}, {Th{\'e}venin}, {Gracia-Abril}, {Portell}, {Teyssier}, {Altmann}, {Andrae}, {Audard}, {Bellas-Velidis}, {Benson}, {Berthier}, {Blomme}, {Burgess}, {Busonero}, {Busso}, {C{\'a}novas}, {Carry}, {Cellino}, {Cheek}, {Clementini}, {Damerdji}, {Davidson}, {de Teodoro}, {Nu{\~n}ez Campos}, {Delchambre}, {Dell'Oro}, {Esquej},
  {Fern{\'a}ndez-Hern{\'a}ndez}, {Fraile}, {Garabato}, {Garc{\'\i}a-Lario}, {Gosset}, {Haigron}, {Halbwachs}, {Hambly}, {Harrison}, {Hern{\'a}ndez}, {Hestroffer}, {Hodgkin}, {Holl}, {Jan{\ss}en}, {Jevardat de Fombelle}, {Jordan}, {Krone-Martins}, {Lanzafame}, {L{\"o}ffler}, {Marchal}, {Marrese}, {Moitinho}, {Muinonen}, {Osborne}, {Pancino}, {Pauwels}, {Recio-Blanco}, {Reyl{\'e}}, {Riello}, {Rimoldini}, {Roegiers}, {Rybizki}, {Sarro}, {Siopis}, {Smith}, {Sozzetti}, {Utrilla}, {van Leeuwen}, {Abbas}, {{\'A}brah{\'a}m}, {Abreu Aramburu}, {Aerts}, {Aguado}, {Ajaj}, {Aldea-Montero}, {Altavilla}, {{\'A}lvarez}, {Alves}, {Anders}, {Anderson}, {Anglada Varela}, {Antoja}, {Baines}, {Baker}, {Balaguer-N{\'u}{\~n}ez}, {Balbinot}, {Balog}, {Barache}, {Barbato}, {Barros}, {Barstow}, {Bartolom{\'e}}, {Bassilana}, {Bauchet}, {Becciani}, {Bellazzini}, {Berihuete}, {Bernet}, {Bertone}, {Bianchi}, {Binnenfeld}, {Blanco-Cuaresma}, {Blazere}, {Boch}, {Bombrun}, {Bossini}, {Bouquillon}, {Bragaglia}, {Bramante}, {Breedt},
  {Bressan}, {Brouillet}, {Brugaletta}, {Bucciarelli}, {Burlacu}, {Butkevich}, {Buzzi}, {Caffau}, {Cancelliere}, {Cantat-Gaudin}, {Carballo}, {Carlucci}, {Carnerero}, {Carrasco}, {Casamiquela}, {Castellani}, {Castro-Ginard}, {Chaoul}, {Charlot}, {Chemin}, {Chiaramida}, {Chiavassa}, {Chornay}, {Comoretto}, {Contursi}, {Cooper}, {Cornez}, {Cowell}, {Crifo}, {Cropper}, {Crosta}, {Crowley}, {Dafonte}, {Dapergolas}, {David}, {David}, {de Laverny}, {De Luise}, {De March}, {De Ridder}, {de Souza}, {de Torres}, {del Peloso}, {del Pozo}, {Delbo}, {Delgado}, {Delisle}, {Demouchy}, {Dharmawardena}, {Di Matteo}, {Diakite}, {Diener}, {Distefano}, {Dolding}, {Edvardsson}, {Enke}, {Fabre}, {Fabrizio}, {Faigler}, {Fedorets}, {Fernique}, {Fienga}, {Figueras}, {Fournier}, {Fouron}, {Fragkoudi}, {Gai}, {Garcia-Gutierrez}, {Garcia-Reinaldos}, {Garc{\'\i}a-Torres}, {Garofalo}, {Gavel}, {Gavras}, {Gerlach}, {Geyer}, {Giacobbe}, {Gilmore}, {Girona}, {Giuffrida}, {Gomel}, {Gomez}, {Gonz{\'a}lez-N{\'u}{\~n}ez},
  {Gonz{\'a}lez-Santamar{\'\i}a}, {Gonz{\'a}lez-Vidal}, {Granvik}, {Guillout}, {Guiraud}, {Guti{\'e}rrez-S{\'a}nchez}, {Guy}, {Hatzidimitriou}, {Hauser}, {Haywood}, {Helmer}, {Helmi}, {Sarmiento}, {Hidalgo}, {Hilger}, {H{\l}adczuk}, {Hobbs}, {Holland}, {Huckle}, {Jardine}, {Jasniewicz}, {Jean-Antoine Piccolo}, {Jim{\'e}nez-Arranz}, {Jorissen}, {Juaristi Campillo}, {Julbe}, {Karbevska}, {Kervella}, {Khanna}, {Kontizas}, {Kordopatis}, {Korn}, {K{\'o}sp{\'a}l}, {Kostrzewa-Rutkowska}, {Kruszy{\'n}ska}, {Kun}, {Laizeau}, {Lambert}, {Lanza}, {Lasne}, {Le Campion}, {Lebreton}, {Lebzelter}, {Leccia}, {Leclerc}, {Lecoeur-Taibi}, {Liao}, {Licata}, {Lindstr{\o}m}, {Lister}, {Livanou}, {Lobel}, {Lorca}, {Loup}, {Madrero Pardo}, {Magdaleno Romeo}, {Managau}, {Mann}, {Manteiga}, {Marchant}, {Marconi}, {Marcos}, {Marcos Santos}, {Mar{\'\i}n Pina}, {Marinoni}, {Marocco}, {Marshall}, {Martin Polo}, {Mart{\'\i}n-Fleitas}, {Marton}, {Mary}, {Masip}, {Massari}, {Mastrobuono-Battisti}, {Mazeh}, {McMillan}, {Messina}, {Michalik},
  {Millar}, {Mints}, {Molina}, {Molinaro}, {Moln{\'a}r}, {Monari}, {Mongui{\'o}}, {Montegriffo}, {Montero}, {Mor}, {Mora}, {Morbidelli}, {Morel}, {Morris}, {Muraveva}, {Murphy}, {Musella}, {Nagy}, {Noval}, {Oca{\~n}a}, {Ogden}, {Ordenovic}, {Osinde}, {Pagani}, {Pagano}, {Palaversa}, {Palicio}, {Pallas-Quintela}, {Panahi}, {Payne-Wardenaar}, {Pe{\~n}alosa Esteller}, {Penttil{\"a}}, {Pichon}, {Piersimoni}, {Pineau}, {Plachy}, {Plum}, {Poggio}, {Pr{\v{s}}a}, {Pulone}, {Racero}, {Ragaini}, {Rainer}, {Raiteri}, {Rambaux}, {Ramos}, {Ramos-Lerate}, {Re Fiorentin}, {Regibo}, {Richards}, {Rios Diaz}, {Ripepi}, {Riva}, {Rix}, {Rixon}, {Robichon}, {Robin}, {Robin}, {Roelens}, {Rogues}, {Rohrbasser}, {Romero-G{\'o}mez}, {Rowell}, {Royer}, {Ruz Mieres}, {Rybicki}, {Sadowski}, {S{\'a}ez N{\'u}{\~n}ez}, {Sagrist{\`a} Sell{\'e}s}, {Sahlmann}, {Salguero}, {Samaras}, {Sanchez Gimenez}, {Sanna}, {Santove{\~n}a}, {Sarasso}, {Schultheis}, {Sciacca}, {Segol}, {Segovia}, {S{\'e}gransan}, {Semeux}, {Shahaf}, {Siddiqui}, {Siebert},
  {Siltala}, {Silvelo}, {Slezak}, {Slezak}, {Smart}, {Snaith}, {Solano}, {Solitro}, {Souami}, {Souchay}, {Spagna}, {Spina}, {Spoto}, {Steele}, {Steidelm{\"u}ller}, {Stephenson}, {S{\"u}veges}, {Surdej}, {Szabados}, {Szegedi-Elek}, {Taris}, {Taylor}, {Teixeira}, {Tolomei}, {Tonello}, {Torra}, {Torra}, {Torralba Elipe}, {Trabucchi}, {Tsounis}, {Turon}, {Ulla}, {Unger}, {Vaillant}, {van Dillen}, {van Reeven}, {Vanel}, {Vecchiato}, {Viala}, {Vicente}, {Voutsinas}, {Weiler}, {Wevers}, {Wyrzykowski}, {Yoldas}, {Yvard}, {Zhao}, {Zorec}, {Zucker}, \& {Zwitter}}]{Gaiadr3}
{Gaia Collaboration}, {Vallenari}, A., {Brown}, A.~G.~A., {et~al.} 2023, \aap, 674, A1, \dodoi{10.1051/0004-6361/202243940}

\bibitem[{{Gaidos} {et~al.}(2023){Gaidos}, {Claytor}, {Dungee}, {Ali}, \& {Feiden}}]{Gaidos2023}
{Gaidos}, E., {Claytor}, Z., {Dungee}, R., {Ali}, A., \& {Feiden}, G.~A. 2023, \mnras, 520, 5283, \dodoi{10.1093/mnras/stad343}

\bibitem[{{Gossage} {et~al.}(2024){Gossage}, {Kiman}, {Monsch}, {Medina}, {Drake}, {Garraffo}, {Yuxi}, {Lu}, {Wing}, \& {Wright}}]{Gossage2024}
{Gossage}, S., {Kiman}, R., {Monsch}, K., {et~al.} 2024, arXiv e-prints, arXiv:2410.20000, \dodoi{10.48550/arXiv.2410.20000}

\bibitem[{{Grand} {et~al.}(2016){Grand}, {Springel}, {G{\'o}mez}, {Marinacci}, {Pakmor}, {Campbell}, \& {Jenkins}}]{Grand2016}
{Grand}, R. J.~J., {Springel}, V., {G{\'o}mez}, F.~A., {et~al.} 2016, \mnras, 459, 199, \dodoi{10.1093/mnras/stw601}

\bibitem[{{Gratton} {et~al.}(2000){Gratton}, {Sneden}, {Carretta}, \& {Bragaglia}}]{Gratton2000}
{Gratton}, R.~G., {Sneden}, C., {Carretta}, E., \& {Bragaglia}, A. 2000, \aap, 354, 169

\bibitem[{{Green}(2018)}]{Green2018}
{Green}, G. 2018, The Journal of Open Source Software, 3, 695, \dodoi{10.21105/joss.00695}

\bibitem[{{Green} {et~al.}(2018){Green}, {Schlafly}, {Finkbeiner}, {Rix}, {Martin}, {Burgett}, {Draper}, {Flewelling}, {Hodapp}, {Kaiser}, {Kudritzki}, {Magnier}, {Metcalfe}, {Tonry}, {Wainscoat}, \& {Waters}}]{Green20182}
{Green}, G.~M., {Schlafly}, E.~F., {Finkbeiner}, D., {et~al.} 2018, \mnras, 478, 651, \dodoi{10.1093/mnras/sty1008}

\bibitem[{{Gruner} {et~al.}(2023){Gruner}, {Barnes}, \& {Janes}}]{Gruner2023}
{Gruner}, D., {Barnes}, S.~A., \& {Janes}, K.~A. 2023, \aap, 675, A180, \dodoi{10.1051/0004-6361/202346590}

\bibitem[{Harris {et~al.}(2020)Harris, Millman, van~der Walt, Gommers, Virtanen, Cournapeau, Wieser, Taylor, Berg, Smith, Kern, Picus, Hoyer, van Kerkwijk, Brett, Haldane, del R{\'{i}}o, Wiebe, Peterson, G{\'{e}}rard-Marchant, Sheppard, Reddy, Weckesser, Abbasi, Gohlke, \& Oliphant}]{Numpy}
Harris, C.~R., Millman, K.~J., van~der Walt, S.~J., {et~al.} 2020, Nature, 585, 357, \dodoi{10.1038/s41586-020-2649-2}

\bibitem[{{Hekker} \& {Johnson}(2019)}]{Hekker2019}
{Hekker}, S., \& {Johnson}, J.~A. 2019, \mnras, 487, 4343, \dodoi{10.1093/mnras/stz1554}

\bibitem[{{Holcomb} {et~al.}(2022){Holcomb}, {Robertson}, {Hartigan}, {Oelkers}, \& {Robinson}}]{Holcomb2022}
{Holcomb}, R.~J., {Robertson}, P., {Hartigan}, P., {Oelkers}, R.~J., \& {Robinson}, C. 2022, \apj, 936, 138, \dodoi{10.3847/1538-4357/ac8990}

\bibitem[{{Howell} {et~al.}(2014){Howell}, {Sobeck}, {Haas}, {Still}, {Barclay}, {Mullally}, {Troeltzsch}, {Aigrain}, {Bryson}, {Caldwell}, {Chaplin}, {Cochran}, {Huber}, {Marcy}, {Miglio}, {Najita}, {Smith}, {Twicken}, \& {Fortney}}]{K2}
{Howell}, S.~B., {Sobeck}, C., {Haas}, M., {et~al.} 2014, \pasp, 126, 398, \dodoi{10.1086/676406}

\bibitem[{{Hunt} \& {Reffert}(2023)}]{Hunt2023}
{Hunt}, E.~L., \& {Reffert}, S. 2023, \aap, 673, A114, \dodoi{10.1051/0004-6361/202346285}

\bibitem[{{Hunt} {et~al.}(2022){Hunt}, {Price-Whelan}, {Johnston}, \& {Darragh-Ford}}]{Hunt:2022}
{Hunt}, J. A.~S., {Price-Whelan}, A.~M., {Johnston}, K.~V., \& {Darragh-Ford}, E. 2022, \mnras, 516, L7, \dodoi{10.1093/mnrasl/slac082}

\bibitem[{Hunter(2007)}]{matplotlib}
Hunter, J.~D. 2007, Computing in Science \& Engineering, 9, 90, \dodoi{10.1109/MCSE.2007.55}

\bibitem[{{Iorio} \& {Belokurov}(2021)}]{Iorio2021}
{Iorio}, G., \& {Belokurov}, V. 2021, \mnras, 502, 5686, \dodoi{10.1093/mnras/stab005}

\bibitem[{{IRSA}(2022{\natexlab{a}})}]{ztfdata}
{IRSA}. 2022{\natexlab{a}}, Zwicky Transient Facility Image Service,  IPAC, \dodoi{10.26131/IRSA539}

\bibitem[{{IRSA}(2022{\natexlab{b}})}]{ztftime}
---. 2022{\natexlab{b}}, Time Series Tool,  IPAC, \dodoi{10.26131/IRSA538}

\bibitem[{{Irwin} {et~al.}(2011){Irwin}, {Berta}, {Burke}, {Charbonneau}, {Nutzman}, {West}, \& {Falco}}]{Irwin2011}
{Irwin}, J., {Berta}, Z.~K., {Burke}, C.~J., {et~al.} 2011, \apj, 727, 56, \dodoi{10.1088/0004-637X/727/1/56}

\bibitem[{{Jofr{\'e}} {et~al.}(2023){Jofr{\'e}}, {Jorissen}, {Aguilera-G{\'o}mez}, {Van Eck}, {Tayar}, {Pinsonneault}, {Zinn}, {Goriely}, \& {Van Winckel}}]{Jofre2023}
{Jofr{\'e}}, P., {Jorissen}, A., {Aguilera-G{\'o}mez}, C., {et~al.} 2023, \aap, 671, A21, \dodoi{10.1051/0004-6361/202244524}

\bibitem[{{Kawaler}(1988)}]{Kawaler1988}
{Kawaler}, S.~D. 1988, \apj, 333, 236, \dodoi{10.1086/166740}

\bibitem[{{Kollmeier} {et~al.}(2017){Kollmeier}, {Zasowski}, {Rix}, {Johns}, {Anderson}, {Drory}, {Johnson}, {Pogge}, {Bird}, {Blanc}, {Brownstein}, {Crane}, {De Lee}, {Klaene}, {Kreckel}, {MacDonald}, {Merloni}, {Ness}, {O'Brien}, {Sanchez-Gallego}, {Sayres}, {Shen}, {Thakar}, {Tkachenko}, {Aerts}, {Blanton}, {Eisenstein}, {Holtzman}, {Maoz}, {Nandra}, {Rockosi}, {Weinberg}, {Bovy}, {Casey}, {Chaname}, {Clerc}, {Conroy}, {Eracleous}, {G{\"a}nsicke}, {Hekker}, {Horne}, {Kauffmann}, {McQuinn}, {Pellegrini}, {Schinnerer}, {Schlafly}, {Schwope}, {Seibert}, {Teske}, \& {van Saders}}]{MWM}
{Kollmeier}, J.~A., {Zasowski}, G., {Rix}, H.-W., {et~al.} 2017, arXiv e-prints, arXiv:1711.03234.
\newblock \doarXiv{1711.03234}

\bibitem[{{Kordopatis} {et~al.}(2023){Kordopatis}, {Schultheis}, {McMillan}, {Palicio}, {de Laverny}, {Recio-Blanco}, {Creevey}, {{\'A}lvarez}, {Andrae}, {Poggio}, {Spitoni}, {Contursi}, {Zhao}, {Oreshina-Slezak}, {Ordenovic}, \& {Bijaoui}}]{Kordopatis2023}
{Kordopatis}, G., {Schultheis}, M., {McMillan}, P.~J., {et~al.} 2023, \aap, 669, A104, \dodoi{10.1051/0004-6361/202244283}

\bibitem[{{Kraft}(1967)}]{Kraft1967}
{Kraft}, R.~P. 1967, \apj, 150, 551, \dodoi{10.1086/149359}

\bibitem[{{Kraft}(1994)}]{Kraft1994}
---. 1994, \pasp, 106, 553, \dodoi{10.1086/133416}

\bibitem[{{Lacey}(1984)}]{Lacey1984}
{Lacey}, C.~G. 1984, \mnras, 208, 687, \dodoi{10.1093/mnras/208.4.687}

\bibitem[{{Li} {et~al.}(2025){Li}, {Huber}, {Ong}, {van Saders}, {Costa}, {Reersted Larsen}, {Basu}, {Bedding}, {Dai}, {Chontos}, {Carmichael}, {Hey}, {Kjeldsen}, {Hon}, {Campante}, {Monteiro}, {Sloth Lundkvist}, {Saunders}, {Isaacson}, {Howard}, {Gibson}, {Halverson}, {Rider}, {Roy}, {Baker}, {Edelstein}, {Smith}, {Fulton}, \& {Walawender}}]{Li2025}
{Li}, Y., {Huber}, D., {Ong}, J.~M.~J., {et~al.} 2025, arXiv e-prints, arXiv:2502.00971, \dodoi{10.48550/arXiv.2502.00971}

\bibitem[{{Loebman} {et~al.}(2011){Loebman}, {Ro{\v{s}}kar}, {Debattista}, {Ivezi{\'c}}, {Quinn}, \& {Wadsley}}]{Loebman2011}
{Loebman}, S.~R., {Ro{\v{s}}kar}, R., {Debattista}, V.~P., {et~al.} 2011, \apj, 737, 8, \dodoi{10.1088/0004-637X/737/1/8}

\bibitem[{{LSST Science Collaboration} {et~al.}(2009){LSST Science Collaboration}, {Abell}, {Allison}, {Anderson}, {Andrew}, {Angel}, {Armus}, {Arnett}, {Asztalos}, {Axelrod}, {Bailey}, {Ballantyne}, {Bankert}, {Barkhouse}, {Barr}, {Barrientos}, {Barth}, {Bartlett}, {Becker}, {Becla}, {Beers}, {Bernstein}, {Biswas}, {Blanton}, {Bloom}, {Bochanski}, {Boeshaar}, {Borne}, {Bradac}, {Brandt}, {Bridge}, {Brown}, {Brunner}, {Bullock}, {Burgasser}, {Burge}, {Burke}, {Cargile}, {Chand rasekharan}, {Chartas}, {Chesley}, {Chu}, {Cinabro}, {Claire}, {Claver}, {Clowe}, {Connolly}, {Cook}, {Cooke}, {Cooray}, {Covey}, {Culliton}, {de Jong}, {de Vries}, {Debattista}, {Delgado}, {Dell'Antonio}, {Dhital}, {Di Stefano}, {Dickinson}, {Dilday}, {Djorgovski}, {Dobler}, {Donalek}, {Dubois-Felsmann}, {Durech}, {Eliasdottir}, {Eracleous}, {Eyer}, {Falco}, {Fan}, {Fassnacht}, {Ferguson}, {Fernandez}, {Fields}, {Finkbeiner}, {Figueroa}, {Fox}, {Francke}, {Frank}, {Frieman}, {Fromenteau}, {Furqan}, {Galaz}, {Gal-Yam}, {Garnavich},
  {Gawiser}, {Geary}, {Gee}, {Gibson}, {Gilmore}, {Grace}, {Green}, {Gressler}, {Grillmair}, {Habib}, {Haggerty}, {Hamuy}, {Harris}, {Hawley}, {Heavens}, {Hebb}, {Henry}, {Hileman}, {Hilton}, {Hoadley}, {Holberg}, {Holman}, {Howell}, {Infante}, {Ivezic}, {Jacoby}, {Jain}, {R}, {Jedicke}, {Jee}, {Garrett Jernigan}, {Jha}, {Johnston}, {Jones}, {Juric}, {Kaasalainen}, {Styliani}, {Kafka}, {Kahn}, {Kaib}, {Kalirai}, {Kantor}, {Kasliwal}, {Keeton}, {Kessler}, {Knezevic}, {Kowalski}, {Krabbendam}, {Krughoff}, {Kulkarni}, {Kuhlman}, {Lacy}, {Lepine}, {Liang}, {Lien}, {Lira}, {Long}, {Lorenz}, {Lotz}, {Lupton}, {Lutz}, {Macri}, {Mahabal}, {Mandelbaum}, {Marshall}, {May}, {McGehee}, {Meadows}, {Meert}, {Milani}, {Miller}, {Miller}, {Mills}, {Minniti}, {Monet}, {Mukadam}, {Nakar}, {Neill}, {Newman}, {Nikolaev}, {Nordby}, {O'Connor}, {Oguri}, {Oliver}, {Olivier}, {Olsen}, {Olsen}, {Olszewski}, {Oluseyi}, {Padilla}, {Parker}, {Pepper}, {Peterson}, {Petry}, {Pinto}, {Pizagno}, {Popescu}, {Prsa}, {Radcka}, {Raddick},
  {Rasmussen}, {Rau}, {Rho}, {Rhoads}, {Richards}, {Ridgway}, {Robertson}, {Roskar}, {Saha}, {Sarajedini}, {Scannapieco}, {Schalk}, {Schindler}, {Schmidt}, {Schmidt}, {Schneider}, {Schumacher}, {Scranton}, {Sebag}, {Seppala}, {Shemmer}, {Simon}, {Sivertz}, {Smith}, {Allyn Smith}, {Smith}, {Spitz}, {Stanford}, {Stassun}, {Strader}, {Strauss}, {Stubbs}, {Sweeney}, {Szalay}, {Szkody}, {Takada}, {Thorman}, {Trilling}, {Trimble}, {Tyson}, {Van Berg}, {Vand en Berk}, {VanderPlas}, {Verde}, {Vrsnak}, {Walkowicz}, {Wand elt}, {Wang}, {Wang}, {Warner}, {Wechsler}, {West}, {Wiecha}, {Williams}, {Willman}, {Wittman}, {Wolff}, {Wood-Vasey}, {Wozniak}, {Young}, {Zentner}, \& {Zhan}}]{LSST}
{LSST Science Collaboration}, {Abell}, P.~A., {Allison}, J., {et~al.} 2009, arXiv e-prints, arXiv:0912.0201.
\newblock \doarXiv{0912.0201}

\bibitem[{{Lu} {et~al.}(2023){Lu}, {Angus}, {Foreman-Mackey}, \& {Hattori}}]{Lu2023gyro}
{Lu}, Y., {Angus}, R., {Foreman-Mackey}, D., \& {Hattori}, S. 2023, arXiv e-prints, arXiv:2310.14990, \dodoi{10.48550/arXiv.2310.14990}

\bibitem[{{Lu} {et~al.}(2021{\natexlab{a}}){Lu}, {Ness}, {Buck}, \& {Zinn}}]{Lu2021_disk}
{Lu}, Y., {Ness}, M., {Buck}, T., \& {Zinn}, J. 2021{\natexlab{a}}, arXiv e-prints, arXiv:2102.12003.
\newblock \doarXiv{2102.12003}

\bibitem[{{Lu} {et~al.}(2024{\natexlab{a}}){Lu}, {Colman}, {Sayeed}, {Amard}, {Buder}, {Manea}, {Hattori}, {Pinsonneault}, {Price-Whelan}, {Bedell}, {Nidever}, {Johnson}, {Ness}, {Angus}, {Claytor}, {Horta}, \& {Behmard}}]{Lu2024_yhalpha}
{Lu}, Y., {Colman}, I.~L., {Sayeed}, M., {et~al.} 2024{\natexlab{a}}, arXiv e-prints, arXiv:2410.02962, \dodoi{10.48550/arXiv.2410.02962}

\bibitem[{{Lu} {et~al.}(2021{\natexlab{b}}){Lu}, {Angus}, {Curtis}, {David}, \& {Kiman}}]{Lu2020_kinage}
{Lu}, Y.~L., {Angus}, R., {Curtis}, J.~L., {David}, T.~J., \& {Kiman}, R. 2021{\natexlab{b}}, \aj, 161, 189, \dodoi{10.3847/1538-3881/abe4d6}

\bibitem[{{Lu} {et~al.}(2022){Lu}, {Curtis}, {Angus}, {David}, \& {Hattori}}]{Lu2022_prot}
{Lu}, Y.~L., {Curtis}, J.~L., {Angus}, R., {David}, T.~J., \& {Hattori}, S. 2022, \aj, 164, 251, \dodoi{10.3847/1538-3881/ac9bee}

\bibitem[{{Lu} {et~al.}(2024{\natexlab{b}}){Lu}, {See}, {Amard}, {Angus}, \& {Matt}}]{Lu2024_abrupt}
{Lu}, Y.~L., {See}, V., {Amard}, L., {Angus}, R., \& {Matt}, S.~P. 2024{\natexlab{b}}, Nature Astronomy, 8, 223, \dodoi{10.1038/s41550-023-02126-2}

\bibitem[{{Lurie} {et~al.}(2017){Lurie}, {Vyhmeister}, {Hawley}, {Adilia}, {Chen}, {Davenport}, {Juri{\'c}}, {Puig-Holzman}, \& {Weisenburger}}]{Lurie2017}
{Lurie}, J.~C., {Vyhmeister}, K., {Hawley}, S.~L., {et~al.} 2017, \aj, 154, 250, \dodoi{10.3847/1538-3881/aa974d}

\bibitem[{{Mackereth} {et~al.}(2019){Mackereth}, {Bovy}, {Leung}, {Schiavon}, {Trick}, {Chaplin}, {Cunha}, {Feuillet}, {Majewski}, {Martig}, {Miglio}, {Nidever}, {Pinsonneault}, {Aguirre}, {Sobeck}, {Tayar}, \& {Zasowski}}]{Mackereth2019}
{Mackereth}, J.~T., {Bovy}, J., {Leung}, H.~W., {et~al.} 2019, Monthly Notices of the RAS, 489, 176, \dodoi{10.1093/mnras/stz1521}

\bibitem[{{Majewski} {et~al.}(2017){Majewski}, {Schiavon}, {Frinchaboy}, {Allende Prieto}, {Barkhouser}, {Bizyaev}, {Blank}, {Brunner}, {Burton}, {Carrera}, {Chojnowski}, {Cunha}, {Epstein}, {Fitzgerald}, {Garc{\'\i}a P{\'e}rez}, {Hearty}, {Henderson}, {Holtzman}, {Johnson}, {Lam}, {Lawler}, {Maseman}, {M{\'e}sz{\'a}ros}, {Nelson}, {Nguyen}, {Nidever}, {Pinsonneault}, {Shetrone}, {Smee}, {Smith}, {Stolberg}, {Skrutskie}, {Walker}, {Wilson}, {Zasowski}, {Anders}, {Basu}, {Beland}, {Blanton}, {Bovy}, {Brownstein}, {Carlberg}, {Chaplin}, {Chiappini}, {Eisenstein}, {Elsworth}, {Feuillet}, {Fleming}, {Galbraith-Frew}, {Garc{\'\i}a}, {Garc{\'\i}a-Hern{\'a}ndez}, {Gillespie}, {Girardi}, {Gunn}, {Hasselquist}, {Hayden}, {Hekker}, {Ivans}, {Kinemuchi}, {Klaene}, {Mahadevan}, {Mathur}, {Mosser}, {Muna}, {Munn}, {Nichol}, {O'Connell}, {Parejko}, {Robin}, {Rocha-Pinto}, {Schultheis}, {Serenelli}, {Shane}, {Silva Aguirre}, {Sobeck}, {Thompson}, {Troup}, {Weinberg}, \& {Zamora}}]{apogee}
{Majewski}, S.~R., {Schiavon}, R.~P., {Frinchaboy}, P.~M., {et~al.} 2017, \aj, 154, 94, \dodoi{10.3847/1538-3881/aa784d}

\bibitem[{{Martig} {et~al.}(2016){Martig}, {Fouesneau}, {Rix}, {Ness}, {M{\'e}sz{\'a}ros}, {Garc{\'\i}a-Hern{\'a}ndez}, {Pinsonneault}, {Serenelli}, {Silva Aguirre}, \& {Zamora}}]{Martig2016}
{Martig}, M., {Fouesneau}, M., {Rix}, H.-W., {et~al.} 2016, \mnras, 456, 3655, \dodoi{10.1093/mnras/stv2830}

\bibitem[{{Masseron} {et~al.}(2017){Masseron}, {Lagarde}, {Miglio}, {Elsworth}, \& {Gilmore}}]{Masseron2017}
{Masseron}, T., {Lagarde}, N., {Miglio}, A., {Elsworth}, Y., \& {Gilmore}, G. 2017, \mnras, 464, 3021, \dodoi{10.1093/mnras/stw2632}

\bibitem[{McKinney {et~al.}(2010)}]{pandas}
McKinney, W., {et~al.} 2010, in Proceedings of the 9th Python in Science Conference, Vol. 445, Austin, TX, 51--56

\bibitem[{{McQuillan} {et~al.}(2014){McQuillan}, {Mazeh}, \& {Aigrain}}]{McQuillan2014}
{McQuillan}, A., {Mazeh}, T., \& {Aigrain}, S. 2014, \apjs, 211, 24, \dodoi{10.1088/0067-0049/211/2/24}

\bibitem[{{Metcalfe} {et~al.}(2024){Metcalfe}, {Strassmeier}, {Ilyin}, {Buzasi}, {Kochukhov}, {Ayres}, {Basu}, {Chontos}, {Finley}, {See}, {Stassun}, {van Saders}, {Sepulveda}, \& {Ricker}}]{Metcalfe2024}
{Metcalfe}, T.~S., {Strassmeier}, K.~G., {Ilyin}, I.~V., {et~al.} 2024, \apjl, 960, L6, \dodoi{10.3847/2041-8213/ad0a95}

\bibitem[{{Metcalfe} {et~al.}(2025){Metcalfe}, {Petit}, {van Saders}, {Ayres}, {Buzasi}, {Kochukhov}, {Stassun}, {Pinsonneault}, {Ilyin}, {Strassmeier}, {Finley}, {Garcia}, {Yuxi}, {Lu}, \& {See}}]{Metcalfe2025}
{Metcalfe}, T.~S., {Petit}, P., {van Saders}, J.~L., {et~al.} 2025, arXiv e-prints, arXiv:2501.19169, \dodoi{10.48550/arXiv.2501.19169}

\bibitem[{{Miglio} {et~al.}(2021){Miglio}, {Chiappini}, {Mackereth}, {Davies}, {Brogaard}, {Casagrande}, {Chaplin}, {Girardi}, {Kawata}, {Khan}, {Izzard}, {Montalb{\'a}n}, {Mosser}, {Vincenzo}, {Bossini}, {Noels}, {Rodrigues}, {Valentini}, \& {Mandel}}]{Miglio2021}
{Miglio}, A., {Chiappini}, C., {Mackereth}, J.~T., {et~al.} 2021, \aap, 645, A85, \dodoi{10.1051/0004-6361/202038307}

\bibitem[{{Minchev} {et~al.}(2014){Minchev}, {Chiappini}, {Martig}, {Steinmetz}, {de Jong}, {Boeche}, {Scannapieco}, {Zwitter}, {Wyse}, {Binney}, {Bland-Hawthorn}, {Bienaym{\'e}}, {Famaey}, {Freeman}, {Gibson}, {Grebel}, {Gilmore}, {Helmi}, {Kordopatis}, {Lee}, {Munari}, {Navarro}, {Parker}, {Quillen}, {Reid}, {Siebert}, {Siviero}, {Seabroke}, {Watson}, \& {Williams}}]{Minchev2014}
{Minchev}, I., {Chiappini}, C., {Martig}, M., {et~al.} 2014, \apjl, 781, L20, \dodoi{10.1088/2041-8205/781/1/L20}

\bibitem[{{Ness} {et~al.}(2016){Ness}, {Hogg}, {Rix}, {Martig}, {Pinsonneault}, \& {Ho}}]{Ness2016}
{Ness}, M., {Hogg}, D.~W., {Rix}, H.~W., {et~al.} 2016, \apj, 823, 114, \dodoi{10.3847/0004-637X/823/2/114}

\bibitem[{{Ochsenbein} {et~al.}(2000){Ochsenbein}, {Bauer}, \& {Marcout}}]{vizier}
{Ochsenbein}, F., {Bauer}, P., \& {Marcout}, J. 2000, \aaps, 143, 23, \dodoi{10.1051/aas:2000169}

\bibitem[{{Pinsonneault} {et~al.}(2018){Pinsonneault}, {Elsworth}, {Tayar}, {Serenelli}, {Stello}, {Zinn}, {Mathur}, {Garc{\'\i}a}, {Johnson}, {Hekker}, {Huber}, {Kallinger}, {M{\'e}sz{\'a}ros}, {Mosser}, {Stassun}, {Girardi}, {Rodrigues}, {Silva Aguirre}, {An}, {Basu}, {Chaplin}, {Corsaro}, {Cunha}, {Garc{\'\i}a-Hern{\'a}ndez}, {Holtzman}, {J{\"o}nsson}, {Shetrone}, {Smith}, {Sobeck}, {Stringfellow}, {Zamora}, {Beers}, {Fern{\'a}ndez-Trincado}, {Frinchaboy}, {Hearty}, \& {Nitschelm}}]{Pinsonneault2018}
{Pinsonneault}, M.~H., {Elsworth}, Y.~P., {Tayar}, J., {et~al.} 2018, Astrophysical Journals, 239, 32, \dodoi{10.3847/1538-4365/aaebfd}

\bibitem[{{Pinsonneault} {et~al.}(2025){Pinsonneault}, {Zinn}, {Tayar}, {Serenelli}, {Garc{\'\i}a}, {Mathur}, {Vrard}, {Elsworth}, {Mosser}, {Stello}, {Bell}, {Bugnet}, {Corsaro}, {Gaulme}, {Hekker}, {Hon}, {Huber}, {Kallinger}, {Cao}, {Johnson}, {Liagre}, {Patton}, {Santos}, {Basu}, {Beck}, {Beers}, {Chaplin}, {Cunha}, {Frinchaboy}, {Girardi}, {Godoy-Rivera}, {Holtzman}, {J{\"o}nsson}, {M{\'e}sz{\'a}ros}, {Reyes}, {Rix}, {Shetrone}, {Smith}, {Spoo}, {Stassun}, \& {Wang}}]{Pinsonneault2024}
{Pinsonneault}, M.~H., {Zinn}, J.~C., {Tayar}, J., {et~al.} 2025, \apjs, 276, 69, \dodoi{10.3847/1538-4365/ad9fef}

\bibitem[{{Price-Whelan}(2017)}]{gala2017}
{Price-Whelan}, A.~M. 2017, The Journal of Open Source Software, 2, 388, \dodoi{10.21105/joss.00388}

\bibitem[{{Price-Whelan} {et~al.}(2018){Price-Whelan}, {Sip{\H{o}}cz}, {G{\"u}nther}, {Lim}, {Crawford}, {Conseil}, {Shupe}, {Craig}, {Dencheva}, {Ginsburg}, {VanderPlas}, {Bradley}, {P{\'e}rez-Su{\'a}rez}, {de Val-Borro}, {Paper Contributors}, {Aldcroft}, {Cruz}, {Robitaille}, {Tollerud}, {Coordination Committee}, {Ardelean}, {Babej}, {Bach}, {Bachetti}, {Bakanov}, {Bamford}, {Barentsen}, {Barmby}, {Baumbach}, {Berry}, {Biscani}, {Boquien}, {Bostroem}, {Bouma}, {Brammer}, {Bray}, {Breytenbach}, {Buddelmeijer}, {Burke}, {Calderone}, {Cano Rodr{\'\i}guez}, {Cara}, {Cardoso}, {Cheedella}, {Copin}, {Corrales}, {Crichton}, {D{\textquoteright}Avella}, {Deil}, {Depagne}, {Dietrich}, {Donath}, {Droettboom}, {Earl}, {Erben}, {Fabbro}, {Ferreira}, {Finethy}, {Fox}, {Garrison}, {Gibbons}, {Goldstein}, {Gommers}, {Greco}, {Greenfield}, {Groener}, {Grollier}, {Hagen}, {Hirst}, {Homeier}, {Horton}, {Hosseinzadeh}, {Hu}, {Hunkeler}, {Ivezi{\'c}}, {Jain}, {Jenness}, {Kanarek}, {Kendrew}, {Kern}, {Kerzendorf}, {Khvalko},
  {King}, {Kirkby}, {Kulkarni}, {Kumar}, {Lee}, {Lenz}, {Littlefair}, {Ma}, {Macleod}, {Mastropietro}, {McCully}, {Montagnac}, {Morris}, {Mueller}, {Mumford}, {Muna}, {Murphy}, {Nelson}, {Nguyen}, {Ninan}, {N{\"o}the}, {Ogaz}, {Oh}, {Parejko}, {Parley}, {Pascual}, {Patil}, {Patil}, {Plunkett}, {Prochaska}, {Rastogi}, {Reddy Janga}, {Sabater}, {Sakurikar}, {Seifert}, {Sherbert}, {Sherwood-Taylor}, {Shih}, {Sick}, {Silbiger}, {Singanamalla}, {Singer}, {Sladen}, {Sooley}, {Sornarajah}, {Streicher}, {Teuben}, {Thomas}, {Tremblay}, {Turner}, {Terr{\'o}n}, {van Kerkwijk}, {de la Vega}, {Watkins}, {Weaver}, {Whitmore}, {Woillez}, {Zabalza}, \& {Contributors}}]{astropy:2018}
{Price-Whelan}, A.~M., {Sip{\H{o}}cz}, B.~M., {G{\"u}nther}, H.~M., {et~al.} 2018, \aj, 156, 123, \dodoi{10.3847/1538-3881/aabc4f}

\bibitem[{{Queiroz} {et~al.}(2023){Queiroz}, {Anders}, {Chiappini}, {Khalatyan}, {Santiago}, {Nepal}, {Steinmetz}, {Gallart}, {Valentini}, {Dal Ponte}, {Barbuy}, {P{\'e}rez-Villegas}, {Masseron}, {Fern{\'a}ndez-Trincado}, {Khoperskov}, {Minchev}, {Fern{\'a}ndez-Alvar}, {Lane}, \& {Nitschelm}}]{Queiroz2023}
{Queiroz}, A. B.~A., {Anders}, F., {Chiappini}, C., {et~al.} 2023, arXiv e-prints, arXiv:2303.09926, \dodoi{10.48550/arXiv.2303.09926}

\bibitem[{{Quinn} {et~al.}(1993){Quinn}, {Hernquist}, \& {Fullagar}}]{Quinn1993}
{Quinn}, P.~J., {Hernquist}, L., \& {Fullagar}, D.~P. 1993, \apj, 403, 74, \dodoi{10.1086/172184}

\bibitem[{{Ricker} {et~al.}(2015){Ricker}, {Winn}, {Vanderspek}, {Latham}, {Bakos}, {Bean}, {Berta-Thompson}, {Brown}, {Buchhave}, {Butler}, {Butler}, {Chaplin}, {Charbonneau}, {Christensen-Dalsgaard}, {Clampin}, {Deming}, {Doty}, {De Lee}, {Dressing}, {Dunham}, {Endl}, {Fressin}, {Ge}, {Henning}, {Holman}, {Howard}, {Ida}, {Jenkins}, {Jernigan}, {Johnson}, {Kaltenegger}, {Kawai}, {Kjeldsen}, {Laughlin}, {Levine}, {Lin}, {Lissauer}, {MacQueen}, {Marcy}, {McCullough}, {Morton}, {Narita}, {Paegert}, {Palle}, {Pepe}, {Pepper}, {Quirrenbach}, {Rinehart}, {Sasselov}, {Sato}, {Seager}, {Sozzetti}, {Stassun}, {Sullivan}, {Szentgyorgyi}, {Torres}, {Udry}, \& {Villasenor}}]{TESS}
{Ricker}, G.~R., {Winn}, J.~N., {Vanderspek}, R., {et~al.} 2015, Journal of Astronomical Telescopes, Instruments, and Systems, 1, 014003, \dodoi{10.1117/1.JATIS.1.1.014003}

\bibitem[{{Roberts} {et~al.}(2024){Roberts}, {Pinsonneault}, {Johnson}, {Zinn}, {Weinberg}, {Vrard}, {Tayar}, {Stello}, {Mosser}, {Johnson}, {Cao}, {Stassun}, {Stringfellow}, {Serenelli}, {Mathur}, {Hekker}, {Garc{\'\i}a}, {Elsworth}, \& {Corsaro}}]{Roberts2024}
{Roberts}, J.~D., {Pinsonneault}, M.~H., {Johnson}, J.~A., {et~al.} 2024, \mnras, 530, 149, \dodoi{10.1093/mnras/stae820}

\bibitem[{{Sagear} {et~al.}(2024){Sagear}, {Price-Whelan}, {Ballard}, {Lu}, {Angus}, \& {Hogg}}]{Sagear2024}
{Sagear}, S., {Price-Whelan}, A.~M., {Ballard}, S., {et~al.} 2024, \apj, 977, 49, \dodoi{10.3847/1538-4357/ad8b26}

\bibitem[{{Santos} {et~al.}(2021){Santos}, {Breton}, {Mathur}, \& {Garc{\'\i}a}}]{Santos2021}
{Santos}, A.~R.~G., {Breton}, S.~N., {Mathur}, S., \& {Garc{\'\i}a}, R.~A. 2021, \apjs, 255, 17, \dodoi{10.3847/1538-4365/ac033f}

\bibitem[{{Saunders} {et~al.}(2024){Saunders}, {van Saders}, {Lyttle}, {Metcalfe}, {Li}, {Davies}, {Hall}, {Ball}, {Townsend}, {Creevey}, \& {Dodds}}]{Saunders2024}
{Saunders}, N., {van Saders}, J.~L., {Lyttle}, A.~J., {et~al.} 2024, \apj, 962, 138, \dodoi{10.3847/1538-4357/ad1516}

\bibitem[{{Schonhut-Stasik} {et~al.}(2024){Schonhut-Stasik}, {Zinn}, {Stassun}, {Pinsonneault}, {Johnson}, {Warfield}, {Stello}, {Elsworth}, {Garc{\'\i}a}, {Mathur}, {Mosser}, {Hon}, {Tayar}, {Stringfellow}, {Beaton}, {J{\"o}nsson}, \& {Minniti}}]{APOK2_1}
{Schonhut-Stasik}, J., {Zinn}, J.~C., {Stassun}, K.~G., {et~al.} 2024, \aj, 167, 50, \dodoi{10.3847/1538-3881/ad0b13}

\bibitem[{{See} {et~al.}(2024){See}, {Lu}, {Amard}, \& {Roquette}}]{See2024}
{See}, V., {Lu}, Y.~L., {Amard}, L., \& {Roquette}, J. 2024, \mnras, 533, 1290, \dodoi{10.1093/mnras/stae1828}

\bibitem[{{Sellwood}(2014)}]{sellwood2014}
{Sellwood}, J.~A. 2014, Reviews of Modern Physics, 86, 1, \dodoi{10.1103/RevModPhys.86.1}

\bibitem[{{Sharma} {et~al.}(2021){Sharma}, {Hayden}, {Bland-Hawthorn}, {Stello}, {Buder}, {Zinn}, {Kallinger}, {Asplund}, {De Silva}, {D'Orazi}, {Freeman}, {Kos}, {Lewis}, {Lin}, {Lind}, {Martell}, {Simpson}, {Wittenmyer}, {Zucker}, {Zwitter}, {Chen}, {Cotar}, {Esdaile}, {Hon}, {Horner}, {Huber}, {Kafle}, {Khanna}, {Ting}, {Nataf}, {Nordlander}, {Saadon}, {Tepper-Garcia}, {Tinney}, {Traven}, {Watson}, {Wright}, \& {Wyse}}]{Sharma2021}
{Sharma}, S., {Hayden}, M.~R., {Bland-Hawthorn}, J., {et~al.} 2021, \mnras, 506, 1761, \dodoi{10.1093/mnras/stab1086}

\bibitem[{{Shetrone} {et~al.}(2019){Shetrone}, {Tayar}, {Johnson}, {Somers}, {Pinsonneault}, {Holtzman}, {Hasselquist}, {Masseron}, {M{\'e}sz{\'a}ros}, {J{\"o}nsson}, {Hawkins}, {Sobeck}, {Zamora}, \& {Garc{\'\i}a-Hern{\'a}ndez}}]{Shetrone2019}
{Shetrone}, M., {Tayar}, J., {Johnson}, J.~A., {et~al.} 2019, \apj, 872, 137, \dodoi{10.3847/1538-4357/aaff66}

\bibitem[{{Silva Aguirre} {et~al.}(2017){Silva Aguirre}, {Lund}, {Antia}, {Ball}, {Basu}, {Christensen-Dalsgaard}, {Lebreton}, {Reese}, {Verma}, {Casagrande}, {Justesen}, {Mosumgaard}, {Chaplin}, {Bedding}, {Davies}, {Handberg}, {Houdek}, {Huber}, {Kjeldsen}, {Latham}, {White}, {Coelho}, {Miglio}, \& {Rendle}}]{Silva2017}
{Silva Aguirre}, V., {Lund}, M.~N., {Antia}, H.~M., {et~al.} 2017, Astrophysical Journal, 835, 173, \dodoi{10.3847/1538-4357/835/2/173}

\bibitem[{{Silva Aguirre} {et~al.}(2020){Silva Aguirre}, {Christensen-Dalsgaard}, {Cassisi}, {Miller Bertolami}, {Serenelli}, {Stello}, {Weiss}, {Angelou}, {Jiang}, {Lebreton}, {Spada}, {Bellinger}, {Deheuvels}, {Ouazzani}, {Pietrinferni}, {Mosumgaard}, {Townsend}, {Battich}, {Bossini}, {Constantino}, {Eggenberger}, {Hekker}, {Mazumdar}, {Miglio}, {Nielsen}, \& {Salaris}}]{SilvaAguirre2020}
{Silva Aguirre}, V., {Christensen-Dalsgaard}, J., {Cassisi}, S., {et~al.} 2020, \aap, 635, A164, \dodoi{10.1051/0004-6361/201935843}

\bibitem[{{Skumanich}(1972)}]{Skumanich1972}
{Skumanich}, A. 1972, \apj, 171, 565, \dodoi{10.1086/151310}

\bibitem[{{Spergel} {et~al.}(2015){Spergel}, {Gehrels}, {Baltay}, {Bennett}, {Breckinridge}, {Donahue}, {Dressler}, {Gaudi}, {Greene}, {Guyon}, {Hirata}, {Kalirai}, {Kasdin}, {Macintosh}, {Moos}, {Perlmutter}, {Postman}, {Rauscher}, {Rhodes}, {Wang}, {Weinberg}, {Benford}, {Hudson}, {Jeong}, {Mellier}, {Traub}, {Yamada}, {Capak}, {Colbert}, {Masters}, {Penny}, {Savransky}, {Stern}, {Zimmerman}, {Barry}, {Bartusek}, {Carpenter}, {Cheng}, {Content}, {Dekens}, {Demers}, {Grady}, {Jackson}, {Kuan}, {Kruk}, {Melton}, {Nemati}, {Parvin}, {Poberezhskiy}, {Peddie}, {Ruffa}, {Wallace}, {Whipple}, {Wollack}, \& {Zhao}}]{Spergel2015}
{Spergel}, D., {Gehrels}, N., {Baltay}, C., {et~al.} 2015, arXiv e-prints, arXiv:1503.03757, \dodoi{10.48550/arXiv.1503.03757}

\bibitem[{{Spitzer} \& {Schwarzschild}(1951)}]{Spitzer1951}
{Spitzer}, Lyman, J., \& {Schwarzschild}, M. 1951, \apj, 114, 385, \dodoi{10.1086/145478}

\bibitem[{{Spoo} {et~al.}(2022){Spoo}, {Tayar}, {Frinchaboy}, {Cunha}, {Myers}, {Donor}, {Majewski}, {Bizyaev}, {Garc{\'\i}a-Hern{\'a}ndez}, {J{\"o}nsson}, {Lane}, {Pan}, {Longa-Pe{\~n}a}, \& {Roman-Lopes}}]{Spoo2022}
{Spoo}, T., {Tayar}, J., {Frinchaboy}, P.~M., {et~al.} 2022, \aj, 163, 229, \dodoi{10.3847/1538-3881/ac5d53}

\bibitem[{{Stone-Martinez} {et~al.}(2024){Stone-Martinez}, {Holtzman}, {Imig}, {Nitschelm}, {Stassun}, \& {Brownstein}}]{StoneMartinez2024}
{Stone-Martinez}, A., {Holtzman}, J.~A., {Imig}, J., {et~al.} 2024, \aj, 167, 73, \dodoi{10.3847/1538-3881/ad12a6}

\bibitem[{{Stone-Martinez} {et~al.}(2025){Stone-Martinez}, {Holtzman}, {Yuxi}, {Lu}, {Hasselquist}, {Imig}, {Griffith}, {Bellinger}, \& {Saydjari}}]{StoneMartinez2025}
{Stone-Martinez}, A., {Holtzman}, J.~A., {Yuxi}, {et~al.} 2025, arXiv e-prints, arXiv:2503.03138, \dodoi{10.48550/arXiv.2503.03138}

\bibitem[{{Sun} {et~al.}(2024){Sun}, {Huang}, {Shen}, {Wang}, {Zhang}, {Tian}, {Liu}, \& {Jiang}}]{Sun2024}
{Sun}, W., {Huang}, Y., {Shen}, H., {et~al.} 2024, \apj, 961, 141, \dodoi{10.3847/1538-4357/ad06ad}

\bibitem[{{Tayar} \& {Joyce}(2025)}]{Tayar2025}
{Tayar}, J., \& {Joyce}, M. 2025, \apjl, 984, L56, \dodoi{10.3847/2041-8213/adcd6f}

\bibitem[{{Tayar} {et~al.}(2017){Tayar}, {Somers}, {Pinsonneault}, {Stello}, {Mints}, {Johnson}, {Zamora}, {Garc{\'\i}a-Hern{\'a}ndez}, {Maraston}, {Serenelli}, {Allende Prieto}, {Bastien}, {Basu}, {Bird}, {Cohen}, {Cunha}, {Elsworth}, {Garc{\'\i}a}, {Girardi}, {Hekker}, {Holtzman}, {Huber}, {Mathur}, {M{\'e}sz{\'a}ros}, {Mosser}, {Shetrone}, {Silva Aguirre}, {Stassun}, {Stringfellow}, {Zasowski}, \& {Roman-Lopes}}]{Tayar2017}
{Tayar}, J., {Somers}, G., {Pinsonneault}, M.~H., {et~al.} 2017, \apj, 840, 17, \dodoi{10.3847/1538-4357/aa6a1e}

\bibitem[{{Ting}(2025)}]{Ting2025}
{Ting}, Y.-S. 2025, arXiv e-prints, arXiv:2506.12230, \dodoi{10.48550/arXiv.2506.12230}

\bibitem[{{Ting} \& {Rix}(2019)}]{Ting2019}
{Ting}, Y.-S., \& {Rix}, H.-W. 2019, \apj, 878, 21, \dodoi{10.3847/1538-4357/ab1ea5}

\bibitem[{{Townsend} \& {Teitler}(2013)}]{Townsend2013}
{Townsend}, R.~H.~D., \& {Teitler}, S.~A. 2013, \mnras, 435, 3406, \dodoi{10.1093/mnras/stt1533}

\bibitem[{{Van-Lane} {et~al.}(2024){Van-Lane}, {Speagle}, {Eadie}, {Douglas}, {Cargile}, {Zucker}, {Yuxi}, {Lu}, \& {Angus}}]{VanLane2024}
{Van-Lane}, P.~R., {Speagle}, J.~S., {Eadie}, G.~M., {et~al.} 2024, arXiv e-prints, arXiv:2412.12244, \dodoi{10.48550/arXiv.2412.12244}

\bibitem[{{van Saders} {et~al.}(2016){van Saders}, {Ceillier}, {Metcalfe}, {Silva Aguirre}, {Pinsonneault}, {Garc{\'\i}a}, {Mathur}, \& {Davies}}]{vansaders2016}
{van Saders}, J.~L., {Ceillier}, T., {Metcalfe}, T.~S., {et~al.} 2016, \nat, 529, 181, \dodoi{10.1038/nature16168}

\bibitem[{{Warfield} {et~al.}(2021){Warfield}, {Zinn}, {Pinsonneault}, {Johnson}, {Stello}, {Elsworth}, {Garc{\'\i}a}, {Kallinger}, {Mathur}, {Mosser}, {Beaton}, \& {Garc{\'\i}a-Hern{\'a}ndez}}]{Warfield_2021}
{Warfield}, J.~T., {Zinn}, J.~C., {Pinsonneault}, M.~H., {et~al.} 2021, \aj, 161, 100, \dodoi{10.3847/1538-3881/abd39d}

\bibitem[{{Weiss} {et~al.}(2025){Weiss}, {Downing}, {Pinsonneault}, {Zinn}, {Stello}, {Bedding}, {Cao}, {Hon}, {Reyes}, {Gaudi}, {Wilson}, {Huber}, \& {Sharma}}]{Weiss2025}
{Weiss}, T.~J., {Downing}, N.~J., {Pinsonneault}, M.~H., {et~al.} 2025, arXiv e-prints, arXiv:2503.04999, \dodoi{10.48550/arXiv.2503.04999}

\bibitem[{{Wenger} {et~al.}(2000){Wenger}, {Ochsenbein}, {Egret}, {Dubois}, {Bonnarel}, {Borde}, {Genova}, {Jasniewicz}, {Lalo{\"e}}, {Lesteven}, \& {Monier}}]{simbad}
{Wenger}, M., {Ochsenbein}, F., {Egret}, D., {et~al.} 2000, \aaps, 143, 9, \dodoi{10.1051/aas:2000332}

\bibitem[{Xiang \& Rix(2022)}]{Xiang2022}
Xiang, M., \& Rix, H.-W. 2022, Nature, 603, 599–603

\bibitem[{{Yu} \& {Liu}(2018)}]{Yu2018}
{Yu}, J., \& {Liu}, C. 2018, \mnras, 475, 1093, \dodoi{10.1093/mnras/stx3204}

\bibitem[{{Zhang} {et~al.}(2021){Zhang}, {Xiang}, {Zhang}, {Ting}, {Rix}, {Wu}, {Huang}, {Sun}, {Tian}, {Wang}, \& {Liu}}]{Zhang2021}
{Zhang}, M., {Xiang}, M., {Zhang}, H.-W., {et~al.} 2021, \apj, 922, 145, \dodoi{10.3847/1538-4357/ac22a5}

\end{thebibliography}
\bibliographystyle{aasjournal}



\end{document}